\documentclass[reprint,amsmath,amssymb,aps]{revtex4-2}
\usepackage[]{graphicx}
\usepackage{float}
\usepackage{dcolumn}
\usepackage{xfrac}
\usepackage{bm}
\usepackage{tikz}
\usepackage{cancel}
\usetikzlibrary{shapes.geometric}
\usetikzlibrary{arrows.meta,arrows}
\usetikzlibrary{quotes,angles}
\usepackage{multirow}
\usepackage{caption}
\UseRawInputEncoding
\usepackage{hyperref}
\hypersetup{
  colorlinks,
  citecolor=blue,
  linkcolor=blue,
  urlcolor=blue}
\usepackage[utf8x]{inputenc}

\DeclareUnicodeCharacter{2212}{\ensuremath{-}}

\begin{document}

\preprint{APS/123-QED}

\title{Asymmetries in invisible Dark Matter mediator production \\ associated  with $t \bar{t}$ final states}

\author{E. Chalbaud$^{1}$}
\email{e.chalbaud@cern.ch}

\author{Rui M. Silva$^{1,2}$}
\email{rui.miguel.silva@coimbra.lip.pt}

\author{António Onofre$^{1,2}$}
\email{antonio.onofre@cern.ch}

\author{Ricardo Gonçalo$^{1}$}
\email{jose.goncalo@cern.ch}

\author{Miguel C. N. Fiolhais$^{1,3,4}$}
\email{miguel.fiolhais@cern.ch}

\affiliation{
{$^1$\sl Laboratório de Instrumentação e Física Experimental de Partículas, Department of Physics, University of Coimbra, 3004-516 Coimbra, Portugal}\\
{$^2$\sl Centro de Física da Universidade do Minho e Universidade do Porto (CF-UM-UP), Universidade do Minho, 4710-057 Braga, Portugal}\\
{$^3$\sl Science Department, Borough of Manhattan Community College, The City University of New York, 199 Chambers St, New York, NY 10007, USA}\\
{$^4$\sl The Graduate School and University Center, The City University of New York, 365 Fifth Avenue, New York, New York 10016, USA}
}

\date{\today}

\begin{abstract}

In this paper, we propose two sets of different CP-sensitive observables inspired by the Higgs production in association with the top quark. We employ a Dark Matter simplified model that couples a scalar mediator with top quarks. The reconstruction of the kinematic variables is presented at NLO accuracy for events associated with this massive scalar particle, which is assumed to be vanishing to invisible decays in a detector such as ATLAS. We build these observables by taking advantage of the similarity between the scalar coupling with the top quark and the factorization theorem in the total scattering amplitude, in order to represent the basis in which the phase space is parameterized. A twofold approach employs the direct implementation of the four-momentum phase space measure in building CP sensitive observables such as $b_{2}$ for the Higgs, and the spin polarization of the top-quark decays in the narrow width approximation for the employed model. We studied the asymmetries of these distributions to test for any improvement in increasing the exclusion region for the $g_{u_{33}}^S-g_{u_{33}}^P$ parameters associated with this vanishing scalar particle. We have found no significant effect in the exclusion limits by using the forward-backward asymmetry distributions and the full-shaped ones. Considering the case of an invisible mediator with mass of 10$^{-2}$~GeV for a luminosity of 300~fb$^{-1}$ expected at the end of Run 3, the best limits for $g_{u_{33}}^S$ and $g_{u_{33}}^P$ at NLO accuracy were obtained using the variables $\tilde{b}_{2}^{\widehat{y}}$ and $b_{2}$ respectively, with corresponding limits set to $[−0.0425, 0.0425] $ and $[−0.83, 0.83]$ at $68\%$ CL. 
\end{abstract}
\maketitle
\section{Introduction}

A plethora of phenomena in astrophysics, such as gravitational lensing, galaxies' rotational curves, collisions of intergalactic clusters, and the evolution of the large-scale structure of the Universe~\cite{drees2012minireview,10.1093/mnras/249.3.523,Corbelli_2000,hoekstra2002nf,rubin1970rotation,moustakas2002iz}, offer persuasive indirect proof for the existence of dark matter (DM). However, the particle nature of DM is still not fully understood. Theoretical enlargements to the Standard Model suggest a "dark sector" comprising particles that interact minimally with known standard model particles, leading to the proposition of weakly interacting massive particles (WIMPs) as a potential DM candidate. For example, the intriguing coincidence that a stable WIMP of a specific mass could naturally explain the observed relic density of DM observed by Planck measurements~\cite{2020}, is a fundamental element in the search for the identification of DM particles. Although the Large Hadron Collider (LHC) offers an unmatched opportunity to generate and analyze DM particles, conventional detection techniques encounter considerable difficulties due to the expected weak interactions of DM with regular matter~\cite{atlascollaboration2023combination}.

Addressing the challenges linked to the weakly interactive character of DM particles, the Large Hadron Collider (LHC) has implemented novel methods for their identification. DM detection is not based on direct observation, but rather on missing transverse momentum in proton-proton collisions~\cite{GonzalezFernandez:2021nbb,Kumar:2022knu,ATLAS:2023tkt}, suggesting the creation of undetectable particles. This approach has led to an emphasis on "mono-X" event topologies, with "X" representing an observable particle like a jet, Z/W boson, or photon, which can create the necessary contrast to the invisible DM particle~\cite{Haisch:2021ugv,Hermann:2021xvs}. To make these studies more manageable, simplified DM models have gained popularity over the effective field theory approach ~\cite{schumann2019direct}. By introducing a limited number of new particles and couplings, these simplified models allow for the examination of specific attributes and interactions of DM~\cite{Milosevic:2021bpv}. This focused method enables a more straightforward link between the experimental findings and the characteristics of the candidates for DM, thus optimizing the search and improving our understanding of where these elusive particles may fit within the cosmic framework~\cite{2020,weinheimer2003laboratory}.

Numerous simplified models have been proposed, indicating that DM could be an isolated heavy entity that only weakly interacts with the recognized particles of the Standard Model (SM). These interactions are theorized to be enabled by a novel particle mediator. At energy scales lower than the mediator mass, its interactions appear as point-like, thus permitting the use of effective field theory (EFT) to detail the interaction through higher-dimensional operators~\cite{Abdallah:2015ter,Degrande:2014vpa}. This method has been popular and validated by experiments such as ATLAS, CMS, and LHCb especially in the context of supersymmetry searches~\cite{Arkani-Hamed:2005qjb,ATLAS:2022ygn}. On the other hand, at the LHC where the energy scales can match or exceed the mediator’s mass, the tangible effects of the mediator require a detailed quantum field theory that explicitly includes the mediator in its particle scope. The ability of the LHC to investigate a wide range of DM and mediator masses, in addition to various intensities and spin configurations of their couplings, is crucial.

The primary objectives of this study are to rigorously investigate CP-violating observables in the context of dileptonic decays of the top quarks, drawn from established literature on the Higgs sector~\cite{Biekotter:2022ckj, Bernreuther_1994, Gunion_1996, Bhupal_Dev_2008, Frederix_2011, Ellis_2014, Khatibi_2014, Demartin_2014, Kobakhidze_2014, Bramante_2014, Boudjema_2015, He_2015, Amor_dos_Santos_2015, Amor_dos_Santos_2017, Gritsan_2016, Dolan_2016, Gon_alves_2016, Gon_alves_2018, Gon_alves_2022, Buckley_2016_v2, Buckley_2016, Mileo_2016, Azevedo_2018, Azevedo_2020, Azevedo_2021, Azevedo_2022, Li_2018, Ferroglia_2019, Faroughy_2020, Cao:2020hhb, Barman_2022, Aguilar_Saavedra_2022, Casler_2019, D_liot_2018, D_liot_2019, Broggio_2017, Bortolato_2021, Milosevic:2021bpv}
, in order to improve the signal-to-background ratio in the context of DM detection. Secondly, a thorough analysis is presented covering the mass spectrum from 10$^{-2}$~GeV to 125~GeV (the Higgs mass scale), delineating a strategic detection window for DM particles. Finally, we determine the sensitivity of these observables with respect to the future luminosity scenarios expected at the LHC, both during Run 3 and at the projected end of its service. This offers a clearer understanding of the potential of LHC to probe the frontiers of DM in the years to come.

The composition of this paper is as follows. We outline the simplified DM model and define the relevant parameters and angular observables in Section~\ref{sec:TH}. The methodology for event generation and simulation is detailed in Section~\ref{sec:simulation}. Section~\ref{sec:cpsensitive} shows the definitions for the CP sensitive observables. Section~\ref{sec:asymmetries} explores the asymmetries of such observables. Exclusion limits are shown in Section~\ref{sec:exclusion_limits} where we have included the results for the total cross-section as a function of the luminosity. We draw our central conclusions in Section~\ref{sec:conclusions}.

\section{Simplified Dark Matter model \label{sec:TH}}

At colliders, the production of DM was searched for by ATLAS and CMS through the $s$ and $t$-channel processes in the context of simplified and beyond simplified models~\cite{ATLAS:2021hza,ATLAS:2019zrq,CMS:2024zqs}. Our choice here was to study the s-channel exchange in simplified models~\cite{kentarou2015higher,Demartin_2014}, which should give a sensible cross-section at the LHC for SM-like couplings, when heavy fermions are present in the final state. However, if couplings are not so strong, it becomes challenging to discriminate the signals from the expected backgrounds originated in SM processes. In the case of scalar-like mediators, the most general Lagrangian density considers a Yukawa coupling structure of the DM mediator ($Y_{0}$) to fermionic DM particles ($X_{D}$) of the form 

\begin{equation}\label{eq:lagrangian_dm}
    \mathcal{L} = \bar{X}_{D}(g^{s} + ig^{p}\gamma^{5})X_{D}Y_{0},
\end{equation}

\noindent
where $g^{s}$ ($g^{p}$) corresponds to the scalar (pseudoscalar) coupling of $Y_{0}$ to $X_{D}$. Additionally, the mediator couplings to the SM particles~\cite{kentarou2015higher} are described by the following Lagrangian density,

\begin{equation}\label{eq:lagrangian_yukawa}
\begin{split}
    \mathcal{L} = \sum_{i,j}\bigg[ & \bar{d}_{i}\frac{y^{d}_{i,j}}{\sqrt{2}}(g^{s}_{d_{ij}} + i g^{p}_{d_{ij}}\gamma^{5}  )d_{j} \\
    & + \bar{u}_{i}\frac{y^{u}_{i,j}}{\sqrt{2}}(g^{s}_{u_{ij}} + i g^{p}_{u_{ij}}\gamma^{5}  )u_{j}\bigg]Y_{0},
\end{split}
\end{equation}

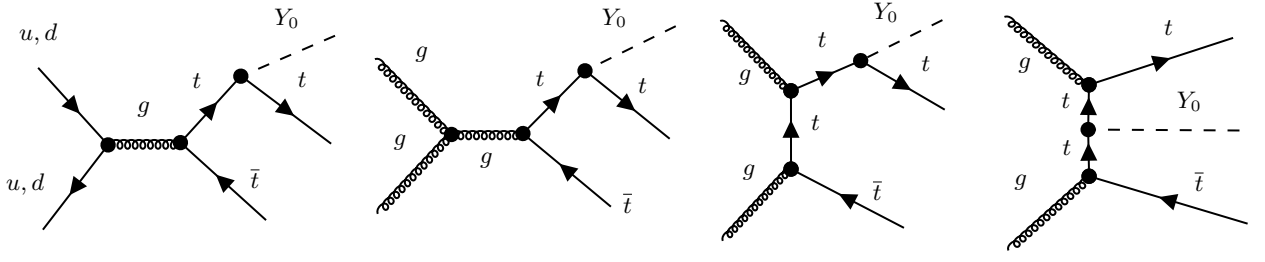
\begin{figure*}
    \centering
\tikzset{every picture/.style={line width=0.75pt}} 

\begin{tikzpicture}[x=0.75pt,y=0.75pt,yscale=-1,xscale=1]

\draw    (28,42.75) -- (63.1,81.46) ;
\draw [shift={(63.1,81.46)}, rotate = 47.8] [color={rgb, 255:red, 0; green, 0; blue, 0 }  ][fill={rgb, 255:red, 0; green, 0; blue, 0 }  ][line width=0.75]      (0, 0) circle [x radius= 3.35, y radius= 3.35]   ;
\draw [shift={(48.91,65.81)}, rotate = 227.8] [fill={rgb, 255:red, 0; green, 0; blue, 0 }  ][line width=0.08]  [draw opacity=0] (8.93,-4.29) -- (0,0) -- (8.93,4.29) -- cycle    ;
\draw    (142.38,119.21) -- (99.4,80.31) ;
\draw [shift={(99.4,80.31)}, rotate = 222.15] [color={rgb, 255:red, 0; green, 0; blue, 0 }  ][fill={rgb, 255:red, 0; green, 0; blue, 0 }  ][line width=0.75]      (0, 0) circle [x radius= 3.35, y radius= 3.35]   ;
\draw [shift={(117.18,96.41)}, rotate = 42.15] [fill={rgb, 255:red, 0; green, 0; blue, 0 }  ][line width=0.08]  [draw opacity=0] (8.93,-4.29) -- (0,0) -- (8.93,4.29) -- cycle    ;
\draw    (99.4,80.31) -- (129.72,46.97) ;
\draw [shift={(129.72,46.97)}, rotate = 312.29] [color={rgb, 255:red, 0; green, 0; blue, 0 }  ][fill={rgb, 255:red, 0; green, 0; blue, 0 }  ][line width=0.75]      (0, 0) circle [x radius= 3.35, y radius= 3.35]   ;
\draw [shift={(117.92,59.94)}, rotate = 132.29] [fill={rgb, 255:red, 0; green, 0; blue, 0 }  ][line width=0.08]  [draw opacity=0] (8.93,-4.29) -- (0,0) -- (8.93,4.29) -- cycle    ;
\draw    (129.96,46.97) -- (175.05,80.85) ;
\draw [shift={(156.5,66.91)}, rotate = 216.92] [fill={rgb, 255:red, 0; green, 0; blue, 0 }  ][line width=0.08]  [draw opacity=0] (8.93,-4.29) -- (0,0) -- (8.93,4.29) -- cycle    ;
\draw  [dash pattern={on 4.5pt off 4.5pt}]  (177,26.66) -- (129.96,46.97) ;
\draw    (63.1,81.46) -- (30.39,124) ;
\draw [shift={(43.7,106.69)}, rotate = 307.56] [fill={rgb, 255:red, 0; green, 0; blue, 0 }  ][line width=0.08]  [draw opacity=0] (8.93,-4.29) -- (0,0) -- (8.93,4.29) -- cycle    ;
\draw   (64.06,81.17) .. controls (64.3,80.07) and (65.25,78.97) .. (67.16,78.97) .. controls (70.98,78.97) and (70.98,83.38) .. (69.07,83.38) .. controls (67.16,83.38) and (67.16,78.97) .. (70.98,78.97) .. controls (74.8,78.97) and (74.8,83.38) .. (72.89,83.38) .. controls (70.98,83.38) and (70.98,78.97) .. (74.8,78.97) .. controls (78.62,78.97) and (78.62,83.38) .. (76.71,83.38) .. controls (74.8,83.38) and (74.8,78.97) .. (78.62,78.97) .. controls (82.44,78.97) and (82.44,83.38) .. (80.53,83.38) .. controls (78.62,83.38) and (78.62,78.97) .. (82.44,78.97) .. controls (86.27,78.97) and (86.27,83.38) .. (84.36,83.38) .. controls (82.44,83.38) and (82.44,78.97) .. (86.27,78.97) .. controls (90.09,78.97) and (90.09,83.38) .. (88.18,83.38) .. controls (86.27,83.38) and (86.27,78.97) .. (90.09,78.97) .. controls (93.91,78.97) and (93.91,83.38) .. (92,83.38) .. controls (90.09,83.38) and (90.09,78.97) .. (93.91,78.97) .. controls (97.73,78.97) and (97.73,83.38) .. (95.82,83.38) .. controls (93.91,83.38) and (93.91,78.97) .. (97.73,78.97) .. controls (97.98,78.97) and (98.22,78.99) .. (98.44,79.03) ;
\draw    (315.45,112.46) -- (271.3,75.74) ;
\draw [shift={(271.3,75.74)}, rotate = 219.75] [color={rgb, 255:red, 0; green, 0; blue, 0 }  ][fill={rgb, 255:red, 0; green, 0; blue, 0 }  ][line width=0.75]      (0, 0) circle [x radius= 3.35, y radius= 3.35]   ;
\draw [shift={(289.53,90.91)}, rotate = 39.75] [fill={rgb, 255:red, 0; green, 0; blue, 0 }  ][line width=0.08]  [draw opacity=0] (8.93,-4.29) -- (0,0) -- (8.93,4.29) -- cycle    ;
\draw    (271.3,75.74) -- (302.45,44.27) ;
\draw [shift={(302.45,44.27)}, rotate = 314.7] [color={rgb, 255:red, 0; green, 0; blue, 0 }  ][fill={rgb, 255:red, 0; green, 0; blue, 0 }  ][line width=0.75]      (0, 0) circle [x radius= 3.35, y radius= 3.35]   ;
\draw [shift={(290.39,56.45)}, rotate = 134.7] [fill={rgb, 255:red, 0; green, 0; blue, 0 }  ][line width=0.08]  [draw opacity=0] (8.93,-4.29) -- (0,0) -- (8.93,4.29) -- cycle    ;
\draw    (302.7,44.27) -- (344.96,78.37) ;
\draw [shift={(327.72,64.46)}, rotate = 218.9] [fill={rgb, 255:red, 0; green, 0; blue, 0 }  ][line width=0.08]  [draw opacity=0] (8.93,-4.29) -- (0,0) -- (8.93,4.29) -- cycle    ;
\draw  [dash pattern={on 4.5pt off 4.5pt}]  (348,21.87) -- (302.45,44.27) ;
\draw   (235,76.56) .. controls (235.25,75.52) and (236.23,74.48) .. (238.19,74.48) .. controls (242.11,74.48) and (242.11,78.64) .. (240.15,78.64) .. controls (238.19,78.64) and (238.19,74.48) .. (242.11,74.48) .. controls (246.04,74.48) and (246.04,78.64) .. (244.08,78.64) .. controls (242.11,78.64) and (242.11,74.48) .. (246.04,74.48) .. controls (249.96,74.48) and (249.96,78.64) .. (248,78.64) .. controls (246.04,78.64) and (246.04,74.48) .. (249.96,74.48) .. controls (253.89,74.48) and (253.89,78.64) .. (251.92,78.64) .. controls (249.96,78.64) and (249.96,74.48) .. (253.89,74.48) .. controls (257.81,74.48) and (257.81,78.64) .. (255.85,78.64) .. controls (253.89,78.64) and (253.89,74.48) .. (257.81,74.48) .. controls (261.73,74.48) and (261.73,78.64) .. (259.77,78.64) .. controls (257.81,78.64) and (257.81,74.48) .. (261.73,74.48) .. controls (265.66,74.48) and (265.66,78.64) .. (263.7,78.64) .. controls (261.73,78.64) and (261.73,74.48) .. (265.66,74.48) .. controls (269.58,74.48) and (269.58,78.64) .. (267.62,78.64) .. controls (265.66,78.64) and (265.66,74.48) .. (269.58,74.48) .. controls (269.85,74.48) and (270.09,74.49) .. (270.32,74.53) ;
\draw   (197.03,38.48) .. controls (197.93,38.02) and (199.35,38.1) .. (200.76,39.54) .. controls (203.59,42.43) and (200.71,44.99) .. (199.3,43.55) .. controls (197.88,42.1) and (200.76,39.54) .. (203.59,42.43) .. controls (206.41,45.31) and (203.53,47.88) .. (202.12,46.43) .. controls (200.71,44.99) and (203.59,42.43) .. (206.41,45.31) .. controls (209.23,48.2) and (206.35,50.76) .. (204.94,49.32) .. controls (203.53,47.88) and (206.41,45.31) .. (209.23,48.2) .. controls (212.05,51.09) and (209.17,53.65) .. (207.76,52.21) .. controls (206.35,50.76) and (209.23,48.2) .. (212.05,51.09) .. controls (214.87,53.97) and (211.99,56.54) .. (210.58,55.09) .. controls (209.17,53.65) and (212.05,51.09) .. (214.87,53.97) .. controls (217.69,56.86) and (214.81,59.42) .. (213.4,57.98) .. controls (211.99,56.54) and (214.87,53.97) .. (217.69,56.86) .. controls (220.51,59.75) and (217.63,62.31) .. (216.22,60.87) .. controls (214.81,59.42) and (217.69,56.86) .. (220.51,59.75) .. controls (223.33,62.63) and (220.45,65.2) .. (219.04,63.75) .. controls (217.63,62.31) and (220.51,59.75) .. (223.33,62.63) .. controls (226.15,65.52) and (223.27,68.08) .. (221.86,66.64) .. controls (220.45,65.2) and (223.33,62.63) .. (226.15,65.52) .. controls (228.97,68.41) and (226.09,70.97) .. (224.68,69.53) .. controls (223.27,68.08) and (226.15,65.52) .. (228.97,68.41) .. controls (231.8,71.29) and (228.92,73.86) .. (227.51,72.41) .. controls (226.09,70.97) and (228.97,68.41) .. (231.8,71.29) .. controls (234.62,74.18) and (231.74,76.74) .. (230.33,75.3) .. controls (228.92,73.86) and (231.8,71.29) .. (234.62,74.18) .. controls (235.47,75.05) and (235.8,75.9) .. (235.81,76.61) ;
\draw   (198.48,116.05) .. controls (198.05,115.1) and (198.15,113.59) .. (199.52,112.12) .. controls (202.28,109.16) and (204.66,112.22) .. (203.29,113.7) .. controls (201.91,115.18) and (199.52,112.12) .. (202.28,109.16) .. controls (205.04,106.21) and (207.42,109.27) .. (206.04,110.74) .. controls (204.66,112.22) and (202.28,109.16) .. (205.04,106.21) .. controls (207.8,103.25) and (210.18,106.31) .. (208.8,107.79) .. controls (207.42,109.27) and (205.04,106.21) .. (207.8,103.25) .. controls (210.55,100.3) and (212.94,103.36) .. (211.56,104.84) .. controls (210.18,106.31) and (207.8,103.25) .. (210.55,100.3) .. controls (213.31,97.34) and (215.69,100.4) .. (214.32,101.88) .. controls (212.94,103.36) and (210.55,100.3) .. (213.31,97.34) .. controls (216.07,94.39) and (218.45,97.45) .. (217.07,98.93) .. controls (215.69,100.4) and (213.31,97.34) .. (216.07,94.39) .. controls (218.82,91.43) and (221.21,94.49) .. (219.83,95.97) .. controls (218.45,97.45) and (216.07,94.39) .. (218.82,91.43) .. controls (221.58,88.48) and (223.97,91.54) .. (222.59,93.02) .. controls (221.21,94.49) and (218.82,91.43) .. (221.58,88.48) .. controls (224.34,85.52) and (226.72,88.58) .. (225.34,90.06) .. controls (223.97,91.54) and (221.58,88.48) .. (224.34,85.52) .. controls (227.1,82.57) and (229.48,85.63) .. (228.1,87.11) .. controls (226.72,88.58) and (224.34,85.52) .. (227.1,82.57) .. controls (229.85,79.61) and (232.24,82.67) .. (230.86,84.15) .. controls (229.48,85.63) and (227.1,82.57) .. (229.85,79.61) .. controls (232.61,76.66) and (234.99,79.72) .. (233.62,81.2) .. controls (232.24,82.67) and (229.85,79.61) .. (232.61,76.66) .. controls (233.44,75.77) and (234.24,75.42) .. (234.92,75.42) ;
\draw    (235.47,76.22) ;
\draw [shift={(235.47,76.22)}, rotate = 0] [color={rgb, 255:red, 0; green, 0; blue, 0 }  ][fill={rgb, 255:red, 0; green, 0; blue, 0 }  ][line width=0.75]      (0, 0) circle [x radius= 3.35, y radius= 3.35]   ;
\draw    (635,120.98) -- (555.44,97.18) ;
\draw [shift={(555.44,97.18)}, rotate = 196.65] [color={rgb, 255:red, 0; green, 0; blue, 0 }  ][fill={rgb, 255:red, 0; green, 0; blue, 0 }  ][line width=0.75]      (0, 0) circle [x radius= 3.35, y radius= 3.35]   ;
\draw [shift={(590.43,107.65)}, rotate = 16.65] [fill={rgb, 255:red, 0; green, 0; blue, 0 }  ][line width=0.08]  [draw opacity=0] (8.93,-4.29) -- (0,0) -- (8.93,4.29) -- cycle    ;
\draw    (555.44,97.18) -- (554.97,73.76) ;
\draw [shift={(554.97,73.76)}, rotate = 268.87] [color={rgb, 255:red, 0; green, 0; blue, 0 }  ][fill={rgb, 255:red, 0; green, 0; blue, 0 }  ][line width=0.75]      (0, 0) circle [x radius= 3.35, y radius= 3.35]   ;
\draw [shift={(555.11,80.47)}, rotate = 88.87] [fill={rgb, 255:red, 0; green, 0; blue, 0 }  ][line width=0.08]  [draw opacity=0] (8.93,-4.29) -- (0,0) -- (8.93,4.29) -- cycle    ;
\draw    (555.39,51.31) -- (627.72,27.74) ;
\draw [shift={(596.31,37.98)}, rotate = 161.95] [fill={rgb, 255:red, 0; green, 0; blue, 0 }  ][line width=0.08]  [draw opacity=0] (8.93,-4.29) -- (0,0) -- (8.93,4.29) -- cycle    ;
\draw  [dash pattern={on 4.5pt off 4.5pt}]  (630.92,74.22) -- (554.97,73.76) ;
\draw    (555.39,51.31) ;
\draw [shift={(555.39,51.31)}, rotate = 0] [color={rgb, 255:red, 0; green, 0; blue, 0 }  ][fill={rgb, 255:red, 0; green, 0; blue, 0 }  ][line width=0.75]      (0, 0) circle [x radius= 3.35, y radius= 3.35]   ;
\draw    (555.44,97.18) ;
\draw [shift={(555.44,97.18)}, rotate = 0] [color={rgb, 255:red, 0; green, 0; blue, 0 }  ][fill={rgb, 255:red, 0; green, 0; blue, 0 }  ][line width=0.75]      (0, 0) circle [x radius= 3.35, y radius= 3.35]   ;
\draw    (554.97,73.76) -- (555.39,51.31) ;
\draw [shift={(555.39,51.31)}, rotate = 271.05] [color={rgb, 255:red, 0; green, 0; blue, 0 }  ][fill={rgb, 255:red, 0; green, 0; blue, 0 }  ][line width=0.75]      (0, 0) circle [x radius= 3.35, y radius= 3.35]   ;
\draw [shift={(555.27,57.54)}, rotate = 91.05] [fill={rgb, 255:red, 0; green, 0; blue, 0 }  ][line width=0.08]  [draw opacity=0] (8.93,-4.29) -- (0,0) -- (8.93,4.29) -- cycle    ;
\draw    (462.53,123.08) -- (405.85,93.59) ;
\draw [shift={(405.85,93.59)}, rotate = 207.49] [color={rgb, 255:red, 0; green, 0; blue, 0 }  ][fill={rgb, 255:red, 0; green, 0; blue, 0 }  ][line width=0.75]      (0, 0) circle [x radius= 3.35, y radius= 3.35]   ;
\draw [shift={(429.75,106.03)}, rotate = 27.49] [fill={rgb, 255:red, 0; green, 0; blue, 0 }  ][line width=0.08]  [draw opacity=0] (8.93,-4.29) -- (0,0) -- (8.93,4.29) -- cycle    ;
\draw    (405.85,93.59) -- (405.82,54.31) ;
\draw [shift={(405.82,54.31)}, rotate = 269.96] [color={rgb, 255:red, 0; green, 0; blue, 0 }  ][fill={rgb, 255:red, 0; green, 0; blue, 0 }  ][line width=0.75]      (0, 0) circle [x radius= 3.35, y radius= 3.35]   ;
\draw [shift={(405.83,68.95)}, rotate = 89.96] [fill={rgb, 255:red, 0; green, 0; blue, 0 }  ][line width=0.08]  [draw opacity=0] (8.93,-4.29) -- (0,0) -- (8.93,4.29) -- cycle    ;
\draw    (405.82,54.31) -- (440.8,39.17) ;
\draw [shift={(427.9,44.75)}, rotate = 156.6] [fill={rgb, 255:red, 0; green, 0; blue, 0 }  ][line width=0.08]  [draw opacity=0] (8.93,-4.29) -- (0,0) -- (8.93,4.29) -- cycle    ;
\draw  [dash pattern={on 4.5pt off 4.5pt}]  (481.83,18.82) -- (440.8,39.17) ;
\draw    (405.82,54.31) ;
\draw [shift={(405.82,54.31)}, rotate = 0] [color={rgb, 255:red, 0; green, 0; blue, 0 }  ][fill={rgb, 255:red, 0; green, 0; blue, 0 }  ][line width=0.75]      (0, 0) circle [x radius= 3.35, y radius= 3.35]   ;
\draw    (405.85,93.59) ;
\draw [shift={(405.85,93.59)}, rotate = 0] [color={rgb, 255:red, 0; green, 0; blue, 0 }  ][fill={rgb, 255:red, 0; green, 0; blue, 0 }  ][line width=0.75]      (0, 0) circle [x radius= 3.35, y radius= 3.35]   ;
\draw    (440.8,39.17) -- (483,63.77) ;
\draw [shift={(466.22,53.99)}, rotate = 210.24] [fill={rgb, 255:red, 0; green, 0; blue, 0 }  ][line width=0.08]  [draw opacity=0] (8.93,-4.29) -- (0,0) -- (8.93,4.29) -- cycle    ;
\draw [shift={(440.8,39.17)}, rotate = 30.24] [color={rgb, 255:red, 0; green, 0; blue, 0 }  ][fill={rgb, 255:red, 0; green, 0; blue, 0 }  ][line width=0.75]      (0, 0) circle [x radius= 3.35, y radius= 3.35]   ;
\draw   (370.31,19.3) .. controls (371.13,18.88) and (372.43,18.96) .. (373.71,20.27) .. controls (376.28,22.9) and (373.66,25.24) .. (372.37,23.92) .. controls (371.09,22.61) and (373.71,20.27) .. (376.28,22.9) .. controls (378.85,25.53) and (376.23,27.87) .. (374.94,26.55) .. controls (373.66,25.24) and (376.28,22.9) .. (378.85,25.53) .. controls (381.42,28.17) and (378.79,30.5) .. (377.51,29.18) .. controls (376.23,27.87) and (378.85,25.53) .. (381.42,28.17) .. controls (383.99,30.8) and (381.36,33.13) .. (380.08,31.82) .. controls (378.79,30.5) and (381.42,28.17) .. (383.99,30.8) .. controls (386.55,33.43) and (383.93,35.76) .. (382.65,34.45) .. controls (381.36,33.13) and (383.99,30.8) .. (386.55,33.43) .. controls (389.12,36.06) and (386.5,38.39) .. (385.22,37.08) .. controls (383.93,35.76) and (386.55,33.43) .. (389.12,36.06) .. controls (391.69,38.69) and (389.07,41.02) .. (387.79,39.71) .. controls (386.5,38.39) and (389.12,36.06) .. (391.69,38.69) .. controls (394.26,41.32) and (391.64,43.65) .. (390.35,42.34) .. controls (389.07,41.02) and (391.69,38.69) .. (394.26,41.32) .. controls (396.83,43.95) and (394.21,46.28) .. (392.92,44.97) .. controls (391.64,43.65) and (394.26,41.32) .. (396.83,43.95) .. controls (399.4,46.58) and (396.78,48.91) .. (395.49,47.6) .. controls (394.21,46.28) and (396.83,43.95) .. (399.4,46.58) .. controls (401.97,49.21) and (399.35,51.54) .. (398.06,50.23) .. controls (396.78,48.91) and (399.4,46.58) .. (401.97,49.21) .. controls (404.54,51.84) and (401.91,54.18) .. (400.63,52.86) .. controls (399.35,51.54) and (401.97,49.21) .. (404.54,51.84) .. controls (405.31,52.64) and (405.62,53.41) .. (405.62,54.05) ;
\draw   (371.63,129.61) .. controls (371.24,128.74) and (371.32,127.37) .. (372.58,126.02) .. controls (375.09,123.33) and (377.26,126.12) .. (376.01,127.47) .. controls (374.75,128.81) and (372.58,126.02) .. (375.09,123.33) .. controls (377.6,120.64) and (379.77,123.43) .. (378.52,124.78) .. controls (377.26,126.12) and (375.09,123.33) .. (377.6,120.64) .. controls (380.11,117.95) and (382.28,120.74) .. (381.03,122.08) .. controls (379.77,123.43) and (377.6,120.64) .. (380.11,117.95) .. controls (382.62,115.25) and (384.79,118.04) .. (383.54,119.39) .. controls (382.28,120.74) and (380.11,117.95) .. (382.62,115.25) .. controls (385.13,112.56) and (387.31,115.35) .. (386.05,116.7) .. controls (384.79,118.04) and (382.62,115.25) .. (385.13,112.56) .. controls (387.64,109.87) and (389.82,112.66) .. (388.56,114.01) .. controls (387.31,115.35) and (385.13,112.56) .. (387.64,109.87) .. controls (390.16,107.18) and (392.33,109.97) .. (391.07,111.31) .. controls (389.82,112.66) and (387.64,109.87) .. (390.16,107.18) .. controls (392.67,104.48) and (394.84,107.27) .. (393.58,108.62) .. controls (392.33,109.97) and (390.16,107.18) .. (392.67,104.48) .. controls (395.18,101.79) and (397.35,104.58) .. (396.09,105.93) .. controls (394.84,107.27) and (392.67,104.48) .. (395.18,101.79) .. controls (397.69,99.1) and (399.86,101.89) .. (398.6,103.24) .. controls (397.35,104.58) and (395.18,101.79) .. (397.69,99.1) .. controls (400.2,96.41) and (402.37,99.2) .. (401.12,100.54) .. controls (399.86,101.89) and (397.69,99.1) .. (400.2,96.41) .. controls (402.71,93.71) and (404.88,96.5) .. (403.63,97.85) .. controls (402.37,99.2) and (400.2,96.41) .. (402.71,93.71) .. controls (403.47,92.9) and (404.2,92.59) .. (404.81,92.58) ;
\draw   (515,134.42) .. controls (514.54,133.56) and (514.64,132.2) .. (516.14,130.87) .. controls (519.13,128.2) and (521.71,130.97) .. (520.22,132.3) .. controls (518.72,133.63) and (516.14,130.87) .. (519.13,128.2) .. controls (522.12,125.53) and (524.7,128.3) .. (523.21,129.63) .. controls (521.71,130.97) and (519.13,128.2) .. (522.12,125.53) .. controls (525.11,122.86) and (527.69,125.63) .. (526.19,126.96) .. controls (524.7,128.3) and (522.12,125.53) .. (525.11,122.86) .. controls (528.09,120.2) and (530.68,122.96) .. (529.18,124.29) .. controls (527.69,125.63) and (525.11,122.86) .. (528.09,120.2) .. controls (531.08,117.53) and (533.67,120.29) .. (532.17,121.63) .. controls (530.68,122.96) and (528.09,120.2) .. (531.08,117.53) .. controls (534.07,114.86) and (536.66,117.62) .. (535.16,118.96) .. controls (533.67,120.29) and (531.08,117.53) .. (534.07,114.86) .. controls (537.06,112.19) and (539.64,114.95) .. (538.15,116.29) .. controls (536.66,117.62) and (534.07,114.86) .. (537.06,112.19) .. controls (540.05,109.52) and (542.63,112.29) .. (541.14,113.62) .. controls (539.64,114.95) and (537.06,112.19) .. (540.05,109.52) .. controls (543.04,106.85) and (545.62,109.62) .. (544.13,110.95) .. controls (542.63,112.29) and (540.05,109.52) .. (543.04,106.85) .. controls (546.03,104.18) and (548.61,106.95) .. (547.12,108.28) .. controls (545.62,109.62) and (543.04,106.85) .. (546.03,104.18) .. controls (549.02,101.51) and (551.6,104.28) .. (550.11,105.61) .. controls (548.61,106.95) and (546.03,104.18) .. (549.02,101.51) .. controls (552.01,98.85) and (554.59,101.61) .. (553.09,102.95) .. controls (551.6,104.28) and (549.02,101.51) .. (552.01,98.85) .. controls (552.91,98.04) and (553.78,97.73) .. (554.51,97.73) ;
\draw   (512.96,17.24) .. controls (513.93,16.82) and (515.47,16.9) .. (517,18.2) .. controls (520.06,20.81) and (516.94,23.12) .. (515.41,21.81) .. controls (513.88,20.51) and (517,18.2) .. (520.06,20.81) .. controls (523.11,23.42) and (520,25.73) .. (518.47,24.42) .. controls (516.94,23.12) and (520.06,20.81) .. (523.11,23.42) .. controls (526.17,26.02) and (523.06,28.33) .. (521.53,27.03) .. controls (520,25.73) and (523.11,23.42) .. (526.17,26.02) .. controls (529.23,28.63) and (526.11,30.94) .. (524.59,29.64) .. controls (523.06,28.33) and (526.17,26.02) .. (529.23,28.63) .. controls (532.29,31.24) and (529.17,33.55) .. (527.64,32.25) .. controls (526.11,30.94) and (529.23,28.63) .. (532.29,31.24) .. controls (535.34,33.85) and (532.23,36.16) .. (530.7,34.85) .. controls (529.17,33.55) and (532.29,31.24) .. (535.34,33.85) .. controls (538.4,36.45) and (535.29,38.76) .. (533.76,37.46) .. controls (532.23,36.16) and (535.34,33.85) .. (538.4,36.45) .. controls (541.46,39.06) and (538.34,41.37) .. (536.82,40.07) .. controls (535.29,38.76) and (538.4,36.45) .. (541.46,39.06) .. controls (544.52,41.67) and (541.4,43.98) .. (539.87,42.68) .. controls (538.34,41.37) and (541.46,39.06) .. (544.52,41.67) .. controls (547.57,44.28) and (544.46,46.59) .. (542.93,45.28) .. controls (541.4,43.98) and (544.52,41.67) .. (547.57,44.28) .. controls (550.63,46.88) and (547.52,49.2) .. (545.99,47.89) .. controls (544.46,46.59) and (547.57,44.28) .. (550.63,46.88) .. controls (553.69,49.49) and (550.57,51.8) .. (549.05,50.5) .. controls (547.52,49.2) and (550.63,46.88) .. (553.69,49.49) .. controls (554.61,50.28) and (554.97,51.04) .. (554.98,51.69) ;

\draw (380.24,90.12) node [anchor=north west][inner sep=0.75pt]    {$g$};
\draw (378.09,42.36) node [anchor=north west][inner sep=0.75pt]    {$g$};
\draw (215.85,30.87) node [anchor=north west][inner sep=0.75pt]    {$g$};
\draw (205.07,74.74) node [anchor=north west][inner sep=0.75pt]    {$g$};
\draw (248.39,83.84) node [anchor=north west][inner sep=0.75pt]    {$g$};
\draw (518.48,35.94) node [anchor=north west][inner sep=0.75pt]    {$g$};
\draw (516.56,95.16) node [anchor=north west][inner sep=0.75pt]    {$g$};
\draw (309.46,11.45) node [anchor=north west][inner sep=0.75pt]    {$Y_{0}$};
\draw (144.98,11.78) node [anchor=north west][inner sep=0.75pt]    {$Y_{0}$};
\draw (597.91,53.18) node [anchor=north west][inner sep=0.75pt]    {$Y_{0}$};
\draw (445.68,8.66) node [anchor=north west][inner sep=0.75pt]    {$Y_{0}$};
\draw (10.52,94.09) node [anchor=north west][inner sep=0.75pt]    {$u,d$};
\draw (17.05,18.92) node [anchor=north west][inner sep=0.75pt]    {$u,d$};
\draw (445.49,97.8) node [anchor=north west][inner sep=0.75pt]    {$\overline{t}$};
\draw (606.95,93.31) node [anchor=north west][inner sep=0.75pt]    {$\overline{t}$};
\draw (133.14,91.73) node [anchor=north west][inner sep=0.75pt]    {$\overline{t}$};
\draw (320.08,103.97) node [anchor=north west][inner sep=0.75pt]    {$\overline{t}$};
\draw (469.78,35.51) node [anchor=north west][inner sep=0.75pt]    {$t$};
\draw (417.85,23.47) node [anchor=north west][inner sep=0.75pt]    {$t$};
\draw (414.52,65.19) node [anchor=north west][inner sep=0.75pt]    {$t$};
\draw (103.9,43.86) node [anchor=north west][inner sep=0.75pt]    {$t$};
\draw (156.86,43.3) node [anchor=north west][inner sep=0.75pt]    {$t$};
\draw (327.61,46.48) node [anchor=north west][inner sep=0.75pt]    {$t$};
\draw (276.29,42.76) node [anchor=north west][inner sep=0.75pt]    {$t$};
\draw (591.69,17.47) node [anchor=north west][inner sep=0.75pt]    {$t$};
\draw (540.05,54.22) node [anchor=north west][inner sep=0.75pt]    {$t$};
\draw (540.65,77.47) node [anchor=north west][inner sep=0.75pt]    {$t$};
\draw (76.42,58.19) node [anchor=north west][inner sep=0.75pt]    {$g$};
\end{tikzpicture}
    \captionsetup{width=0.95\textwidth} 
    \caption{Representative Feynman diagrams for the spin 0 mediator production, with the top quarks in the final state employed in the UFO model.}
    \label{fig:born_level}
\end{figure*}

\noindent
where $y^{d(u)}_{i,j}$ denotes the elements of the quark Yukawa-like coupling matrix with the scalar mediator, $g^{s}_{d(u)_{ij}}$ and $g^{p}_{d(u)_{ij}}$. Note that we can create different versions of the model according to the coupling choices, including the off-diagonal elements of the coupling matrix. In our analysis, however, we looked at top quarks in the final state coming from gluon splitting. This makes the couplings of the $Y_{0}$ to the top quark the only ones that are relevant for our study. Following~\cite{kentarou2015higher}, the $Y_{0}$ couplings to the SM particles are proportional to their masses ($m_{f}$) and were normalized to the SM value by fixing  $y_{ii} = \sqrt{2}m_{f}/v$ ($v$=246~GeV), setting as well, all off-diagonal terms of the coupling matrix to zero. Although scalar mediator models that employ 4 and 5 flavor schemes have been known in the literature, NLO computations of the cross-section do not change significantly irrespectively from the scheme employed. Even though the renormalizability of the coupling with the b-quark is expected to be challenging, due to the higher order uncertainties when switching schemes, regularization appears to be formally complete~\cite{kentarou2015higher,Degrande:2014vpa,Botje:2011sn}. As mentioned before, given the fact that final states involve only top quarks produced in association with $Y_{0}$, the only relevant contribution for the Lagrangian is the one involving these quarks, simplifying equation~\ref{eq:lagrangian_yukawa} to,

\begin{equation}\label{eq:lagrangian_main}
    \mathcal{L} = \frac{y^{}_{3,3}}{\sqrt{2}} \bar{t}(g^{s}_{} + i g^{p}_{}\gamma^{5} )t Y_{0},
\end{equation}

\noindent
where $y^{}_{3,3}$ is the corresponding coupling constant with the top quark, and the $g^{s}_{}$ denotes the CP-even (scalar) coupling with the mediator and $g^{p}_{}$ the corresponding CP-odd (pseudoscalar) interaction. No additional new physics is considered here, apart from the DM mediator $Y_{0}$. In this work, we employ an implementation of the Lagrangian in equation~\ref{eq:lagrangian_main}, including LO and NLO corrections in the total cross-sections, by using the \texttt{FeynRules} package~\cite{Alloul:2013bka}. In order to allow the $Y_{0}$ to escape detection, we have assumed no $Y_{0}$ decays by setting its decay width ($\Gamma$) to $\Gamma$=0. This construction is analogous to the freely available DM simplified model \texttt{DMsimp} for scalar mediators in the s-channel interaction, but including the 5 flavor scheme. The main topologies obtained in the model for interaction of the top quark with the particle mediator are given in Figure~\ref{fig:born_level}, where corresponding NLO diagrams are the same as those obtained in~\cite{kentarou2015higher} for the \texttt{DMsimp} package.

Due to the similar nature of the scalar coupling for the Higgs boson, and by assuming the narrow width approximation for the top and antitop quarks, we can approximate and define the unpolarized differential cross-section for the process $g g \rightarrow t\left(\rightarrow b \ell^{+} \nu_{\ell}\right) \bar{t}\left(\rightarrow \bar{b} \ell^{-} \bar{\nu}_{\ell}\right)Y$ in a "factorized" form:

\begin{equation}\label{eq:factorization}
\begin{split}
d \sigma = \sum_{\substack{b \ell^{+} \nu_n \\ \text { spins spins } \\ \ell^{-} \bar{\nu}_{\ell}}}& \left(\frac{2}{\Gamma_t}\right)^2 d \sigma\left(g g \rightarrow t\left(n_t\right) \bar{t}\left(n_{\bar{t}}\right) Y \right) \\
& \times d \Gamma\left(t \rightarrow b \ell^{+} \nu_{\ell}\right) d \Gamma\left(\bar{t} \rightarrow \bar{b} \ell^{-} \bar{\nu}_{\ell}\right),
\end{split}
\end{equation}

\noindent
where $d \sigma\left(g g \rightarrow t\left(n_t\right) \bar{t}\left(n_{\bar{t}}\right) Y\right)$ is the differential cross-section for the production of a top and antitop quarks, where the spin vectors $n_t$ and $n_{\bar{t}}$ are introduced for enhancing the relative sign sensitivity for the pseudoscalar component of the model in addition of the quadratic enhancement proposed in~\cite{Gunion_1996}. 

The factorized terms of the form: $d \Gamma\left(t \rightarrow b \ell^{+} \nu_{\ell}\right)$ and $d \Gamma\left(\bar{t} \rightarrow \bar{b} \ell^{-} \bar{\nu}_{\ell}\right)$ are the partial differential decay widths for an unpolarized top and anti-top quark and such factorized form is valid within the narrow-width approximation. The spin basis is defined in such a way that $t, \bar{t}, \ell^{+}$and $\ell^{-}$, normalize 
 the 4-vectors:

\begin{equation}
\begin{aligned}\label{eq:polarization}
& n^{}_t=-\frac{p_t}{m_t}+\frac{m_t}{\left(p_t \cdot p_{\ell^{+}}\right)} p_{\ell^{+}} \\
& n^{}_{\bar{t}}=\frac{p_{\bar{t}}}{m_t}-\frac{m_t}{\left(p_{\bar{t}} \cdot p_{\ell^{-}}\right)} p_{\ell^{-}},
\end{aligned}
\end{equation}
\noindent
where $p_{\ell^{\pm}}$ correspond to the momentum of the outgoing lepton coming from the top quark decay. The spatial components of these variables are used to define equations~\ref{eq:observables}. Given that the representation of the total differential cross-section is analogue to the CP-sensitive one, we can in principle construct analogue operators as in the Higgs case.

\section{Simulation \label{sec:simulation}}

In this section, we show the event generation and simulation from an independent implementation of the \texttt{DMsimp} DM simplified model~\cite{kentarou2015higher}. We have produced 1 million parton-level events using \texttt{MadGraph5\_aMC@NLO} \cite{Alwall:2014hca} for both signal ($p p \rightarrow t \bar{t} Y_0$) and SM background events at $\sqrt{s}= 14$~TeV. We have explored DM mediator masses set to $m_{Y_0}=10^{-2},1,10,125$~GeV and 1~TeV. For background processes we have considered the most relevant ones with similar signal final state topologies. While $t \bar{t}$, $t \bar{t} V$ ($V$=$W$,$Z$), $t \bar{t} H$, $W/Z$+jets and diboson $(W W, Z Z, W Z)$ processes were generated at NLO, single top quark production ($t$-, $s$- and $W t$-channels) was generated at LO, up to one jet.

The top quark ($m_t$) and $W$ boson ($m_W$) masses were set to 172.5~GeV and 80.4~GeV, respectively, and we used the \texttt{NNPDF2.3} parton distribution functions (PDFs)~\cite{Ball:2012cx}. The renormalization and factorization scales were set to their default value i.e., the sum of the transverse masses of all particles and partons in the final state~\cite{Azevedo_2020}.

Particle decays were handled by \texttt{MadSpin}~\cite{Artoisenet:2012st} to maintain spin correlations for both signal and background events, linked to their heavy parent resonances. All decays of the top quark were assumed to be into the leptonic channel. The parton shower and hadronization processes were conducted using \texttt{Pythia}~\cite{Sjostrand:2006za}. For the LO generation, the jets after parton-shower evolution and jet clustering were matched to the original partons using the MLM merging scheme~\cite{Alwall:2014hca}. \texttt{Delphes}~\cite{deFavereau:2013fsa} was utilized for a fast, parameterized simulation of an LHC-like detector, with the default ATLAS parameter card. The analysis of signal and background events was performed within the MadAnalysis5 framework  \cite{Conte:2012fm}. Jet reconstruction was managed by \texttt{FastJet}~\cite{Cacciari:2011ma} using the anti-$k_t$ algorithm with a cone size of $\Delta R=0.4$.
In the event generation, jets and leptons were both required to have transverse momentum ($p_T$) above 10~\text{GeV}. In addition, jets (leptons) were required to have pseudorapidity ($\eta$) below 5.0 (2.5). 

The analysis of signal and background events was performed within the MadAnalysis5 framework~\cite{Conte:2012fm}. The event selection was based on jet and lepton identification with specific constraints ($\eta < 2.5$ and $p_T>20$~GeV), focusing on events with a pair of jets and isolated leptons. Reconstruction of the system $(t \bar{t})$ involved assigning jets to their corresponding b-quarks and utilizing multivariate analysis for improved accuracy. This analysis included various likelihood discrimination estimators as those used in~\cite{Azevedo_2022,Azevedo:2023xuc}, where the Boosted Decision Tree with Gradient boost (BDTG) showed superiority in classifying the events. 

\begin{figure*}
    \centering
    \includegraphics[width=0.48\linewidth]{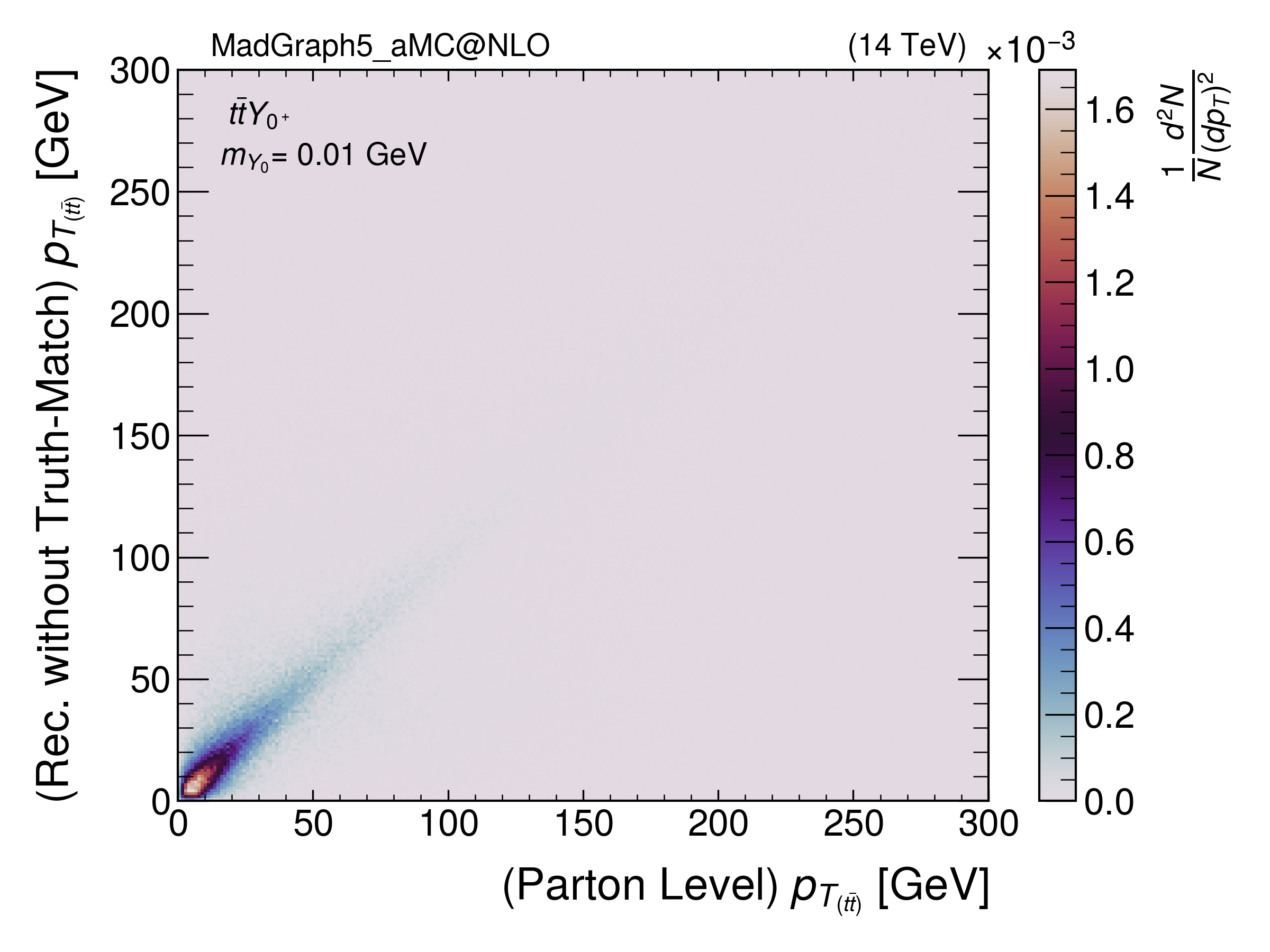}\hfill
    \includegraphics[width=0.48\linewidth]{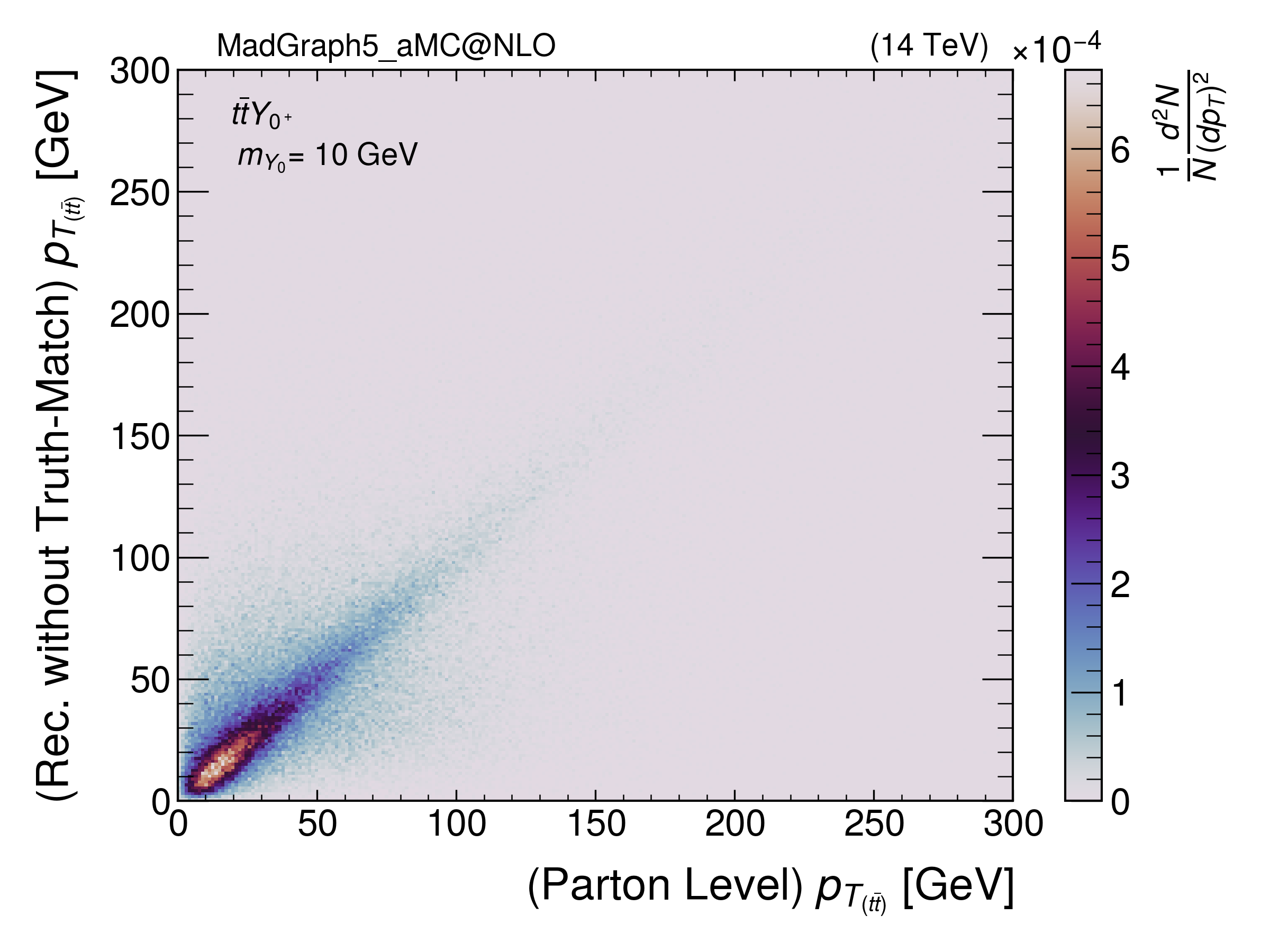}
    \vspace{3mm}
    \includegraphics[width=0.48\linewidth]{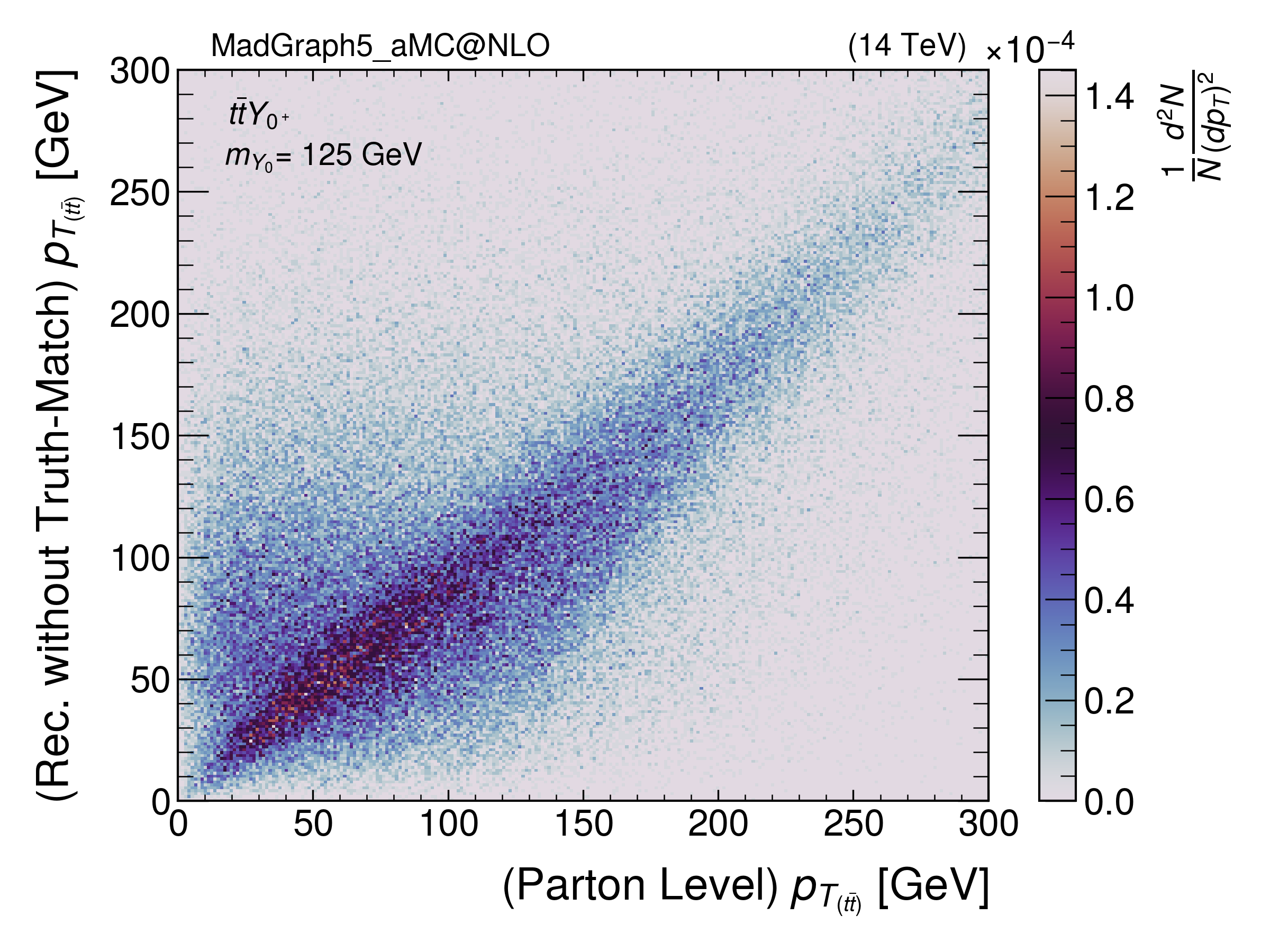}\hfill
    \includegraphics[width=0.48\linewidth]{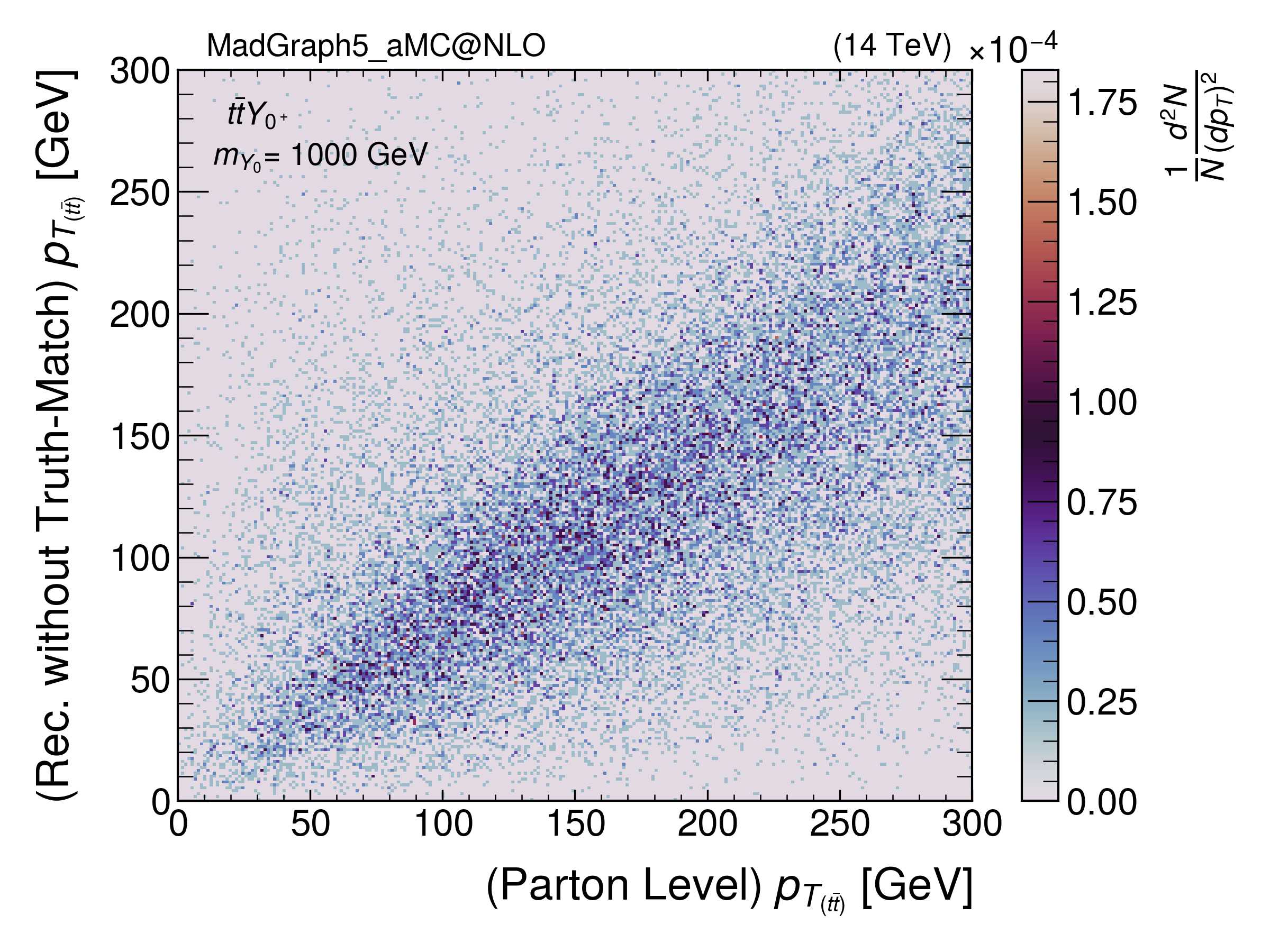}
    \captionsetup{width=0.95\textwidth}
    \caption{Two-dimensional distributions of $p_T$ in $t\Bar{t} Y_0^{+}$ events. The impact of the mediator mass on the kinematic reconstruction of the $t\bar{t}$ system for mediator masses of $10^{-2}$~GeV (upper-left), $10$~GeV (upper-right), $125$~GeV (lower-left), and $10^{3}$~GeV (lower-right).}
    \label{fig:reconstruction}
\end{figure*}

For the kinematic fit used in the reconstruction of the differential distributions, we have followed the same approach as in~\cite{Azevedo_2020}, but including NLO corrections in both signal and background events.
In the fit, we reconstructed the $t\bar{t}$ system, considering that the top quarks decayed through leptons, i.e.~$ t \bar{t} \rightarrow b W^{+} \bar{b} W^{-} \rightarrow b \ell^{+} \nu_{\ell} \bar{b} \ell^{-} \bar{\nu}_{\ell}$, using the four-momentum information from their respective decay products. The sum of the neutrino~$\nu$ and corresponding lepton four-momentum squared should combine to give the $W$ boson mass squared. Moreover, by combining the $W$ bosons with the corresponding $b$-quarks, the top quarks are reconstructed, using similar mass constraints for both top quarks.
The total missing transverse energy ($\cancel{E_T}$) is assumed to be fully accounted for by the undetected neutrinos. No constraints were applied on the mediator (the invisible part of the events).
In Figure~\ref{fig:reconstruction}, it is shown the reconstruction of the $t\bar{t}Y$ events, presenting the impact in the reconstruction when a mediator particle with different masses is considered in the analysis.
By using the mass and $\cancel{E_T}$ constraints described above~\cite{Azevedo_2020}, we showed that heavy mediator particles with masses of the order of magnitude of 1~TeV contribute more to the total missing energy of the reconstruction, normally associated to the neutrinos, translating in an increasing bias in the reconstruction of the kinematic variables.

\section{CP sensitive obervables \label{sec:cpsensitive}}

In the study, the search of sensitive CP-observables starts by considering the total cross-section coming from a Yukawa interaction term in the DM simplified model. For scalar particles, when performing the spin-averaged cross-section calculation, taking into account the $gg$ and $q\bar{q}$ contributions at LO, no terms proportional to the product of scalar and pseudo-scalar couplings exist~\cite{Gunion_1996}. While CP observables were explored already at LO~\cite{Azevedo:2023xuc} here we extend the studies to NLO by considering their expected value calculated in the usual way,

\begin{equation}\label{eq:phase_space}
\alpha\left[\mathcal{O}_{C P}\right] \equiv \int\left[\mathcal{O}_{C P}\right]\{d \sigma(p p \rightarrow t \bar{t} Y) / \sigma d P S\} d P S \ .
\end{equation}
\noindent
The integral over the phase space $PS$, for a given differential observable $\mathcal{O}_{C P}$ is such that in the case of the Higgs boson leads to the definition of the $b_2$ observable that can be  written in terms of the projections of the 4-momenta for the outgoing top quarks in a given event:
\begin{equation}\label{eq:observables}
\begin{aligned}
& b_2 = \frac{\left(\vec{p}_t \times \widehat{z}\right) \cdot\left(\vec{p}_t \times \widehat{z}\right)}{\left|\vec{p}_{{t}}\right|\left|\vec{p}_{\bar{t}}\right|}
\quad 
\tilde{b}_{2}^{\widehat{y}} = \frac{\left(\vec{p}_t \times \widehat{y}\right) \cdot\left(\vec{p}_t \times \widehat{y}\right)}{\left|\vec{p}_{{t}}\right|\left|\vec{p}_{\bar{t}}\right|} 
\quad 
\\
&
\tilde{b}_{2}^{\widehat{d}} = \frac{(\vec{p}_t \times \widehat{d}) \cdot(\vec{p}_t \times \widehat{d} )}{\left|\vec{p}_{{t}}\right|\left|\vec{p}_{\bar{t}}\right|}  
\quad \text{where }
\widehat{d} = \frac{(1,1,0)}{\sqrt{2}}
\quad 
\\
&
n_4^{t \bar{t} Y}=\frac{\left(n_t^z \cdot n_{\bar{t}}^z\right) }{\left(\left|\vec{n}_t\right| \cdot\left|\vec{n}_{\bar{t}}\right|\right)}
\quad 
n_2^{t \bar{t} Y}=\frac{(\vec{n}_t \times \hat{k}_z) \cdot(\vec{n}_{\bar{t}} \times \hat{k}_z) }{\left(\left|\vec{n}_t\right| \cdot\left|\vec{n}_{\bar{t}}\right|\right)},
\end{aligned}
\end{equation}
\noindent
where $\hat{n}$ corresponds to the projection direction of $\vec{p}$ following the same definitions in~\cite{Mileo_2016}. This observable has been studied in the context of minimal pseudoscalar extensions of the Higgs boson, where it appears to have competitive sensitivity following this construction. However, given the nature of the DM simplified coupling with the scalar mediator, we can exploit this definition and its rotational symmetry to build two new angular observables. More specifically, a new approach involving orthogonal projections with respect to different axis. To test the hypothesis of these projections improving sensitivity for CP-odd components in the model, we extend the projections of the $b_2$ variable from the $\hat{z}$ axis, to the normalized $(0,1,0)$ and $(0,\sfrac{1}{\sqrt{2}},\sfrac{1}{\sqrt{2}})$ directions of projection.

\begin{figure*}
\begin{center}
\begin{tabular}{ccc}
\hspace*{-5mm}\includegraphics[height=6.5cm]{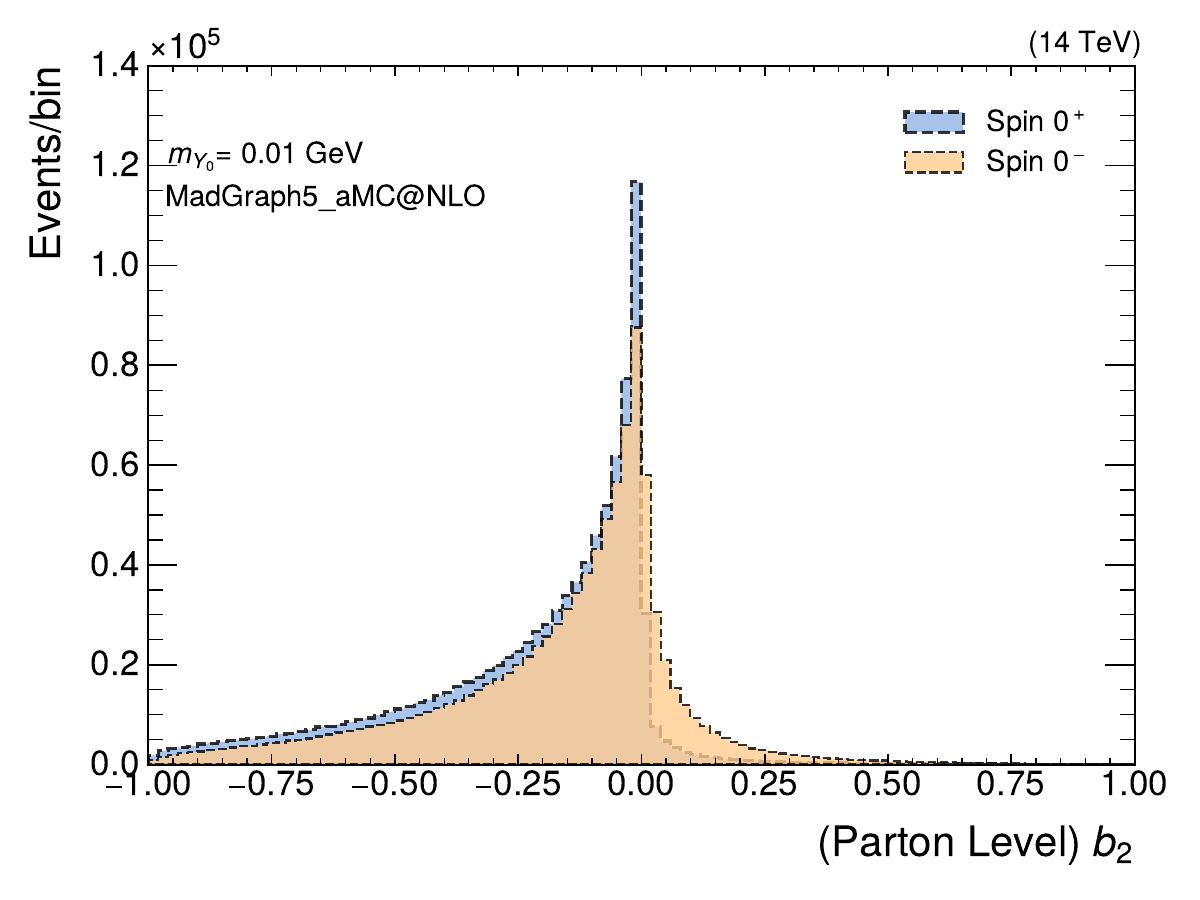}
\hspace*{5mm}\includegraphics[height=6.5cm]{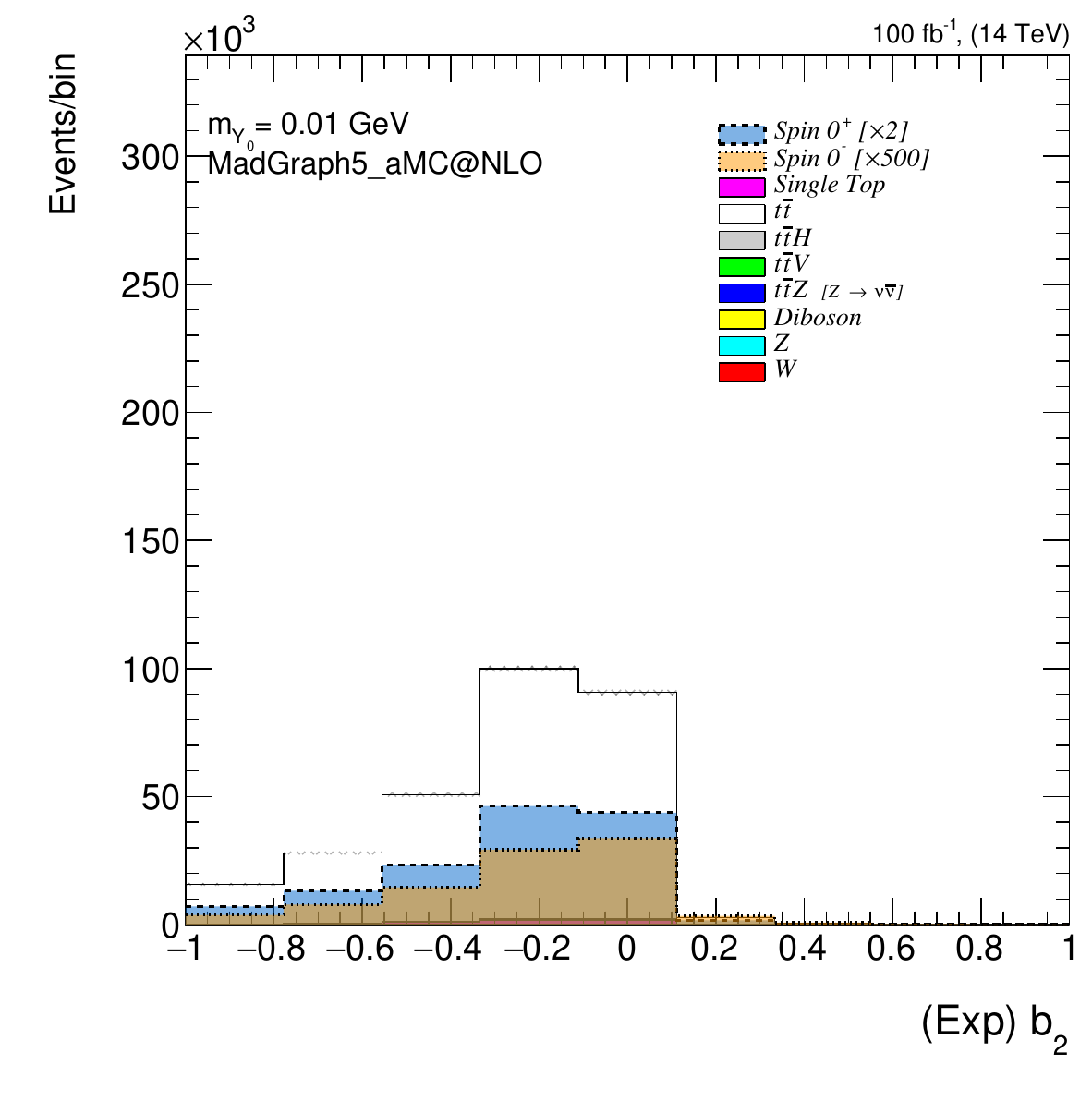}
\\
\hspace*{-5mm}\includegraphics[height=6.5cm]{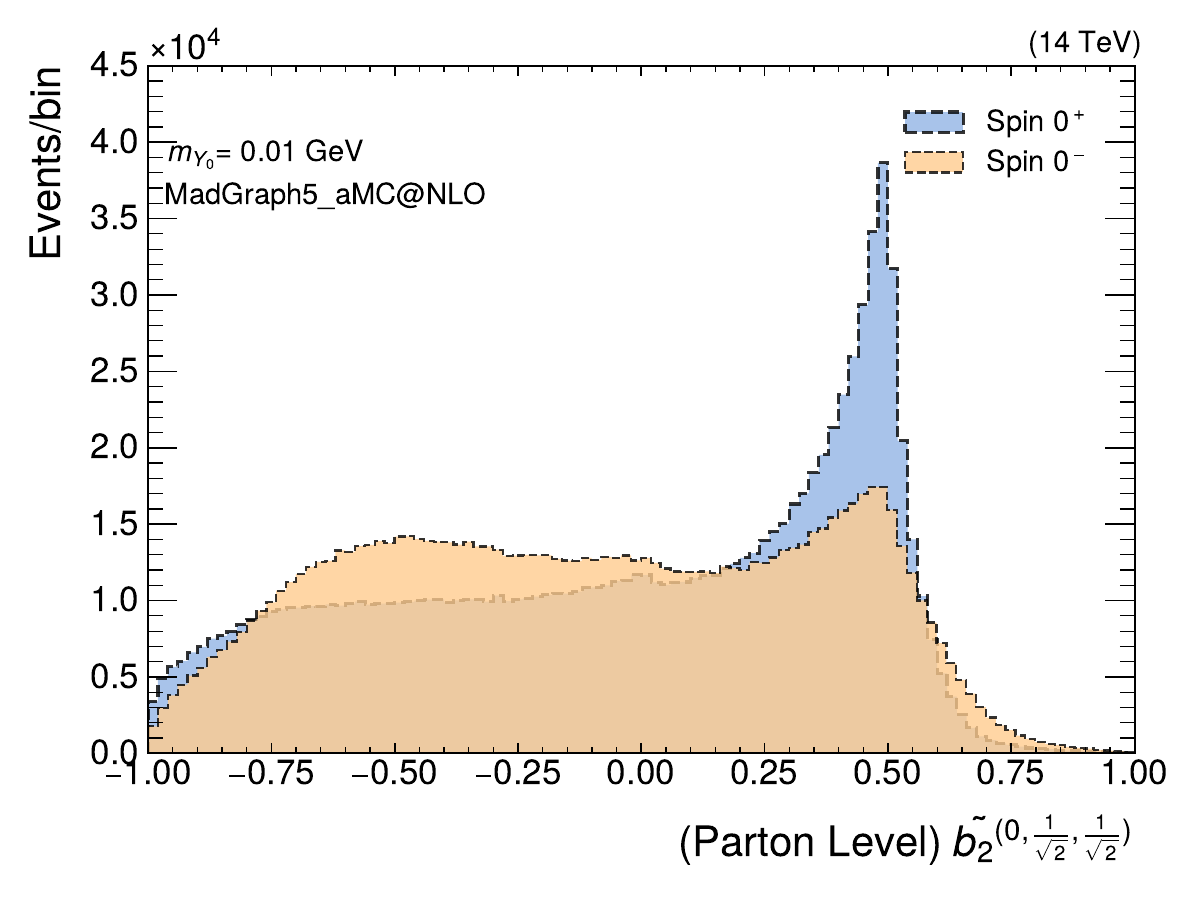}
\hspace*{5mm}\includegraphics[height=6.5cm]{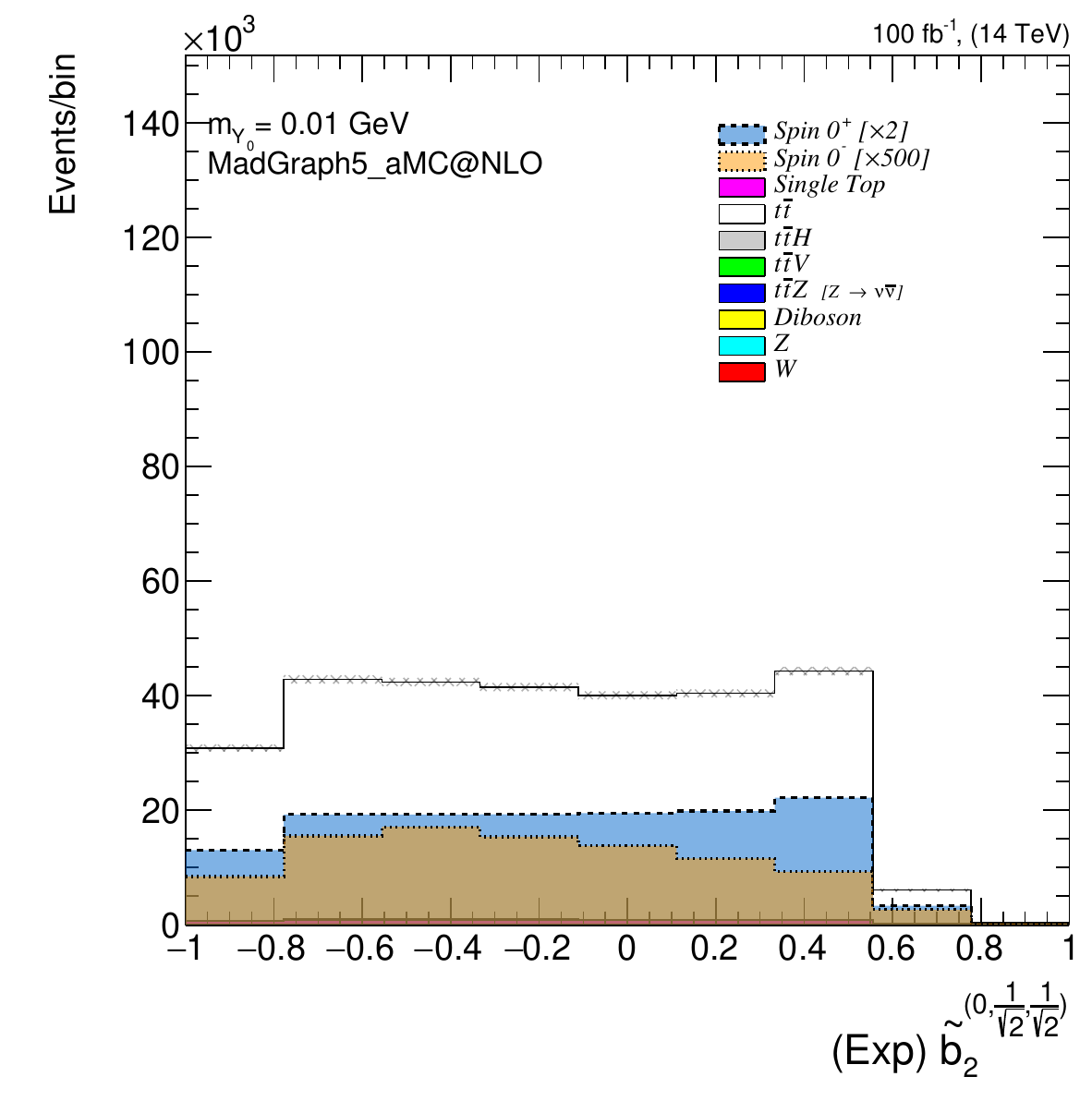}
\\
\hspace*{-5mm}\includegraphics[height=6.5cm]{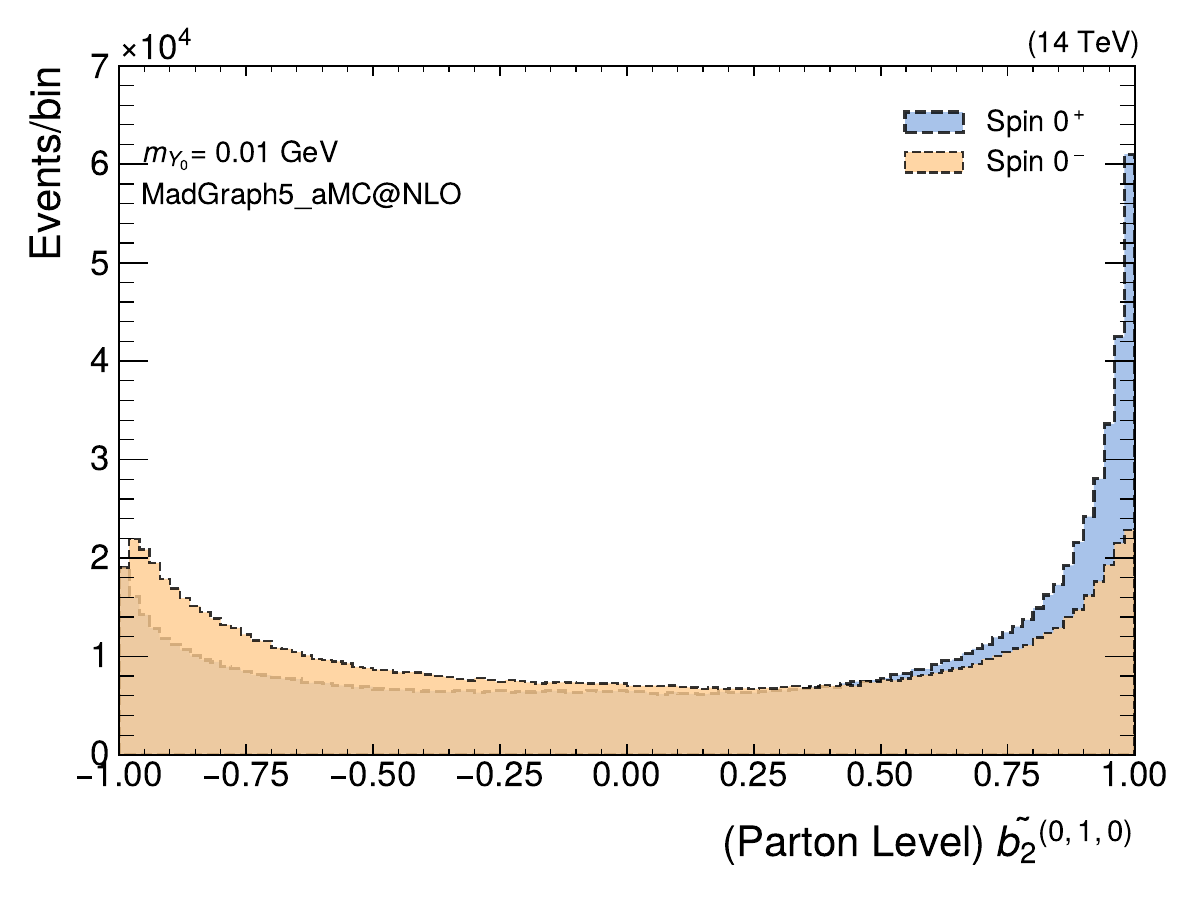}
\hspace*{5mm}\includegraphics[height=6.5cm]{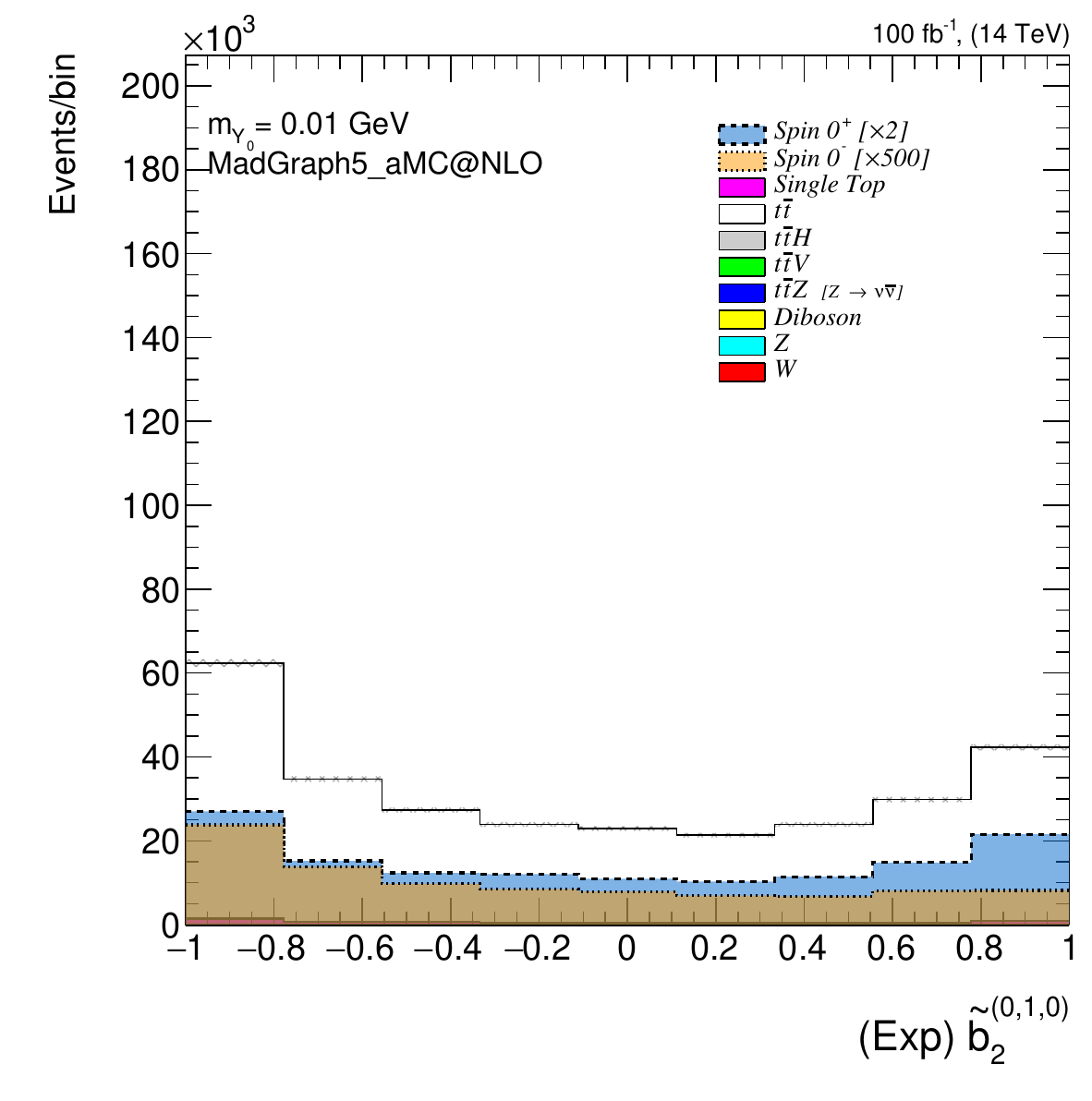}
\end{tabular}
\end{center}
\caption{Angular distributions at parton level before selection cuts (left) and after final selection cuts and full kinematic reconstruction, considering all the dominant backgrounds in $t\bar{t}$ production (right) of the: (top) $b_2$, (middle) $\tilde{b}_{2}^{\widehat{d}}$ and (bottom) $\tilde{b}_{2}^{\widehat{y}}$ distributions. Mass reference is $10^{-2}$~GeV for a luminosity of 100~fb$^{-1}$.}
\label{fig:stack_vs_parton_b2}
\end{figure*}

\begin{figure*}
\begin{center}
\begin{tabular}{ccc}
\hspace*{-5mm}\includegraphics[height=6.5cm]{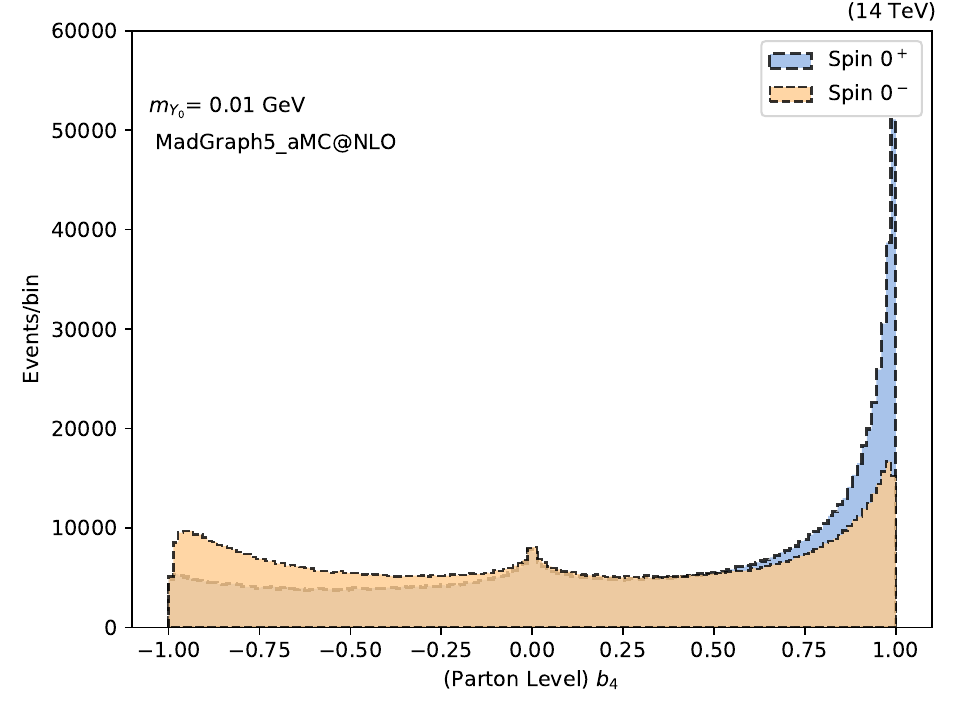}
\hspace*{5mm}\includegraphics[height=6.5cm]{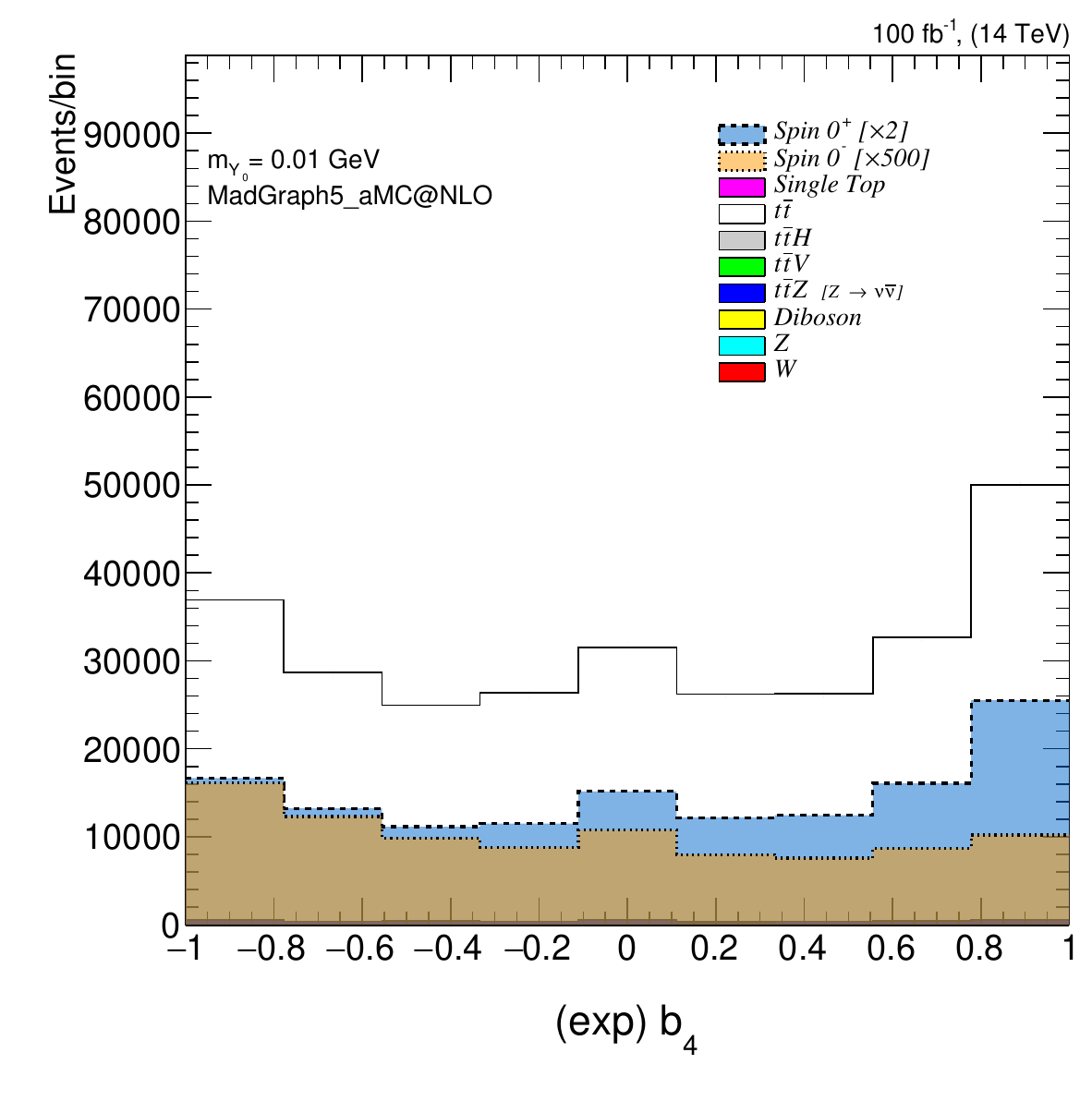}
\\
\hspace*{-5mm}\includegraphics[height=6.5cm]{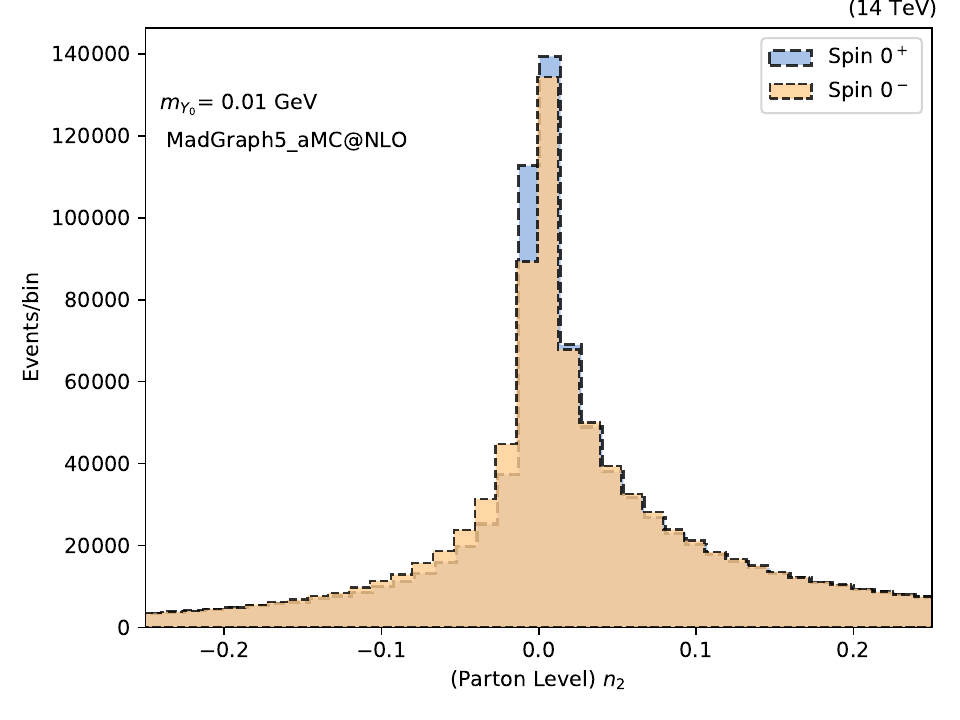}
\hspace*{5mm}\includegraphics[height=6.5cm]{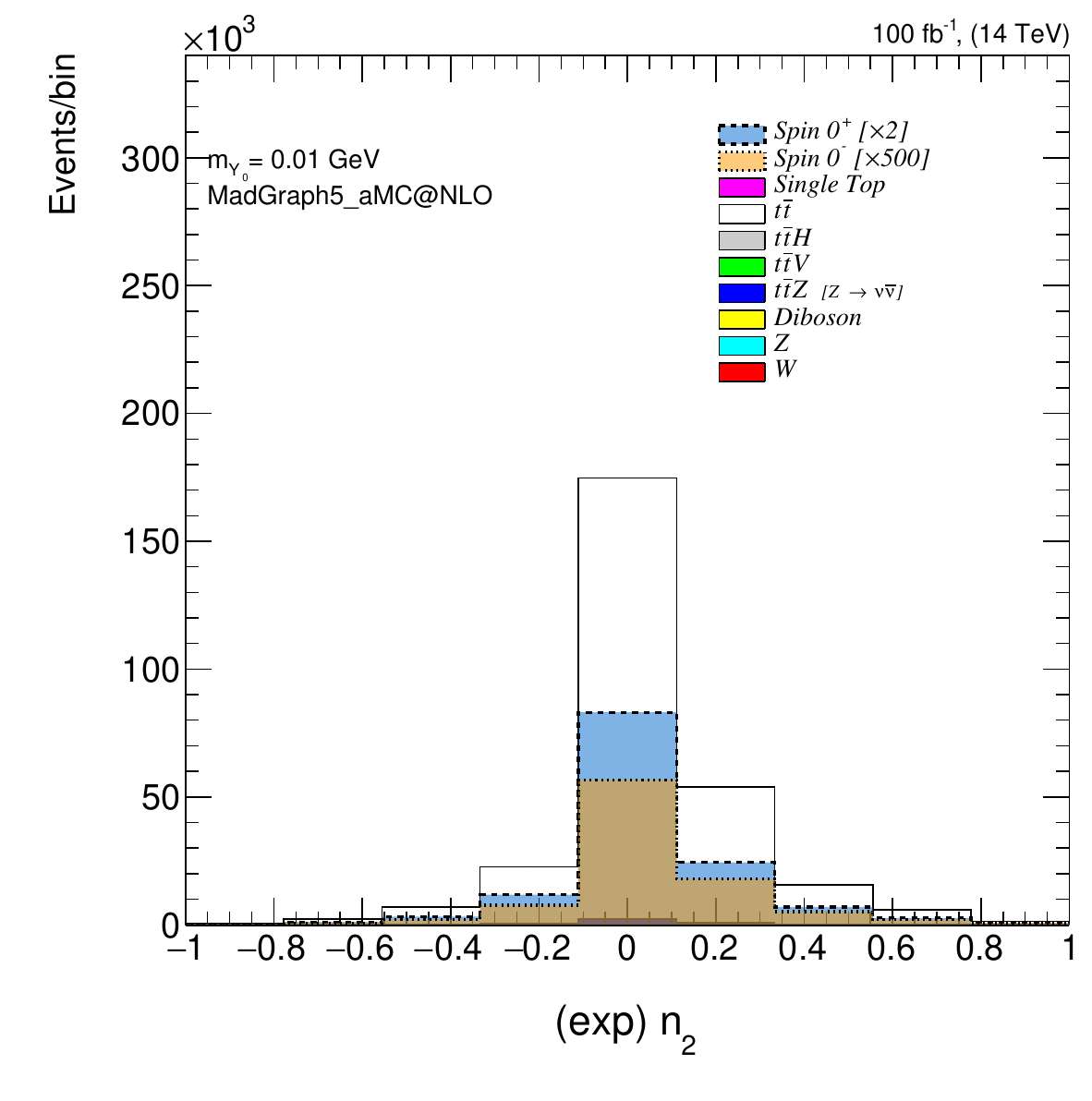}
\\
\hspace*{-5mm}\includegraphics[height=6.5cm]{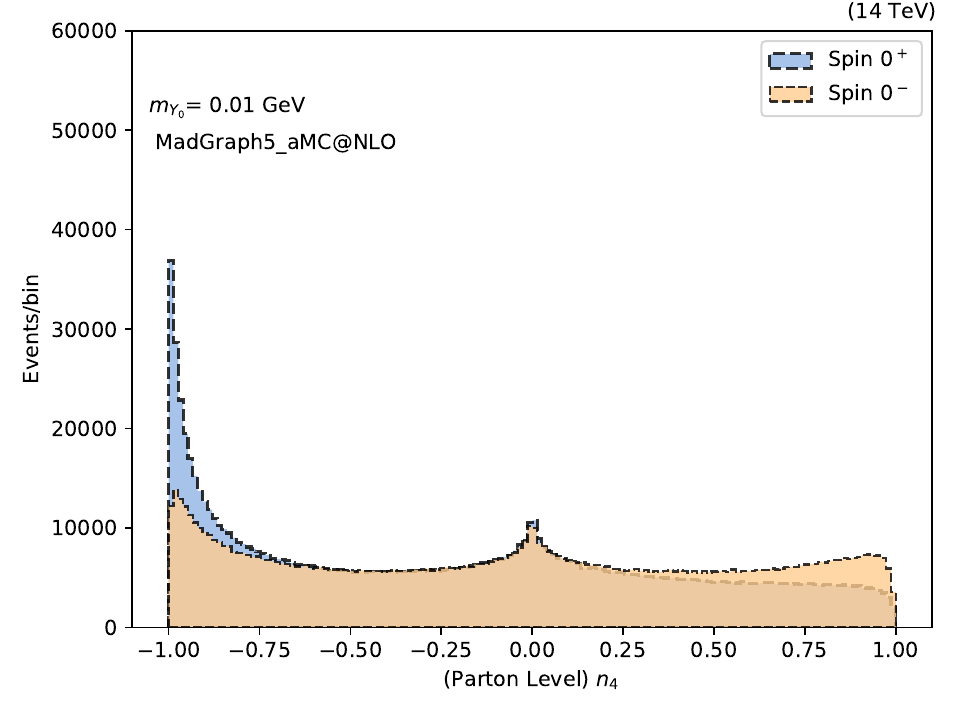}
\hspace*{5mm}\includegraphics[height=6.5cm]{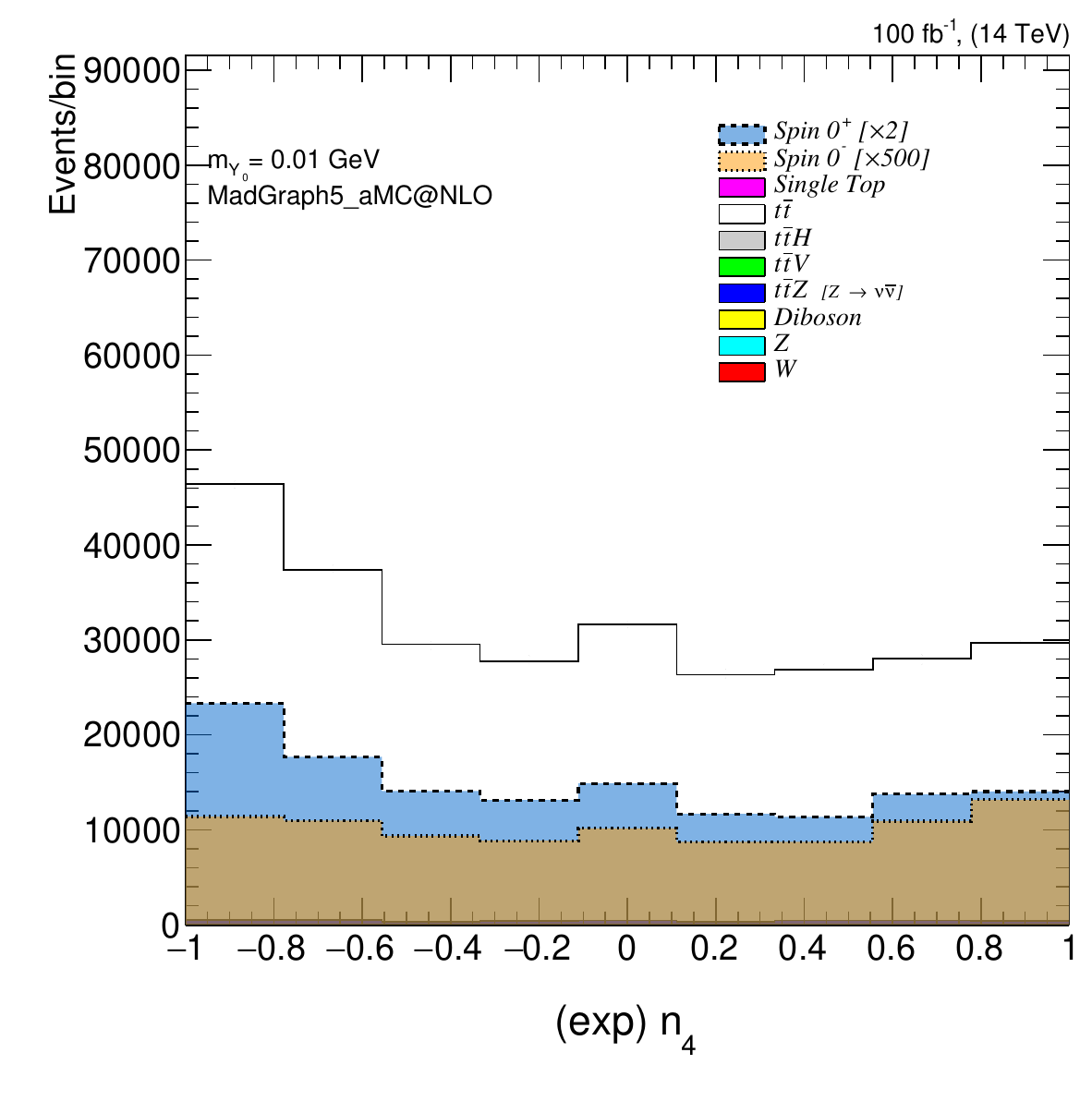}
\end{tabular}
\end{center}
\caption{Angular distributions at parton-level before selection cuts (left) and after final selection cuts and full kinematic reconstruction, considering all the dominant backgrounds in $t\bar{t}$ production (right) of the: (top)~ $b_4$, (middle) $n_2$ and (bottom) $n_4$ distributions. Mass reference is $10^{-2}$~GeV for a luminosity of~100~fb$^{-1}$.}
\label{fig:stack_vs_parton_spin}
\end{figure*}

In Figures~\ref{fig:stack_vs_parton_b2} and~\ref{fig:stack_vs_parton_spin}, we show all the corresponding parton-level distributions (left) and after event selection and kinematic reconstruction (right), for a reference luminosity of $100 \mathrm{fb}^{-1}$ for all the considered observables. All SM backgrounds events are included. The $t \bar{t} Y_0$ scalar and pseudoscalar signals, with $m_{Y_0}= 10^{-2} \ \mathrm{GeV}$, are shown as well, scaled by factors of 2 and 500, respectively, for convenience. After event selection and kinematic reconstruction, we have found that the $t \bar{t}$ is the main SM dominant background. All other backgrounds are effectively negligible, as expected for a $t \bar{t}$ dileptonic final state analysis, given the smallness of their respective total cross-sections.

Differences in shapes of the background distributions can also be noticed when compared with the signals. In addition, selection cuts heavily impact the relative bin value of the distribution but preserve the shape of the parton-level distributions. In Figure~\ref{fig:stack_vs_parton_b2}, both the scalar (in blue) and pseudoscalar (in orange) signals show a relatively even distribution around the central values of the $b_2$ observable. However, it is notable that the scalar signal displays a slight predominance in populating these central values compared to the pseudoscalar signal. In Figure~\ref{fig:stack_vs_parton_spin}, for instance, the $b_4$ distribution for the scalar signal is more populated in positive values rather than in the negative ones. This behavior is inverted for the pseudoscalar case.

\section{Asymmetries
\label{sec:asymmetries}}

As it is explored in $t\bar{t}$ with invisible particles production studies \cite{Azevedo_2022}, an asymmetry in the angular distributions of the top and antitop quark production can provide an additional source of information for the study of the properties of DM mediators. In particular, the measurement of the forward-backward asymmetry, in differential angular distributions $(x_{\alpha[\mathcal{O}_{CP}]})$, which involved the calculation of the kinematics in several center of mass systems in the $t\bar{t}h$ searches~\cite{Amor_dos_Santos_2015}  were particularly useful. They were defined using the number of events in the corresponding angular distribution above and below zero (as defined in eq.~\ref{eq:a9}). Furthermore, the study of the spin correlation in the top-antitop production, which is sensitive to the relative orientation of the spins of the top and antitop quarks, can provide additional insights into the production and decay dynamics of these particles.

Detailed studies of these asymmetries can provide valuable tests of the standard-model predictions and potentially reveal evidence of new physics and the reduced impact that systematic effects have in such measurements. As an example, the Higgs boson production in association with the top quark has been reported in the literature ~\cite{Amor_dos_Santos_2015} showing interesting results in constraining the minimal extensions of the Higgs sector. In this work, the same forward-backward asymmetries approach is employed for testing the impact on the exclusions limits. The asymmetry in the distribution is defined as:

\begin{eqnarray}
\ensuremath{A^{\alpha[\mathcal{O}_{CP}]}_{FB}= \frac  {    \sigma(x_{\alpha[\mathcal{O}_{CP}]}>0)-\sigma(x_{\alpha[\mathcal{O}_{CP}]}<0)   }{ \sigma( x_{\alpha[\mathcal{O}_{CP}]}>0)+\sigma(x_{\alpha[\mathcal{O}_{CP}]}<0) } },
\label{eq:a9}
\end{eqnarray}
\noindent
where $\sigma(x_{\alpha[\mathcal{O}_{CP}]})$ represents the cross-section associated to the $x_{\alpha[\mathcal{O}_{CP}]}$ observable, evaluate below and above zero in equation~\ref{eq:a9}. In Table~\ref{tab:asymmetries1} (Table~\ref{tab:asymmetries2}) the asymmetries for several observables and 5 benchmarks masses are shown at parton level (after event selection and kinematic reconstruction). It is noticeable that the application of the selection cuts and the kinematic reconstruction distorts the asymmetry observed in the differential distributions at parton-level. Remarkably, these distortions exhibit minimal variation with the mass of the mediator particle, indicating a consistent impact across different mass scales, even when the observables show a different behavior. However, the variation observed between the different observables presented in the Tables~\ref{tab:asymmetries1} and~\ref{tab:asymmetries2} is substantial. This significant disparity among the observables could have critical implications for the exclusion limits, a topic that will be explored in detail in the forthcoming section.

\begin{table*}
    \centering
    \begin{ruledtabular}
        \begin{tabular}{cccccc}
    \multirow{2}{*}{Observable} & \multicolumn{1}{c}{10$^{-2}$GeV}     & \multicolumn{1}{c}{1~GeV} & \multicolumn{1}{c}{10~GeV}  & \multicolumn{1}{c}{100~GeV} & \multicolumn{1}{c}{125~GeV}\\
         
         & $t\bar{t}Y^{+}/t\bar{t}Y^{-}$ & $t\bar{t}Y^{+}/t\bar{t}Y^{-}$ & $t\bar{t}Y^{+}/t\bar{t}Y^{-}$ & $t\bar{t}Y^{+}/t\bar{t}Y^{-}$ & $t\bar{t}Y^{+}/t\bar{t}Y^{-}$ \\
     \hline
         $b_2$ & -0.839/-0.579  & -0.834/-0.579  & -0.819/-0.568  & -0.703/-0.409  & -0.674/-0.377 \\
    
         $\tilde{b}_{2}^{\widehat{y}}$ & +0.222/-0.042  & +0.219/-0.041  & +0.217/-0.049   & +0.211/-0.156  & +0.199/-0.180  \\
  
         $\tilde{b}_{2}^{\widehat{d}}$ & +0.098/-0.110  & +0.092/-0.109  & +0.086/-0.116  & +0.061/-0.185  & +0.046/-0.199 \\
   
         $n_4$ & -0.243/-0.061 & -0.245/-0.062  & -0.246/-0.056  & -0.250/+0.0298 & -0.244/+0.050  \\

         $n_2$ & +0.278/+0.278 & +0.242/+0.221 & +0.234/+0.149 & +0.262/0.117 & +0.212/+0.178  \\
    \end{tabular}
    \end{ruledtabular}
    \caption{ Parton level asymmetry values computed for the 5 benchmark masses employed in the model.}
    \label{tab:asymmetries1}
\end{table*}

\begin{table*}
    \centering
    \begin{ruledtabular}
        \begin{tabular}{cccccc}
         
        \multirow{2}{*}{Observable} & \multicolumn{1}{c}{10$^{-2}$GeV} & \multicolumn{1}{c}{1~GeV} & \multicolumn{1}{c}{10~GeV} & \multicolumn{1}{c}{100~GeV} & \multicolumn{1}{c}{125~GeV} \\
        
        & $t\bar{t}Y^{+}/t\bar{t}Y^{-}$  & $t\bar{t}Y^{+}/t\bar{t}Y^{-}$ & $t\bar{t}Y^{+}/t\bar{t}Y^{-}$ & $t\bar{t}Y^{+}/t\bar{t}Y^{-}$ & $t\bar{t}Y^{+}/t\bar{t}Y^{-}$ \\
        \hline
        $b_2$ & -0.896/-0.697  & -0.888/-0.697  & -0.874/-0.690  & -0.753/-0.512  & -0.724/-0.454\\
       
        $\tilde{b}_{2}^{\widehat{y}}$ & -0.066/-0.281  & -0.073/-0.276  & -0.078/-0.293  & -0.079/-0.363 & -0.051/-0.369 \\
        
        $\tilde{b}_{2}^{\widehat{d}}$ & -0.191/-0.360 & -0.205/-0.349 & -0.212/-0.370  & -0.203/-0.386 & -0.180/-0.383\\
        
        $n_4$ & +0.132/-0.008 & +0.131/-0.031 &  +0.140/-0.036  & +0.129/-0.104 & +0.122/-0.125  \\
        
        $n_2$ &  -0.285/-0.286 & -0.269/-0.283 & -0.292/-0.270 &  -0.332/-0.228 &  -0.323/-0.222 \\
    \end{tabular}
    \end{ruledtabular}
    \caption{ Asymmetry values after applying the selection criteria and kinematic reconstruction computed for the same benchmark masses.}
    \label{tab:asymmetries2}
\end{table*}

\section{Exclusion limits
\label{sec:exclusion_limits}}

Confidence levels (CLs) for excluding the scalar and pseudoscalar nature of the top quark couplings to the DM mediator are computed in two distinct scenarios:
\begin{itemize}
    \item \textit{Scenario 1} : Exclusion of the SM plus the addition of a new CP-mixed DM mediator, assuming the SM. In this instance, $H_0$ is the SM-only hypothesis, while $H_1$ is the SM plus a new CP-mixed DM mediator;
    \item \textit{Scenario 2} : Exclusion of the SM plus the addition of a new CP-mixed DM mediator, assuming the SM plus a new pure CP-even DM mediator which has already been discovered. Here, $H_0$ is the SM plus a new CP-even DM mediator signal hypothesis, while $H_1$ is the SM plus a new CP-mixed DM mediator signal.   
\end{itemize}

For each scenario, these confidence levels (CLs) are computed for two different luminosities: one roughly corresponding to the expected integrated luminosity at the end of the LHC Run 3, and the other to the High-Luminosity phase of the LHC (HL-LHC). In each analyzed scenario, one hundred thousand pseudo-experiments were generated by applying bin-by-bin Poisson fluctuations in the observables distributions used to evaluate the CLs (i.e. $b_2$, $\tilde{b}_{2}^{\widehat{y}}$ and $\tilde{b}_{2}^{\widehat{d}}$, $n_{2}$ and $n_4$). The 2-binned distributions that are utilized to define the asymmetries and the full shape distributions, were used. These fluctuations were performed using the contents of each distribution's bins as the expected mean value~\cite{Azevedo_2020}. Evaluating whether a null hypothesis ($H_0$) or an alternative hypothesis ($H_1$ i.e., the signal hypothesis) can effectively explain the pseudo-experiment involves assessing the probabilities associated with each hypothesis. The likelihood ratio between the signal and the null hypothesis is used as our test statistics to compute the confidence level with which hypothesis $H_1$ is excluded, assuming $H_0$ is true.

In Table~\ref{tab:asymmetries_vs_full_shapes}, the impact of using the asymmetries or the full shape of the differential distributions in the couplings limits, is shown. It is remarkable to observe the consistency in exclusion limits across most observables, implying the robustness of these measurements despite computing such values using full shape distribution and their corresponding asymmetric 2-binned distribution. Notably, the minimal variations in the exclusion limits, particularly for the $\tilde{b}_{2}^{\widehat{y}}$ and $n_{2}$ observables, reinforce the stability of these parameters against the applied cuts. However, the observable $n_{4}$ stands out, showing a slight increase on the allowed values of the couplings when using the 2-binned distribution associated to the asymmetries.

\begin{figure*}
\begin{center}
\begin{tabular}{ccc}
\hspace*{-5mm}\includegraphics[height=6.1cm]{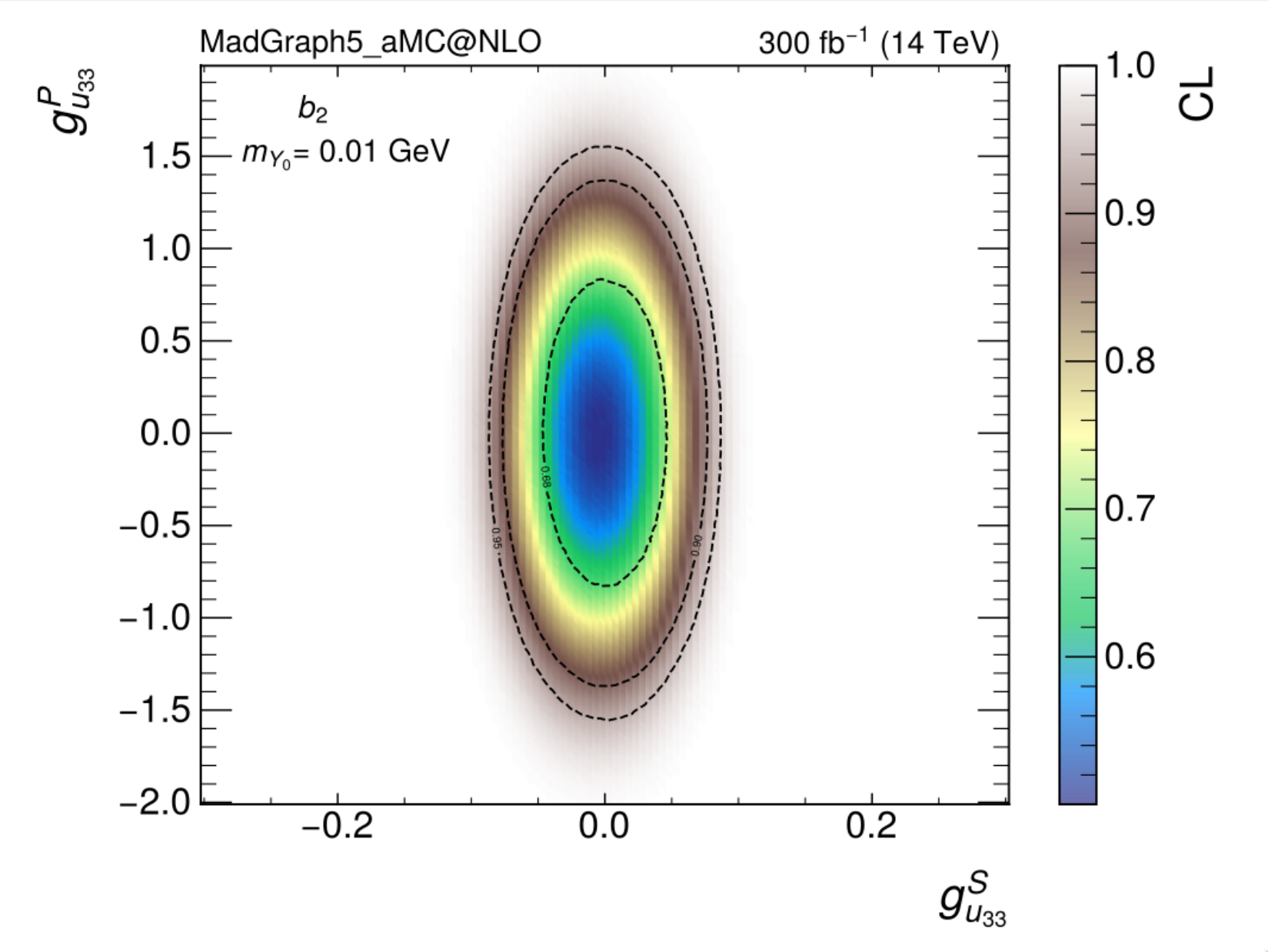}
\hspace*{0mm}\includegraphics[height=6.1cm]{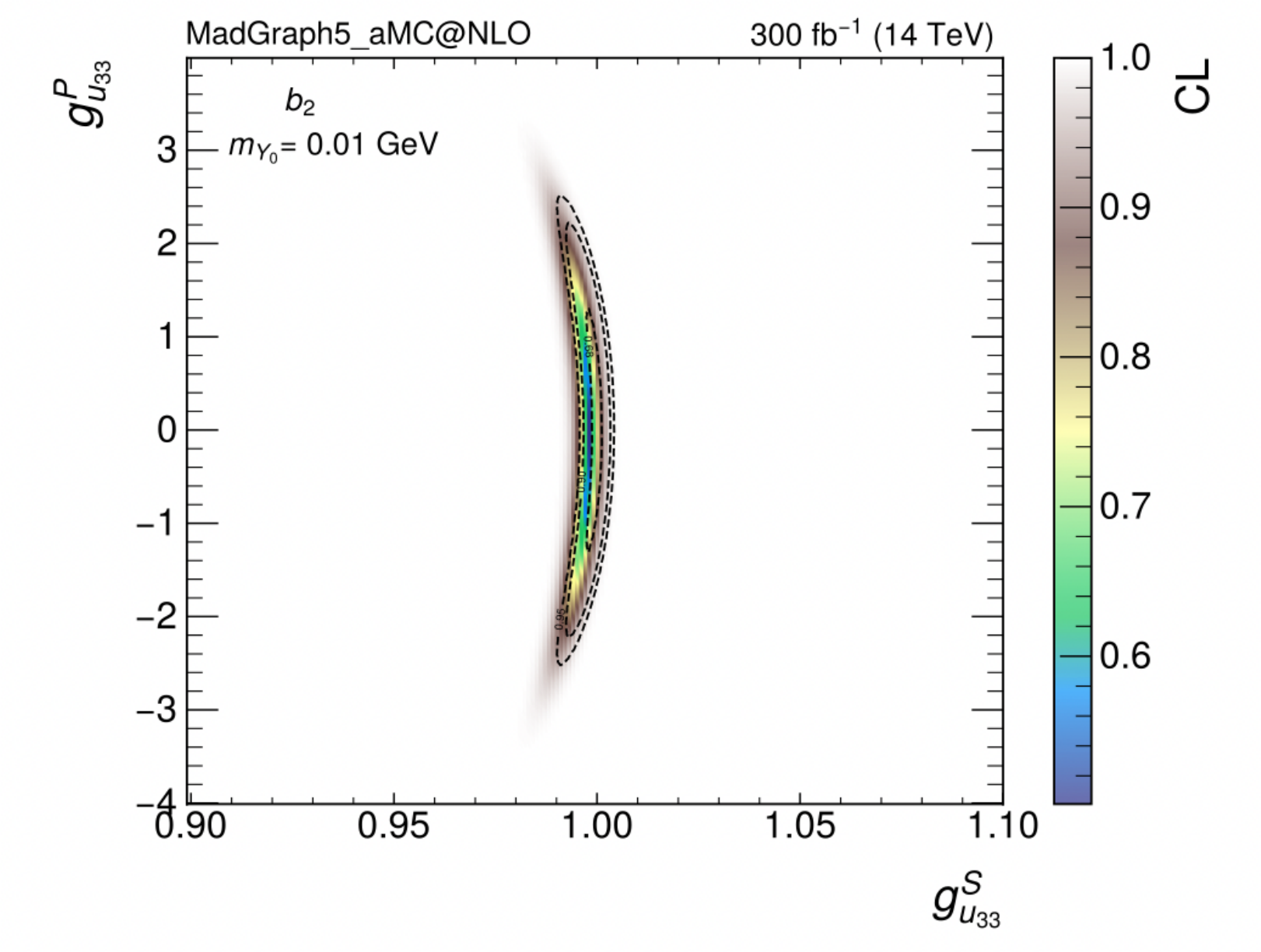}
\\
\hspace*{-5mm}\includegraphics[height=6.1cm]{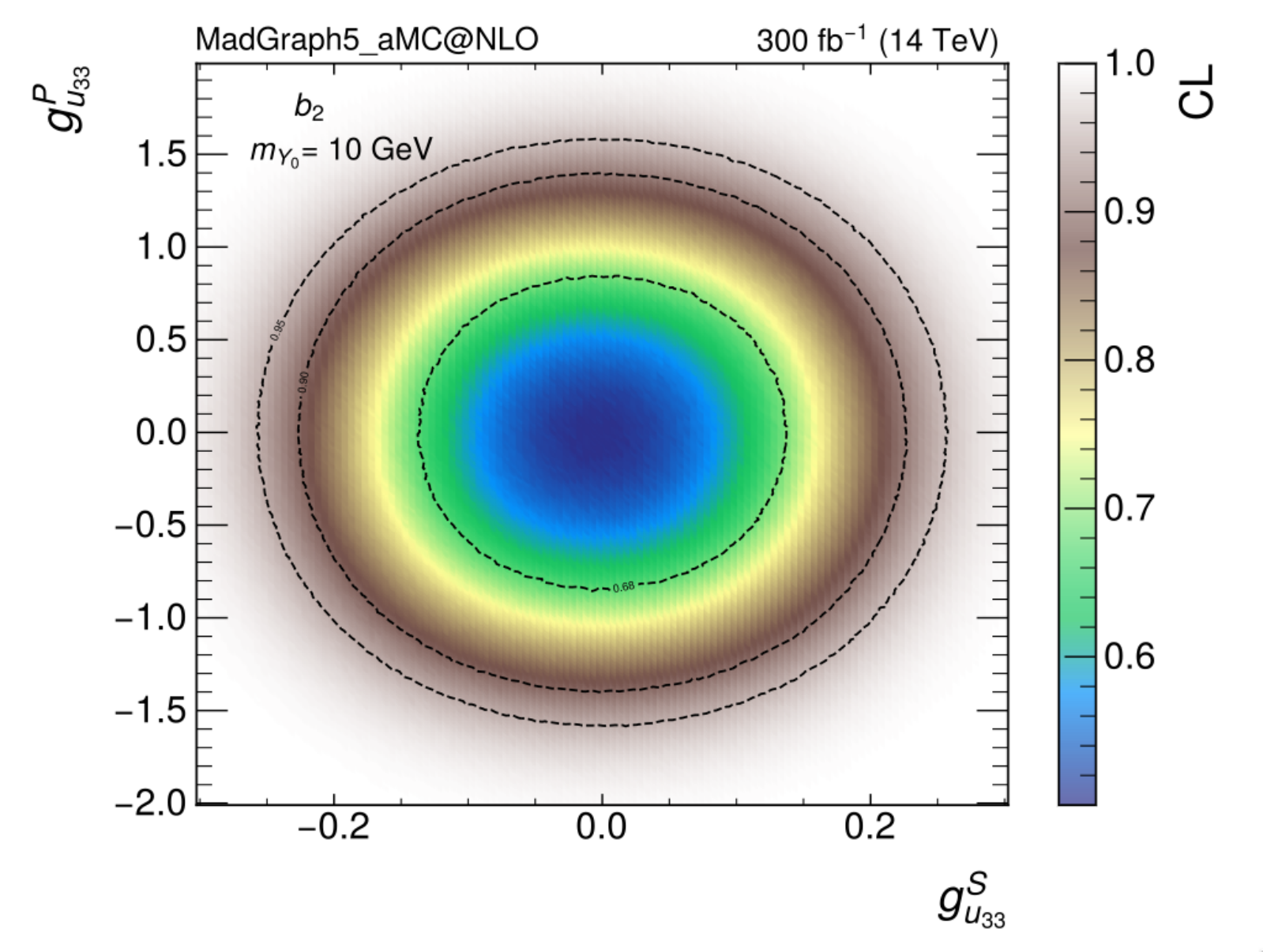}
\hspace*{0mm}\includegraphics[height=6.1cm]{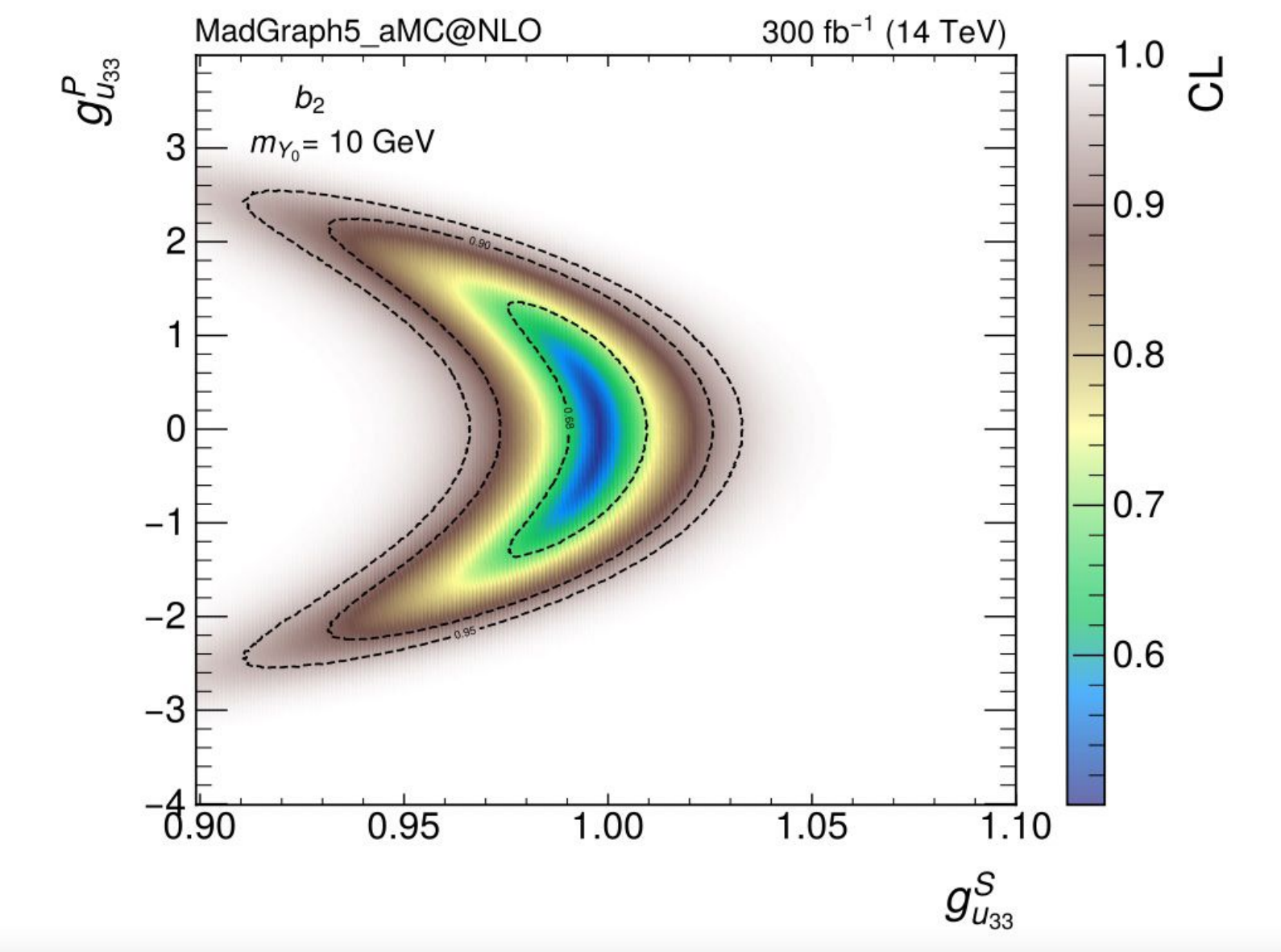}
\\
\hspace*{-5mm}\includegraphics[height=6.1cm]{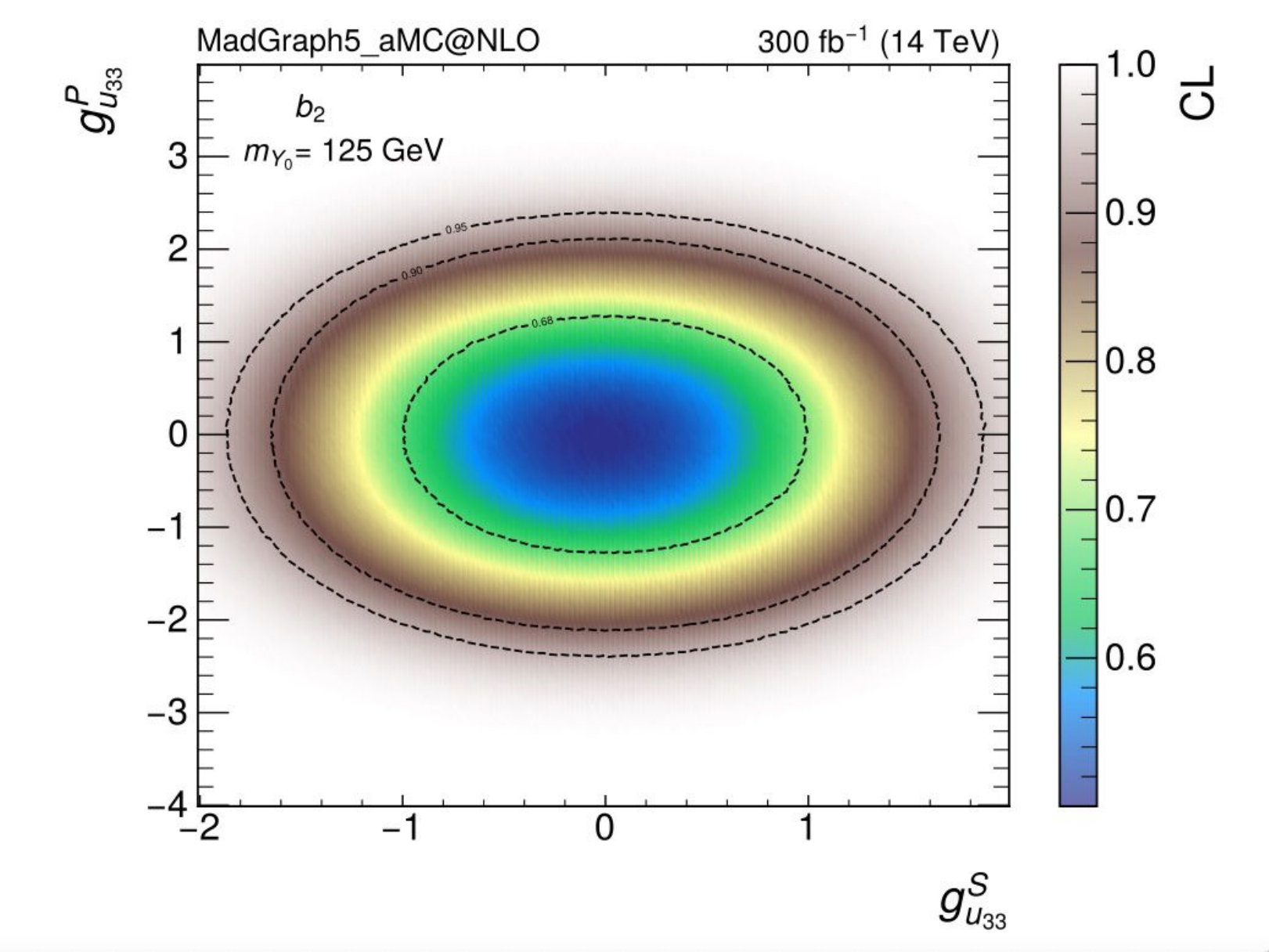}
\hspace*{0mm}\includegraphics[height=6.1cm]{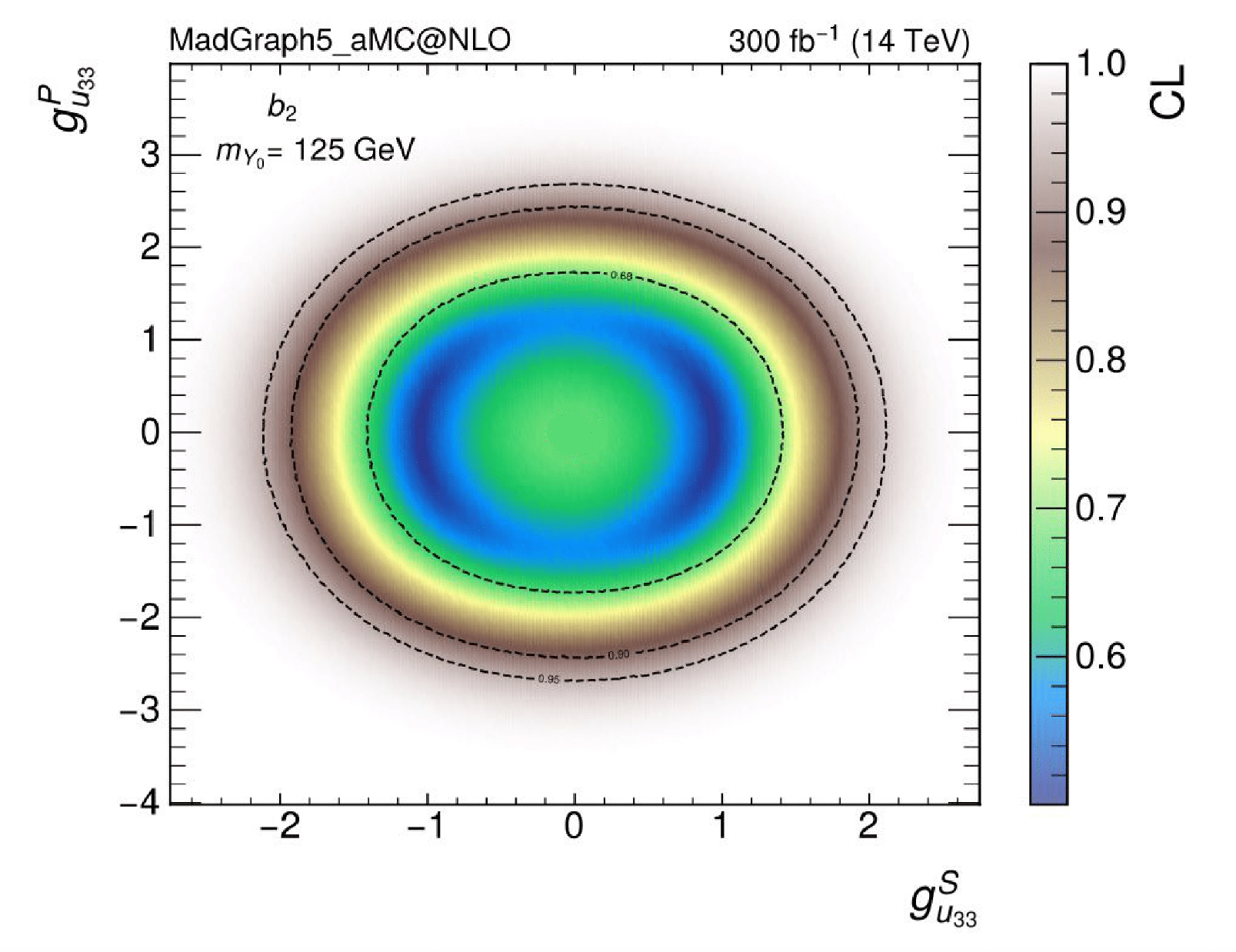}

\end{tabular}
\end{center}
\caption{Exclusion limits for the $\mathrm{SM}$ with a massive DM mediator in the low mass region, $Y_0\left(m_{Y_0}=10^{-2},10 \right.$ and the Higgs benchmark case of $125 \mathrm{GeV}$ in the top, middle, and bottom rows, respectively). We have included the mixed scalar and pseudoscalar couplings against the SM as null hypothesis (left), for the spin observable $b_{2}$ and their corresponding exclusion limit obtained from the SM plus a mixed DM mediator against the SM plus a pure scalar mediator as a null hypothesis (right). Limits are shown for the expected luminosity of the Run 3 of the LHC of $L=300~\mathrm{fb}^{-1}$.}
\label{fig:exclusion_limits_b2}
\end{figure*}

\begin{figure*}
\begin{center}
\begin{tabular}{ccc}
\hspace*{-5mm}\includegraphics[height=6.1cm]{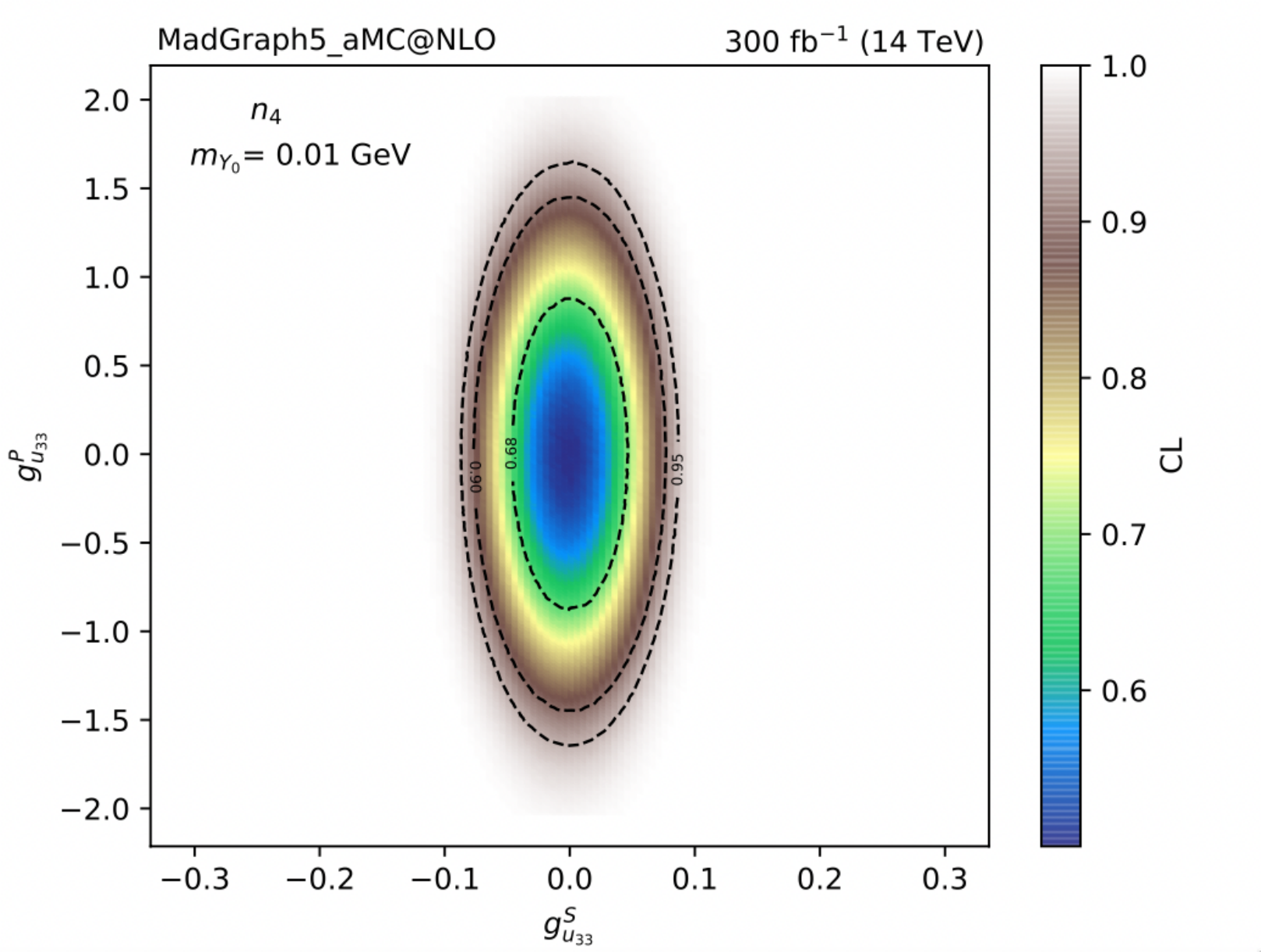}
\hspace*{0mm}\includegraphics[height=6.1cm]{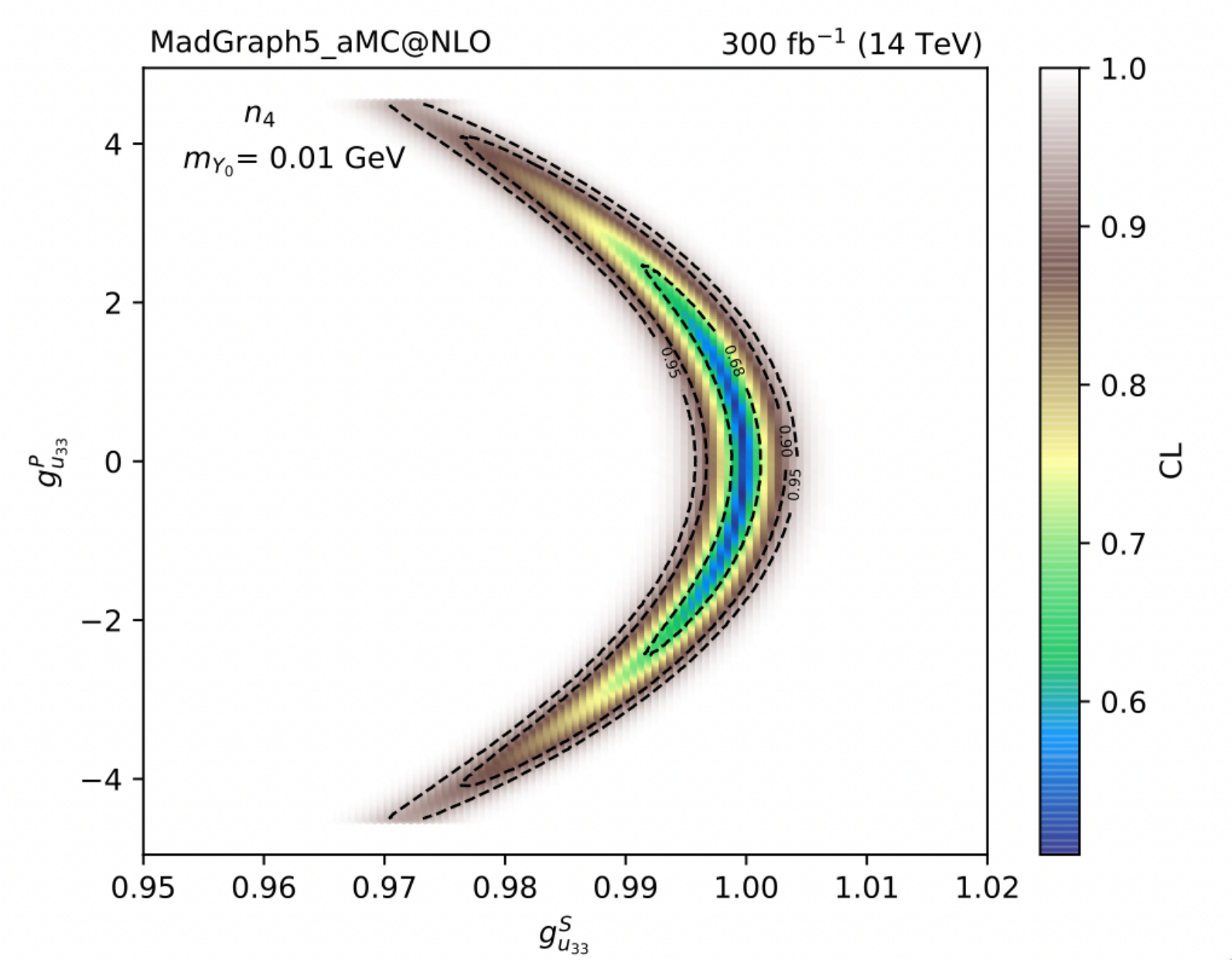}
\\
\hspace*{-5mm}\includegraphics[height=6.1cm]{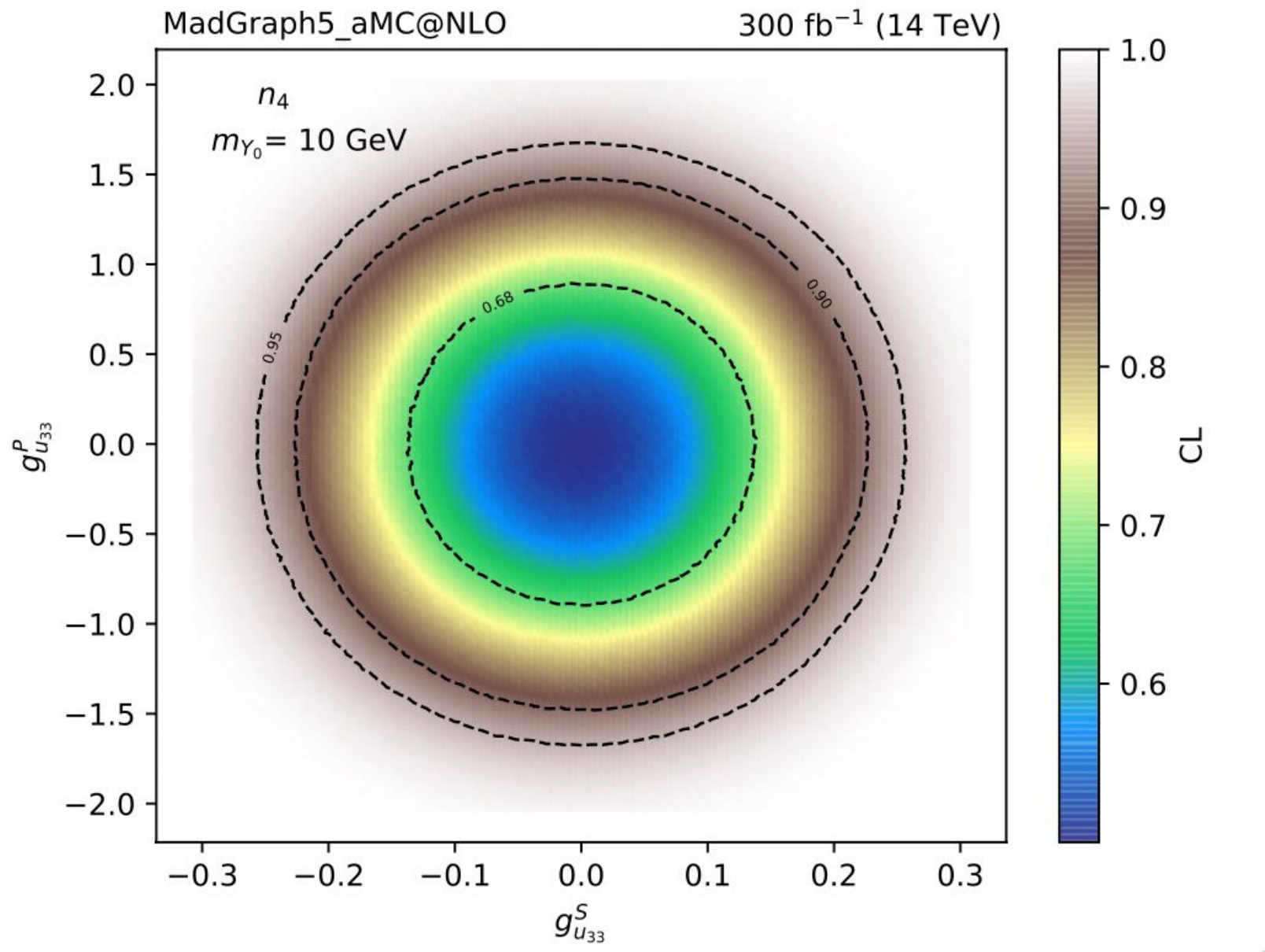}
\hspace*{0mm}\includegraphics[height=6.1cm]{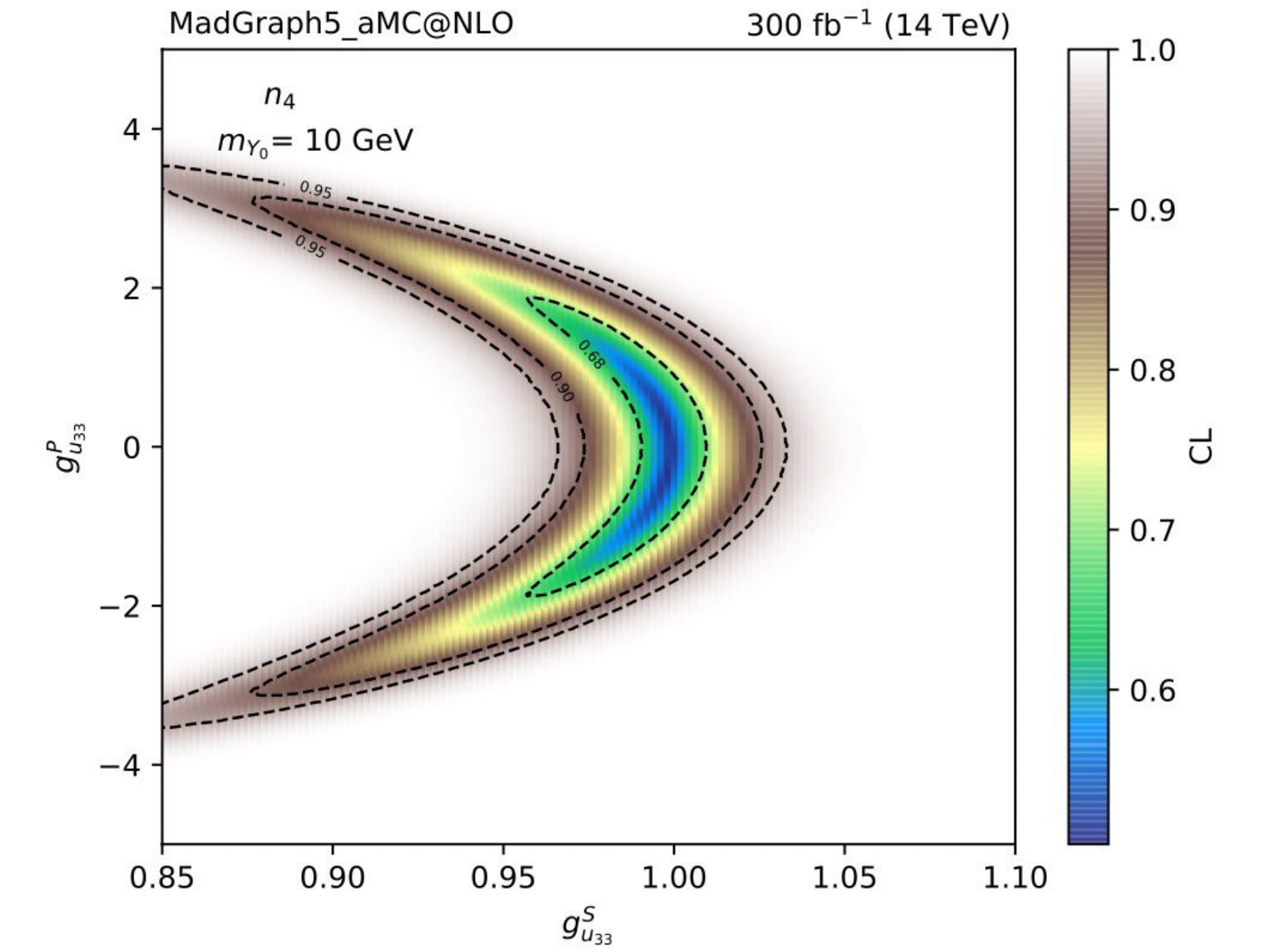}
\\
\hspace*{-5mm}\includegraphics[height=6.1cm]{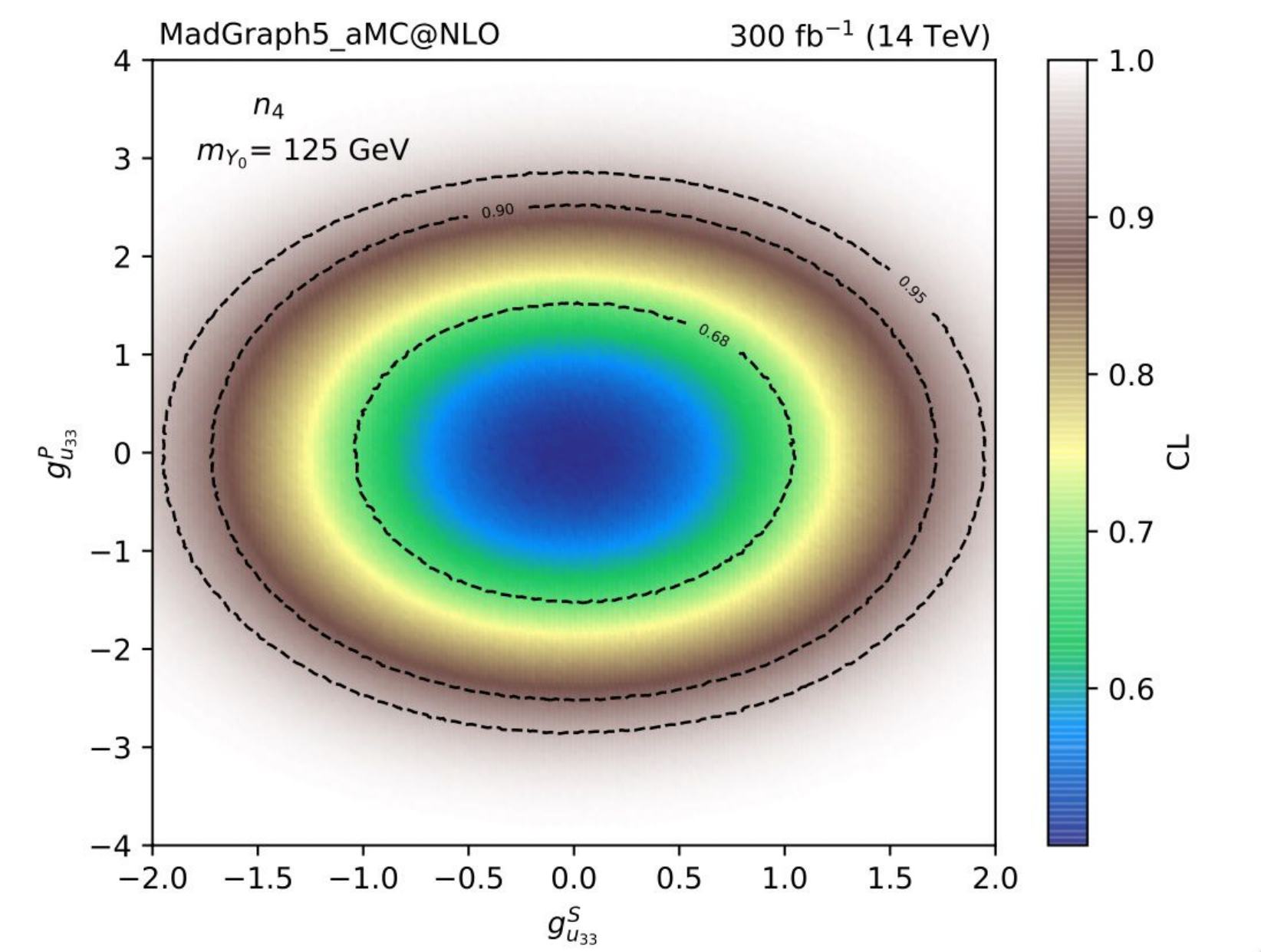}
\hspace*{0mm}\includegraphics[height=6.1cm]{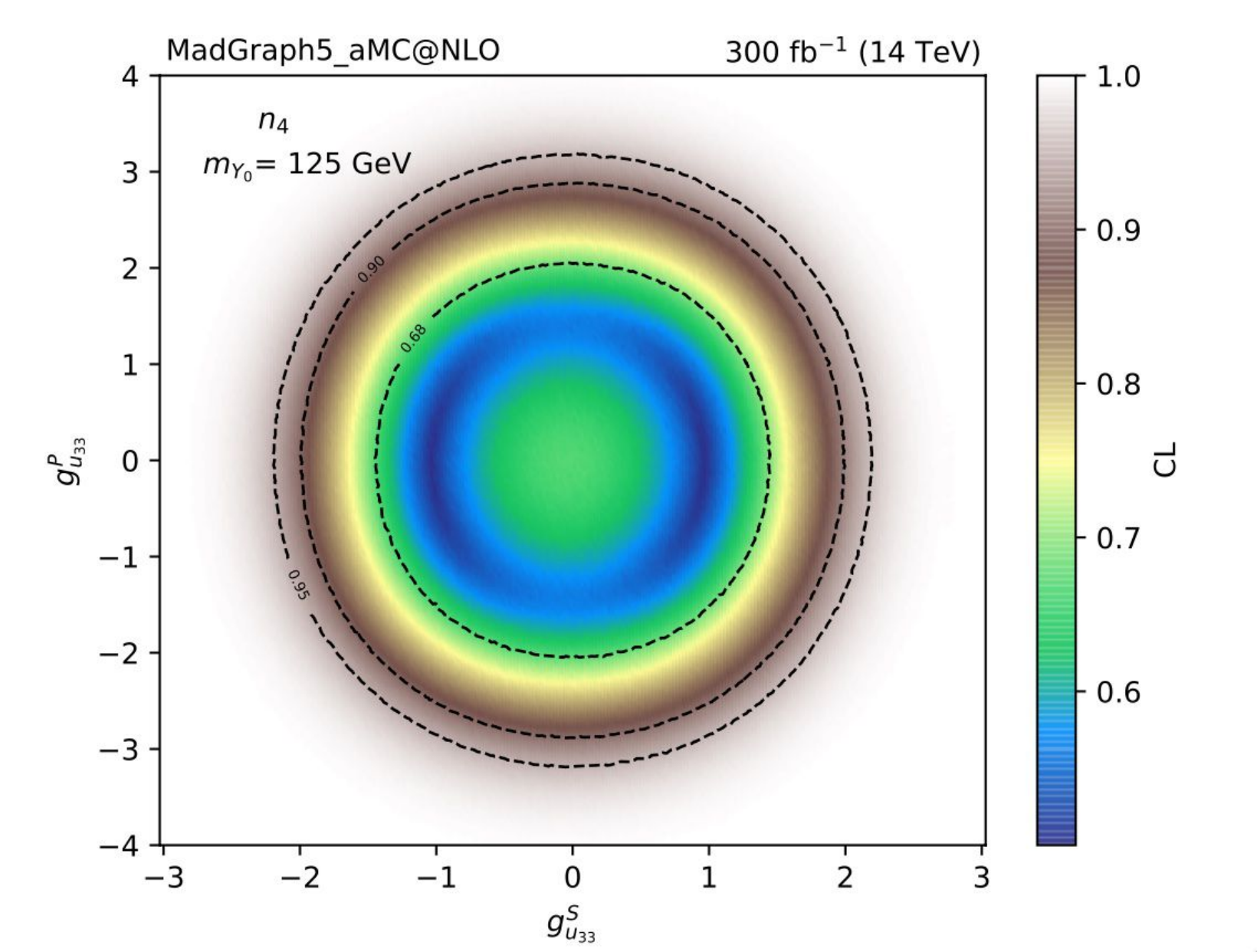}

\end{tabular}
\end{center}
\caption{Exclusion limits for the SM with a massive DM mediator in the low mass region, $Y_0 (m_{Y_0}=10^{-2},10).$ and the Higgs benchmark case of $125 \mathrm{GeV}$ in the top, middle, and bottom rows, respectively). We have included the mixed scalar and pseudoscalar couplings, against the SM as null hypothesis (left), for the spin observable $n_{4}$ and their corresponding exclusion limit obtained from the SM plus a mixed DM mediator against the SM plus a pure scalar mediator as a nul hypothesis (right). Limits are shown for a luminosity of the Run 3 of the LHC of $L=300~ \mathrm{fb}^{-1}$.}
\label{fig:exclusion_limits_n4}
\end{figure*}

The computation of the exclusion limits in the $g_{u_{33}}^S-g_{u_{33}}^P$ plane are obtained using the 2-binned distribution associated to the asymmetries. In Figure~\ref{fig:exclusion_limits_b2}, it is presented the exclusion limits for the $b_2$ observable. An interesting trend is observed: in \textit{Scenario 1} there is a notable improvement in the constraint of the pseudoscalar component as the mass of the mediator decreases, even without trying to reconstruct the mass of the mediator. This contrasts sharply with the results for \textit{Scenario 2}, where no such improvement in the exclusion limits for the pseudoscalar component is evident. Similarly, Figure~\ref{fig:exclusion_limits_n4}, focusing on the $n_4$ observable, mirrors this pattern, with \textit{Scenario 1} showing improved constraints on the pseudoscalar component with decreasing mediator mass, while \textit{Scenario 2} does not exhibit this enhancement.

\begin{table*}
\begin{ruledtabular}
        \begin{tabular}{ccccc}
        \multirow{2}{*}{Observable ($\alpha[\mathcal{O}_{CP}]$)} & \multicolumn{2}{c}{Full-shape ($\alpha[\mathcal{O}_{CP}]$)} & \multicolumn{2}{c}{Asymmetry $(\ensuremath{A^{\alpha[\mathcal{O}_{CP}]}}_{FB})$}  \\
        & $g_{u_{33}}^S$ & $g_{u_{33}}^P$  & $g_{u_{33}}^S$ & $g_{u_{33}}^P$ \\
          \hline       
        $b_2$ & $[-0.0475, 0.0425]$  & $[-0.87, 0.87]$  & $[-0.0425, 0.0475]$  & $[-0.83, 0.83]$   \\
        $\tilde{b}_{2}^{\widehat{y}}$ & $[-0.0425, 0.0425]$  & $[-0.87, 0.87]$ & $[-0.0425, 0.0425] $ & $[-0.87, 0.87]$ \\
        $\tilde{b}_{2}^{\widehat{d}}$ & $[-0.0475, 0.0425] $ & $[-0.89, 0.87]$ & $[-0.0425, 0.0475] $ & $[-0.89, 0.87]$ \\
        $n_{2}$ & $[-0.0425, 0.0475]$ & $[-0.89, 0.89]$ & $[-0.0425, 0.0475]$ &  $[-0.89, 0.89]$  \\
        $n_{4}$ & $[-0.0425, 0.0425]$ & $[-0.87, 0.89]$  & $[-0.0450, 0.0450]$ &  $[-0.89, 0.89]$  \\
    \end{tabular}
\end{ruledtabular}
    \caption{ Exclusion limits at 68\% for two different distribution shape scenarios studied in this paper: 9-bin distributions (Full-shape) compared with the 2-binned one post cut application in the distributions, namely symmetrically applied cut (Forward-Backwards). The values refer to a particle mediator mass of 10$^{-2}$~GeV.}
    \label{tab:asymmetries_vs_full_shapes}
\end{table*}

In Table~\ref{tab:exclusion_limits_b2FB}, which presents exclusion limits for the coupling constant $g^{S,P}_{33}$ from $b_2$ asymmetries in forward-backward (FB) scenarios, it is observed, for both the scalar and pseudoscalar couplings, a sensitivity improvement with increasing luminosity, from 300~fb$^{-1}$ to 3000~fb$^{-1}$ and decreasing mediator mass, as expected. For instance, at 95\% confidence level (CL), the exclusion limits for $g_{u_{33}}^S$ at $10^{-2}$~GeV and 125~GeV changes from $[-0.0475, 0.0475]$ to $[-1.0575, 1.0575]$ for 3000~fb$^{-1}$.

\begin{table*}
  \centering
  \begin{ruledtabular}
      \begin{tabular}{cccccc}
  \multicolumn{6}{c}{\textbf{Exclusion Limits from $b_2$ Asymmetries FB}} \\
    \multicolumn{2}{c}{} & \multicolumn{2}{c}{300~fb$^{-1}$} & \multicolumn{2}{c}{3000~fb$^{-1}$} \\
    $m_{Y_0}$ & & \texttt{(68\% CL)} & \texttt{(95\% CL)} & \texttt{(68\% CL)} & \texttt{(95\% CL)} \\
    \hline
    \multirow{2}{*}{$10^{-2}$GeV} & $g_{u_{33}}^S \in$ & $[-0.0425, 0.0475]$ & $[-0.0875, 0.0875]$ & $[-0.0225, 0.0225]$ & $[-0.0475, 0.0475]$ \\
    & $g_{u_{33}}^P \in$ & $[-0.83, 0.83]$ & $[-1.57, 1.57]$ & $[-0.4725, 0.4575]$ & $[-0.8775, 0.8925]$ \\
    \multirow{2}{*}{$10$~GeV} & $g_{u_{33}}^S \in$ & $[-0.1375, 0.1375]$ & $[-0.2575, 0.2625]$ & $[-0.0775, 0.0775]$ & $[-0.1425, 0.1475]$\\
    & $g_{u_{33}}^P \in$ & $[-0.85, 0.85]$ & $[-1.61, 1.61]$ & $[-0.4725, 0.4725]$ & $[-0.8925, 0.8925]$  \\
    \multirow{2}{*}{$125$~GeV} & $g_{u_{33}}^S \in$ & $[-1.01, 1.015]$ & $[-1.885, 1.89]$ & $[-0.5625, 0.5625]$ & $[-1.0575, 1.0575]$ \\
    & $g_{u_{33}}^P \in$ & $[-1.29, 1.27]$  & $[-2.41, 2.43]$  & $[-0.725, 0.725]$ & $[-1.35, 1.375]$ \\
    \end{tabular}
  \end{ruledtabular}
    \caption{Exclusion limits for the $t\Bar{t} Y_0$ CP-couplings, considering the CP-mixed case against irreducible backgrounds, are depicted for various $Y_0$ masses under fixed luminosities of 300~fb$^{-1}$ and 3000~fb$^{-1}$. The limits are presented at 68\% and 95\% confidence levels for the $b_2$ observable.}
    \label{tab:exclusion_limits_b2FB}
\end{table*}

Additionally, Table~\ref{tab:exclusion_limits_b2KDFB2} details the exclusion limits obtained using the 2-binned distributions associated with the asymmetries as above, for the $n_4$ observable. These results are consistent with the previous ones for the $b_2$ observable. 

\begin{table*}
  \centering
  \begin{ruledtabular}
      \begin{tabular}{cccccc}
  \multicolumn{6}{c}{\textbf{Exclusion Limits from $n_4$ Asymmetries FB}} \\
    \multicolumn{2}{c}{} & \multicolumn{2}{c}{300~fb$^{-1}$} & \multicolumn{2}{c}{3000~fb$^{-1}$} \\
    $m_{Y_0}$ & & \texttt{(68\% CL)} & \texttt{(95\% CL)} & \texttt{(68\% CL)} & \texttt{(95\% CL)} \\
    \hline
    \multirow{2}{*}{$10^{-2}$~GeV} & $g_{u_{33}}^S \in$ & $[-0.0450, 0.0450]$ & $[-0.0875, 0.0875]$ & $[-0.0225, 0.0225]$ & $[-0.0475, 0.0475]$ \\
    & $g_{u_{33}}^P \in$ & $[-0.8900, 0.8900]$ & $[-1.6700, 1.6700]$ & $[-0.5025, 0.5025]$  & $[-0.9375, 0.9375]$ \\
    \multirow{2}{*}{$10$~GeV} & $g_{u_{33}}^S \in$ & $[-0.135, 0.135]$ & $[-0.255, 0.255]$ & $[-0.075, 0.075]$ & $[-0.1475, 0.1425]$ \\
    & $g_{u_{33}}^P \in$ & $[-0.9100, 0.8900]$ & $[-1.7100, 1.6900]$ & $[-0.5025, 0.5025]$ & $[-0.9525, 0.9525]$ \\
    \multirow{2}{*}{$125$~GeV} & $g_{u_{33}}^S \in$ & $[-1.05, 1.05]$ & $[-1.965, 1.965]$ & $[-0.593, 0.593]$ & $[-1.118, 1.118]$ \\
    & $g_{u_{33}}^P \in$ & $[-1.53, 1.55]$ & $[-2.89, 2.87]$ & $[-0.85, 0.85]$ &  $[-1.63, 1.63]$\\
    \end{tabular}
  \end{ruledtabular}
    \caption{Exclusion limits for the $t\Bar{t} Y_0$ CP-couplings, considering the CP-mixed case against irreducible backgrounds, are depicted for various $Y_0$ masses under fixed luminosities of 300~fb$^{-1}$ and 3000~fb$^{-1}$. The limits are presented at 68\% and 95\% confidence levels for the $n_4$ observable.}
    \label{tab:exclusion_limits_b2KDFB2}
\end{table*}

For completeness, we have checked that for the $\tilde{b}_{2}^{\widehat{y}}$, $\tilde{b}_{2}^{\widehat{d}}$ and $n_2$ observables lead to similar results on the exclusion limits of the couplings. In Appendix~\ref{app:A} we show these limits in the ($g^{S}_{33}$,$g^{P}_{33}$) plane, obtained using the full distribution shapes of the $b_2$, $\tilde{b}_{2}^{\widehat{y}}$ and $\tilde{b}_{2}^{\widehat{d}}$ (Figure~\ref{fig:comparison_observables_1}) and $n_4$ and $n_2$ (Figure~\ref{fig:comparison_observables_2}), respectively. The corresponding 1D limits are also shown in Tables~\ref{tab:exclusion_limits_b2},~\ref{tab:exclusion_limits_b2Y},~\ref{tab:exclusion_limits_b2KD},~\ref{tab:exclusion_limits_n4}, and~\ref{tab:exclusion_limits_n2} of Appendix~\ref{app:B}. In Appendix~\ref{app:C}, we show for all variables the 1D limits using the 2-binned distributions associated to the asymmetries.

\subsection{Exclusion limits in the total cross-section
\label{sec:total_cross_section}}

\begin{figure*}
\begin{center}
\includegraphics[height=6.5cm]{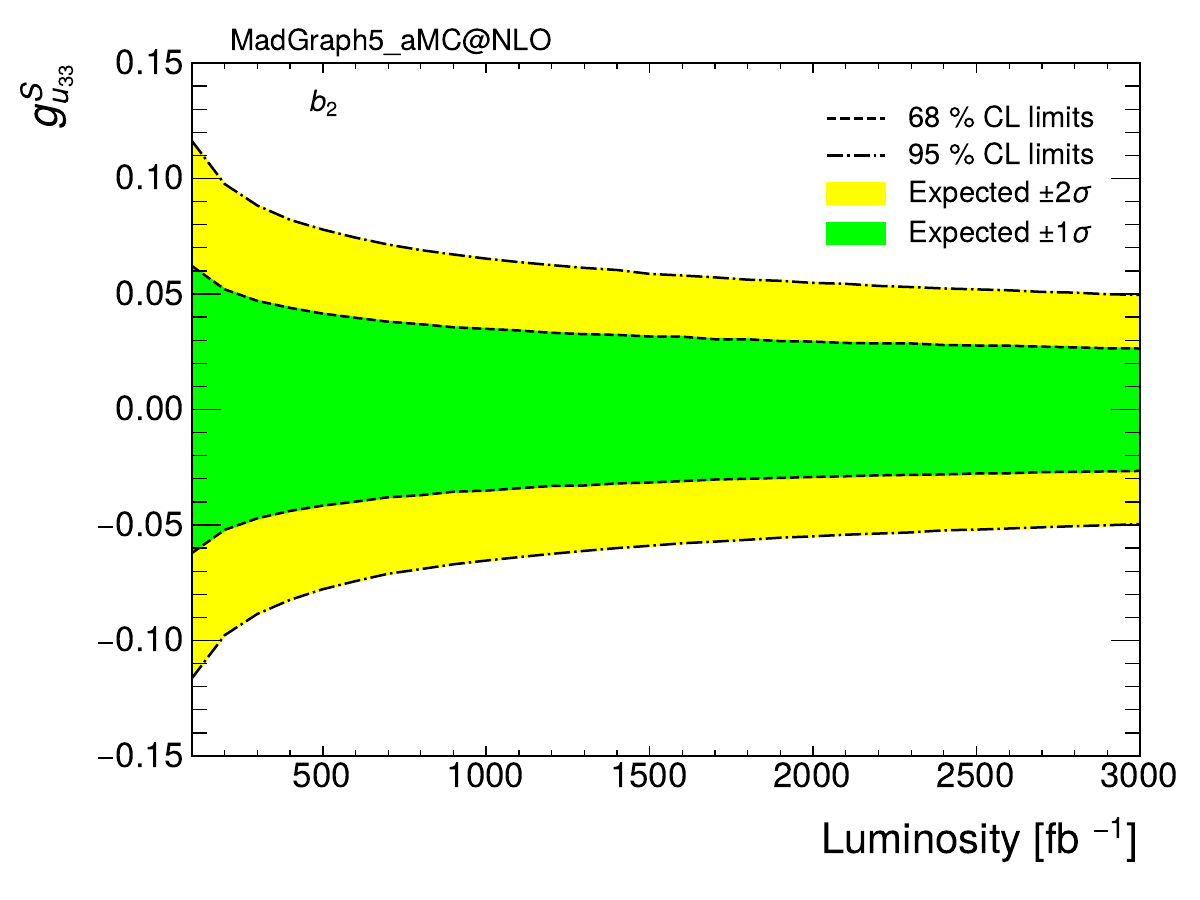}
\includegraphics[height=6.5cm]{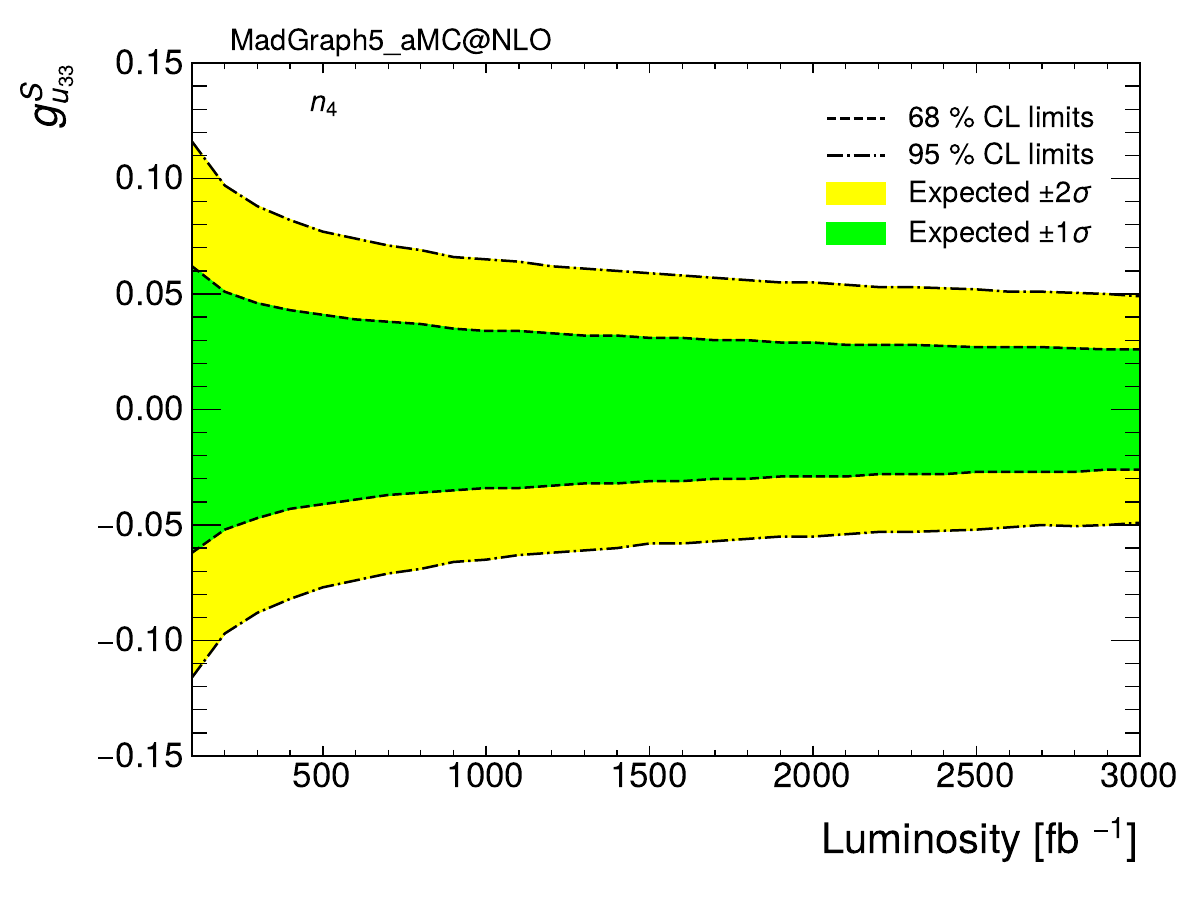}
\end{center}
\caption{Expected exclusion limits at 95\% and 68\% CL for the $g_{u_{33}}^S$ scalar coupling assuming vanishing pseudoscalar component $g_{u_{33}}^P = 0$ as a function of the luminosity in fb$^{-1}$ assuming the SM as the null hypothesis for the most sensitive observables in this work $b_{2}$ (left) and $n_{4}$ (right).}
\label{fig:total_cross_section}
\end{figure*}

In addition to the CLs for the different benchmark cases, the sensitivity of the analysis to the coupling constants and cross sections with the luminosity, is investigated. As a first approach, this sensitivity is studied by fixing the pseudoscalar coupling constant to a vanishing value ($g^{P}_{33}=0$) and assuming the null hypothesis is the SM background. The focus is on determining exclusion limits in a scenario where a pure scalar mediator is considered. Figure~\ref{fig:total_cross_section} shows the dependence of the expected 95$\%$ and 68$\%$ CLs as a function of the luminosity for a mediator mass of 1$\times10^{-2}$~GeV, for the most sensitive observables, i.e. $b_{2}$ (left) and $n_{4}$ (right). It is noticeable that, as the luminosity increases, the improvement on the $g^{S}_{33}$ CLs can be as high as a factor of two when the luminosity increases from $100$~fb$^{-1}$ to $3000$~fb$^{-1}$, at the end of the HL-LHC.

In addition, a flattening is observed in the behavior of the exclusion limits as the luminosity approaches the end of the HL-LHC for both variables. In fact, no significant gains are expected after, roughly, 2000~fb$^{-1}$.

\begin{figure*}
\begin{center}
\includegraphics[height=6.5cm]{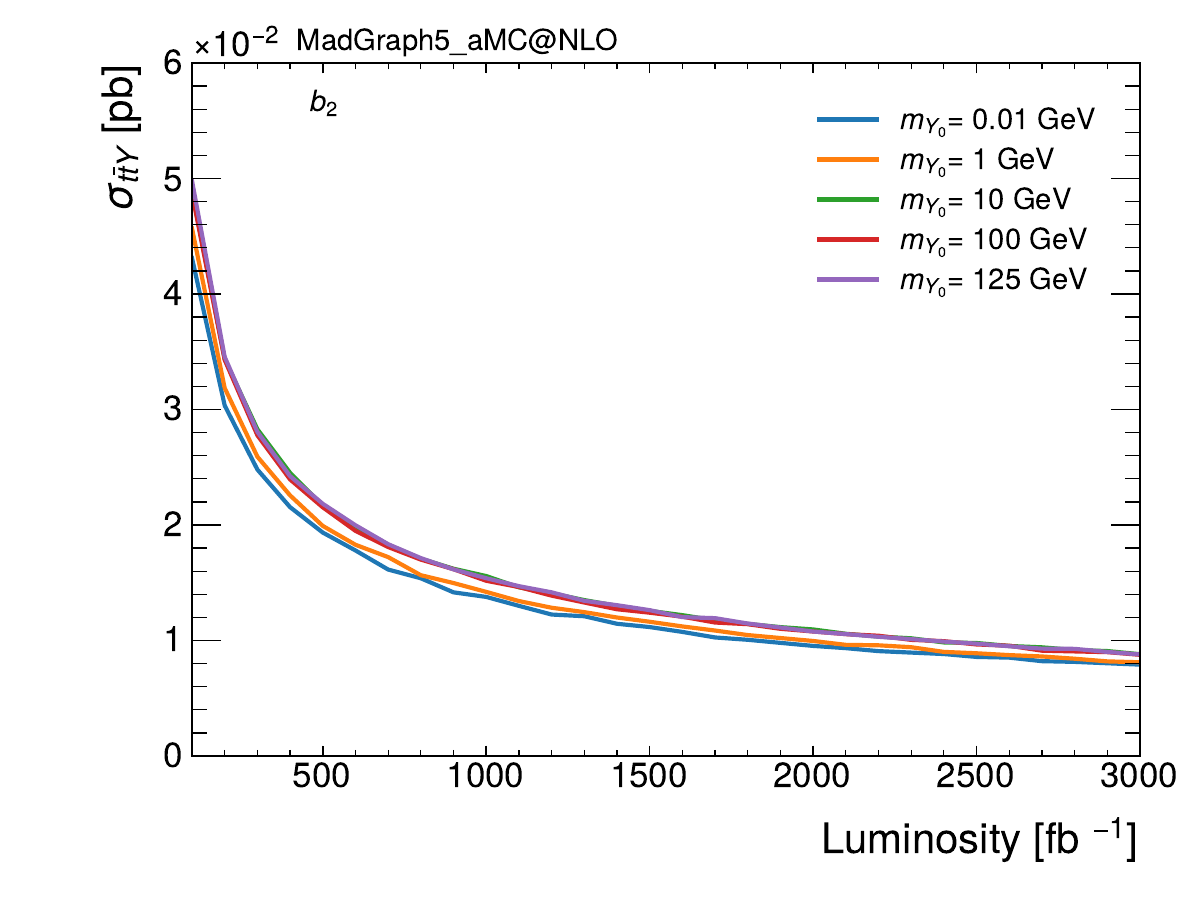}
\includegraphics[height=6.5cm]{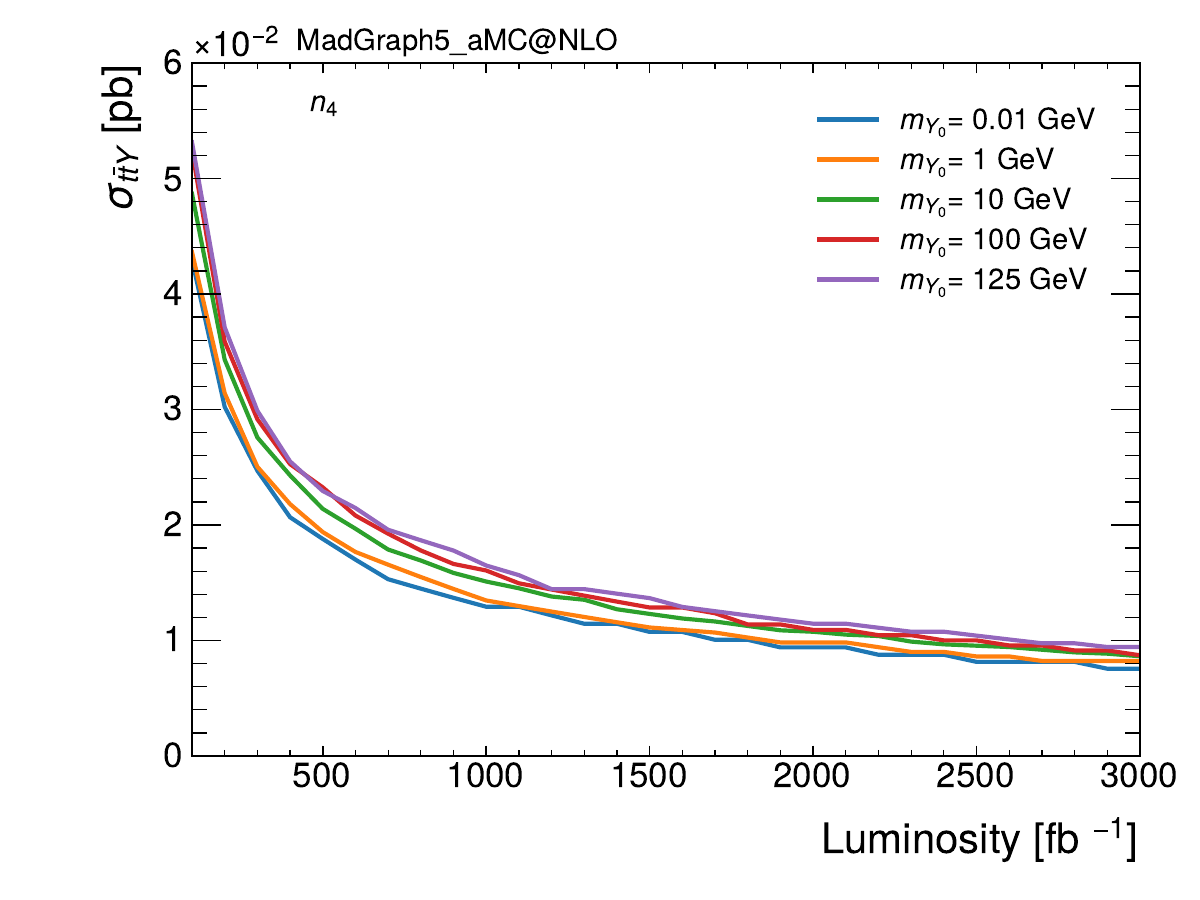}
\end{center}
\caption{Expected exclusion limits at 68\% CL for the total cross-section $\sigma_{ttY}$ for production of a pure scalar mediator particle $Y_{0}$ as a function of the luminosity in fb$^{-1}$ assuming the SM as the null hypothesis for the most sensitive observables in this work $b_{2}$ (left) and $n_{4}$ (right).}
\label{fig:running_lumin}
\end{figure*}

The $b_{2}$ observable shows the small impact of the mass in the exclusion limits for the scalar mediator production cross-section, as a function of the luminosity (Figure~\ref{fig:running_lumin}, left). We have computed the cross-sections, including NLO corrections, for the range of mass values considered in this paper. Despite increasing the luminosity beyond 2000~fb$^{-1}$, the enhancement in the exclusion limit of the total cross-section is not as noticeable as the one observed in the interval from 300~fb$^{-1}$ to 600~fb$^{-1}$. This finding suggests that a pure increase in luminosity is not sufficient to improve these limits. The most sensitive spin basis observable shows an identical result as the $b_{2}$ observable (Figure~\ref{fig:running_lumin}, right).

As expected, the profile of the distribution is irrelevant in the improvement of the total cross-section exclusion limit, as it has already been shown from the forward-backwards asymmetry study for \textit{Scenario 1}. The observed consistency across different mediator masses reinforces the robustness of the analysis and suggests that the understanding of the underlying physics is on solid ground. his opens up the possibility of exploring a wider mass range of the DM mediator with the current analysis, even in the case no attempt is performed to reconstruct the mediator.

\section{Conclusions \label{sec:conclusions}}

In this paper, we study the possibility of using spin motivated angular distributions to search for the presence of DM mediators in $t\bar{t}Y$ events at the LHC, with luminosities that range from $ 100~\mathrm{fb}^{-1}$ up to the end of the HL-LHC ($3000~\mathrm{fb}^{-1}$). Signal events from $t\bar{t}Y_0$ associated production are searched for, in dileptonic final states of the $t\bar{t}$ system. In this study we performed the reconstruction of the $t\bar{t}$ system, but no attempt to reconstruct the DM mediator was tried. We assume that the total transverse missing energy was due to the undetected neutrinos from the top-quarks decays. This approximation has proven to be quite reasonable until the mass of the DM mediator gets closer to the Higgs boson mass (i.e. 125~GeV). For higher mass scales (around 1~TeV) tails on the distributions are observed and the approximation is no longer valid. Five different DM mediator benchmark masses were used in our studies, i.e. 1$\times10^{-2}$, 1, 10, 100 and 125~GeV.

By introducing a suite of novel observables, $b_2$, $\tilde{b}_{2}^{\widehat{y}}$, $\tilde{b}_{2}^{\widehat{d}}$, $n_2$ and $ n_4 $, we have extracted exclusion limits in the signal couplings plane ($g^{S}_{u_{33}}, g^{P}_{u_{33}}$), in two different scenarios, i.e. assuming the SM as the null-hypothesis (\textit{Scenario 1}), and assuming that a scalar DM mediator was already discovered (\textit{Scenario 2}). For \textit{Scenario 1} no significant changes are expected on the exclusion limits across different variables. However, for \textit{Scenario 2} the shape of distributions is quite important and improved results are expected when using the variables $b_2$ and $n_4$. Similar results were obtained by using 2-binned distributions utilized to define FB asymmetries on the different variables, showing indeed that asymmetries can be used to study the sensitivity to this new physics.

Remarkably, our studies demonstrated that asymmetries derived from angular CP observables can indeed be used to probe DM mediators in $t\bar{t}Y_0$ events at the LHC, without the need of the full-shape distributions. Asymmetries associated to the top-Higgs coupling which depend on potential interference terms were proposed in the literature~\cite{Mileo_2016}. However, given the fact that the sensitivity to interference terms, even at the HL-LHC, is quite negligible~\cite{Azevedo_2022} for $t\bar{t}H$, the measurement of these asymmetries may be challenging at the LHC. The asymmetries we proposed in this paper for the $b_2$, $\tilde{b}_{2}^{\widehat{y}}$ and $\tilde{b}_{2}^{\widehat{d}}$, $n_2$ and $ n_4 $, (\ref{tab:asymmetries1}) are significant, roughly an order of magnitude bigger at parton level when compare with the ones proposed in~\cite{Mileo_2016}, and do not change significantly with the mass of the DM mediator.

Our results can be compared with those in the context of SUSY and 2HDM-a searches, with similar final states topologies ($t\bar{t}$~X)~\cite{DeRoeck:2024fjq,CMS:2024zqs,ATLAS:2024rcx} as all scenarios lead to transverse missing energy. Our study's approach, which incorporates novel observables such as the $b_2$, $\tilde{b}_{2}^{\widehat{y}}$ and $\tilde{b}_{2}^{\widehat{d}}$, $n_2$ and $ n_4 $, extend the search explored, to the low mass region down to $1\times10^{-2}$~GeV with clear improvement in the cross-sections limits. We should however stress that in our analysis, no systematic uncertainties were considered. Although angular distributions seem to be an excellent tool to address spin dependent DM mediators, the full study of these observables should be considered by the LHC experiments in order to better address the sensitivity in the low mass region.

\begin{acknowledgments}
E. C. was supported by Fundação para a Ciência e Technologia (FCT), under the Contracts No. PRT/BD/ 154189/2022 and No. CERN/FIS-PAR/0037/2021 funded project. 
R.M.S. was partially supported by FCT Grant No CERN/FIS-PAR/0037/2021. The work of M.C.N.F. was supported by the PSC-CUNY Awards 65071-00 53 and 66073-00 54.
A.O. was partially supported by FCT, under the contract
CERN/FIS-PAR/0037/2021.
R.G  was partially supported by FCT, under the Contract CERN/FIS-PAR/0026/2021. The Monte Carlo \texttt{MadGraph5\_aMC@NLO} simulation we developed and employed in our analysis were run on the computer cluster at LIP.
\end{acknowledgments}

\newpage
\bibliographystyle{h-physrev}
\bibliography{references}

\begin{thebibliography}{10}

\bibitem{drees2012minireview}
M.~Drees and G.~Gerbier,
\newblock {Mini-Review of Dark Matter: 2012}, 2012, 1204.2373.

\bibitem{10.1093/mnras/249.3.523}
K.~G. Begeman, A.~H. Broeils, and R.~H. Sanders,
\newblock Mon. Not. Roy. Astron. Soc. {\bf 249}, 523 (1991).

\bibitem{Corbelli_2000}
E.~Corbelli and P.~Salucci,
\newblock Mon. Not. Roy. Astron. Soc. {\bf 311}, 441 (2000), astro-ph/9909252.

\bibitem{hoekstra2002nf}
H.~Hoekstra, H.~Yee, and M.~Gladders,
\newblock New Astron. Rev. {\bf 46}, 767 (2002), astro-ph/0205205.

\bibitem{rubin1970rotation}
V.~C. Rubin and W.~K. Ford, Jr.,
\newblock Astrophys. J. {\bf 159}, 379 (1970).

\bibitem{moustakas2002iz}
L.~A. Moustakas and R.~B. Metcalf,
\newblock Mon. Not. Roy. Astron. Soc. {\bf 339}, 607 (2003), astro-ph/0206176.

\bibitem{2020}
Planck, N.~Aghanim {\em et~al.},
\newblock Astron. Astrophys. {\bf 641}, A1 (2020), 1807.06205.

\bibitem{atlascollaboration2023combination}
ATLAS, G.~Aad {\em et~al.},
\newblock {Combination and summary of ATLAS dark matter searches interpreted in
  a 2HDM with a pseudo-scalar mediator using 139 fb$^{-1}$ of $\sqrt{s} = 13$
  TeV $pp$ collision data}, 2023, 2306.00641.

\bibitem{GonzalezFernandez:2021nbb}
ATLAS, S.~Gonz\'alez~Fern\'andez,
\newblock PoS {\bf ICHEP2020}, 658 (2021).

\bibitem{Kumar:2022knu}
CMS, D.~Kumar,
\newblock PoS {\bf ICHEP2022}, 260 (2022).

\bibitem{ATLAS:2023tkt}
ATLAS, G.~Aad {\em et~al.},
\newblock Phys. Lett. B {\bf 842}, 137963 (2023), 2301.10731.

\bibitem{Haisch:2021ugv}
U.~Haisch, G.~Polesello, and S.~Schulte,
\newblock JHEP {\bf 09}, 206 (2021), 2107.12389.

\bibitem{Hermann:2021xvs}
J.~Hermann and M.~Worek,
\newblock Eur. Phys. J. C {\bf 81}, 1029 (2021), 2108.01089.

\bibitem{schumann2019direct}
M.~Schumann,
\newblock Journal of Physics G: Nuclear and Particle Physics {\bf 46}, 103003
  (2019).

\bibitem{Milosevic:2021bpv}
CMS, V.~Milo\v{s}evi\'c,
\newblock PoS {\bf ICHEP2020}, 070 (2021).

\bibitem{weinheimer2003laboratory}
C.~Weinheimer,
\newblock Springer Tracts in Modern Physics {\bf 190}, 25 (2003).

\bibitem{Abdallah:2015ter}
J.~Abdallah {\em et~al.},
\newblock Phys. Dark Univ. {\bf 9-10}, 8 (2015), 1506.03116.

\bibitem{Degrande:2014vpa}
C.~Degrande,
\newblock Comput. Phys. Commun. {\bf 197}, 239 (2015), 1406.3030.

\bibitem{Arkani-Hamed:2005qjb}
N.~Arkani-Hamed, G.~L. Kane, J.~Thaler, and L.-T. Wang,
\newblock JHEP {\bf 08}, 070 (2006), hep-ph/0512190.

\bibitem{ATLAS:2022ygn}
ATLAS, G.~Aad {\em et~al.},
\newblock Eur. Phys. J. C {\bf 83}, 503 (2023), 2211.05426.

\bibitem{Biekotter:2022ckj}
T.~Biek\"otter and M.~Pierre,
\newblock Eur. Phys. J. C {\bf 82}, 1026 (2022), 2208.05505.

\bibitem{Bernreuther_1994}
W.~Bernreuther and A.~Brandenburg,
\newblock Phys. Rev. D {\bf 49}, 4481 (1994), hep-ph/9312210.

\bibitem{Gunion_1996}
J.~F. Gunion and X.-G. He,
\newblock Phys. Rev. Lett. {\bf 76}, 4468 (1996), hep-ph/9602226.

\bibitem{Bhupal_Dev_2008}
P.~S. Bhupal~Dev, A.~Djouadi, R.~M. Godbole, M.~M. Muhlleitner, and S.~D.
  Rindani,
\newblock Phys. Rev. Lett. {\bf 100}, 051801 (2008), 0707.2878.

\bibitem{Frederix_2011}
R.~Frederix {\em et~al.},
\newblock Phys. Lett. B {\bf 701}, 427 (2011), 1104.5613.

\bibitem{Ellis_2014}
J.~Ellis, D.~S. Hwang, K.~Sakurai, and M.~Takeuchi,
\newblock JHEP {\bf 04}, 004 (2014), 1312.5736.

\bibitem{Khatibi_2014}
S.~Khatibi and M.~Mohammadi~Najafabadi,
\newblock Phys. Rev. D {\bf 90}, 074014 (2014), 1409.6553.

\bibitem{Demartin_2014}
F.~Demartin, F.~Maltoni, K.~Mawatari, B.~Page, and M.~Zaro,
\newblock Eur. Phys. J. C {\bf 74}, 3065 (2014), 1407.5089.

\bibitem{Kobakhidze_2014}
A.~Kobakhidze, L.~Wu, and J.~Yue,
\newblock JHEP {\bf 10}, 100 (2014), 1406.1961.

\bibitem{Bramante_2014}
J.~Bramante, A.~Delgado, and A.~Martin,
\newblock Phys. Rev. D {\bf 89}, 093006 (2014), 1402.5985.

\bibitem{Boudjema_2015}
F.~Boudjema, R.~M. Godbole, D.~Guadagnoli, and K.~A. Mohan,
\newblock Phys. Rev. D {\bf 92}, 015019 (2015), 1501.03157.

\bibitem{He_2015}
X.-G. He, G.-N. Li, and Y.-J. Zheng,
\newblock Int. J. Mod. Phys. A {\bf 30}, 1550156 (2015), 1501.00012.

\bibitem{Amor_dos_Santos_2015}
S.~P. Amor~dos Santos {\em et~al.},
\newblock Phys. Rev. D {\bf 92}, 034021 (2015), 1503.07787.

\bibitem{Amor_dos_Santos_2017}
S.~Amor Dos~Santos {\em et~al.},
\newblock Phys. Rev. D {\bf 96}, 013004 (2017), 1704.03565.

\bibitem{Gritsan_2016}
A.~V. Gritsan, R.~R\"ontsch, M.~Schulze, and M.~Xiao,
\newblock Phys. Rev. D {\bf 94}, 055023 (2016), 1606.03107.

\bibitem{Dolan_2016}
M.~J. Dolan, M.~Spannowsky, Q.~Wang, and Z.-H. Yu,
\newblock Phys. Rev. D {\bf 94}, 015025 (2016), 1606.00019.

\bibitem{Gon_alves_2016}
D.~Gonçalves and D.~Lopez-Val,
\newblock Phys. Rev. D {\bf 94}, 095005 (2016), 1607.08614.

\bibitem{Gon_alves_2018}
D.~Gon\c{c}alves, K.~Kong, and J.~H. Kim,
\newblock JHEP {\bf 06}, 079 (2018), 1804.05874.

\bibitem{Gon_alves_2022}
D.~Gon\c{c}alves, J.~H. Kim, K.~Kong, and Y.~Wu,
\newblock JHEP {\bf 01}, 158 (2022), 2108.01083.

\bibitem{Buckley_2016_v2}
M.~R. Buckley and D.~Goncalves,
\newblock Phys. Rev. D {\bf 93}, 034003 (2016), 1511.06451.

\bibitem{Buckley_2016}
M.~R. Buckley and D.~Goncalves,
\newblock Phys. Rev. Lett. {\bf 116}, 091801 (2016), 1507.07926.

\bibitem{Mileo_2016}
N.~Mileo, K.~Kiers, A.~Szynkman, D.~Crane, and E.~Gegner,
\newblock JHEP {\bf 07}, 056 (2016), 1603.03632.

\bibitem{Azevedo_2018}
D.~Azevedo, A.~Onofre, F.~Filthaut, and R.~Gon\c{c}alo,
\newblock Phys. Rev. D {\bf 98}, 033004 (2018), 1711.05292.

\bibitem{Azevedo_2020}
D.~Azevedo, R.~Capucha, A.~Onofre, and R.~Santos,
\newblock JHEP {\bf 06}, 155 (2020), 2003.09043.

\bibitem{Azevedo_2021}
D.~Azevedo, R.~Capucha, E.~Gouveia, A.~Onofre, and R.~Santos,
\newblock JHEP {\bf 04}, 077 (2021), 2012.10730.

\bibitem{Azevedo_2022}
D.~Azevedo, R.~Capucha, A.~Onofre, and R.~Santos,
\newblock JHEP {\bf 09}, 246 (2022), 2208.04271.

\bibitem{Li_2018}
J.~Li, Z.-g. Si, L.~Wu, and J.~Yue,
\newblock Phys. Lett. B {\bf 779}, 72 (2018), 1701.00224.

\bibitem{Ferroglia_2019}
A.~Ferroglia, M.~C.~N. Fiolhais, E.~Gouveia, and A.~Onofre,
\newblock Phys. Rev. D {\bf 100}, 075034 (2019), 1909.00490.

\bibitem{Faroughy_2020}
D.~A. Faroughy, J.~F. Kamenik, N.~Ko\v{s}nik, and A.~Smolkovi\v{c},
\newblock JHEP {\bf 02}, 085 (2020), 1909.00007.

\bibitem{Cao:2020hhb}
Q.-H. Cao, K.-P. Xie, H.~Zhang, and R.~Zhang,
\newblock Chin. Phys. C {\bf 45}, 023117 (2021), 2008.13442.

\bibitem{Barman_2022}
R.~K. Barman, D.~Gon\c{c}alves, and F.~Kling,
\newblock Phys. Rev. D {\bf 105}, 035023 (2022), 2110.07635.

\bibitem{Aguilar_Saavedra_2022}
J.~A. Aguilar-Saavedra, M.~C.~N. Fiolhais, P.~Mart\'\i{}n-Ramiro, J.~M. Moreno,
  and A.~Onofre,
\newblock Eur. Phys. J. C {\bf 82}, 134 (2022), 2111.10394.

\bibitem{Casler_2019}
H.~Casler, M.~Manganel, M.~C.~N. Fiolhais, A.~Ferroglia, and A.~Onofre,
\newblock Phys. Rev. D {\bf 99}, 054011 (2019), 1902.01976.

\bibitem{D_liot_2018}
F.~D\'eliot {\em et~al.},
\newblock Phys. Rev. D {\bf 97}, 013007 (2018), 1711.04847.

\bibitem{D_liot_2019}
F.~D\'eliot, M.~C.~N. Fiolhais, and A.~Onofre,
\newblock Mod. Phys. Lett. A {\bf 34}, 1950142 (2019), 1811.02492.

\bibitem{Broggio_2017}
A.~Broggio, A.~Ferroglia, M.~C.~N. Fiolhais, and A.~Onofre,
\newblock Phys. Rev. D {\bf 96}, 073005 (2017), 1707.01803.

\bibitem{Bortolato_2021}
B.~Bortolato, J.~F. Kamenik, N.~Ko\v{s}nik, and A.~Smolkovi\v{c},
\newblock Nucl. Phys. B {\bf 964}, 115328 (2021), 2006.13110.

\bibitem{ATLAS:2021hza}
ATLAS, G.~Aad {\em et~al.},
\newblock JHEP {\bf 04}, 165 (2021), 2102.01444.

\bibitem{ATLAS:2019zrq}
ATLAS, M.~Aaboud {\em et~al.},
\newblock Eur. Phys. J. C {\bf 80}, 754 (2020), 1903.07570.

\bibitem{CMS:2024zqs}
CMS, A.~Hayrapetyan {\em et~al.},
\newblock {Dark sector searches with the CMS experiment}, 2024, 2405.13778.

\bibitem{kentarou2015higher}
M.~Backovi\'c {\em et~al.},
\newblock Eur. Phys. J. C {\bf 75}, 482 (2015), 1508.05327.

\bibitem{Botje:2011sn}
M.~Botje {\em et~al.},
\newblock {The PDF4LHC Working Group Interim Recommendations}, 2011, 1101.0538.

\bibitem{Alloul:2013bka}
A.~Alloul, N.~D. Christensen, C.~Degrande, C.~Duhr, and B.~Fuks,
\newblock Comput. Phys. Commun. {\bf 185}, 2250 (2014), 1310.1921.

\bibitem{Alwall:2014hca}
J.~Alwall {\em et~al.},
\newblock JHEP {\bf 07}, 079 (2014), 1405.0301.

\bibitem{Ball:2012cx}
R.~D. Ball {\em et~al.},
\newblock Nucl. Phys. B {\bf 867}, 244 (2013), 1207.1303.

\bibitem{Artoisenet:2012st}
P.~Artoisenet, R.~Frederix, O.~Mattelaer, and R.~Rietkerk,
\newblock JHEP {\bf 03}, 015 (2013), 1212.3460.

\bibitem{Sjostrand:2006za}
T.~Sjostrand, S.~Mrenna, and P.~Z. Skands,
\newblock JHEP {\bf 05}, 026 (2006), hep-ph/0603175.

\bibitem{deFavereau:2013fsa}
DELPHES 3, J.~de~Favereau {\em et~al.},
\newblock JHEP {\bf 02}, 057 (2014), 1307.6346.

\bibitem{Conte:2012fm}
E.~Conte, B.~Fuks, and G.~Serret,
\newblock Comput. Phys. Commun. {\bf 184}, 222 (2013), 1206.1599.

\bibitem{Cacciari:2011ma}
M.~Cacciari, G.~P. Salam, and G.~Soyez,
\newblock Eur. Phys. J. C {\bf 72}, 1896 (2012), 1111.6097.

\bibitem{Azevedo:2023xuc}
D.~Azevedo {\em et~al.},
\newblock JHEP {\bf 11}, 125 (2023), 2308.00819.

\bibitem{DeRoeck:2024fjq}
A.~De~Roeck,
\newblock Nucl. Phys. B {\bf 1003}, 116480 (2024).

\bibitem{ATLAS:2024rcx}
ATLAS, G.~Aad {\em et~al.},
\newblock JHEP {\bf 03}, 139 (2024), 2401.13430.

\end{thebibliography}

\raggedbottom 
\cleardoublepage
\begin{widetext}
\appendix

\section{2D Exclusion Limits}\label{app:A}

Figures~\ref{fig:comparison_observables_1} and~\ref{fig:comparison_observables_2} illustrate the 2D exclusion limits contour plots in the ($g^{S}_{33}$,$g^{P}_{33}$) plane, obtained using the full distribution shapes of the $b_2$, $\tilde{b}_{2}^{\widehat{y}}$ and $\tilde{b}_{2}^{\widehat{d}}$ (Figure~\ref{fig:comparison_observables_1}) and $n_4$ and $n_2$ (Figure~\ref{fig:comparison_observables_2}), respectively. The corresponding 1D limits are shown in Tables~\ref{tab:exclusion_limits_b2},~\ref{tab:exclusion_limits_b2Y},~\ref{tab:exclusion_limits_b2KD},~\ref{tab:exclusion_limits_n4} and~\ref{tab:exclusion_limits_n2}, for different mediator masses.

\begin{figure}[H]
\begin{center}
\begin{tabular}{ccc}
\hspace*{-10mm}\includegraphics[height=5cm]{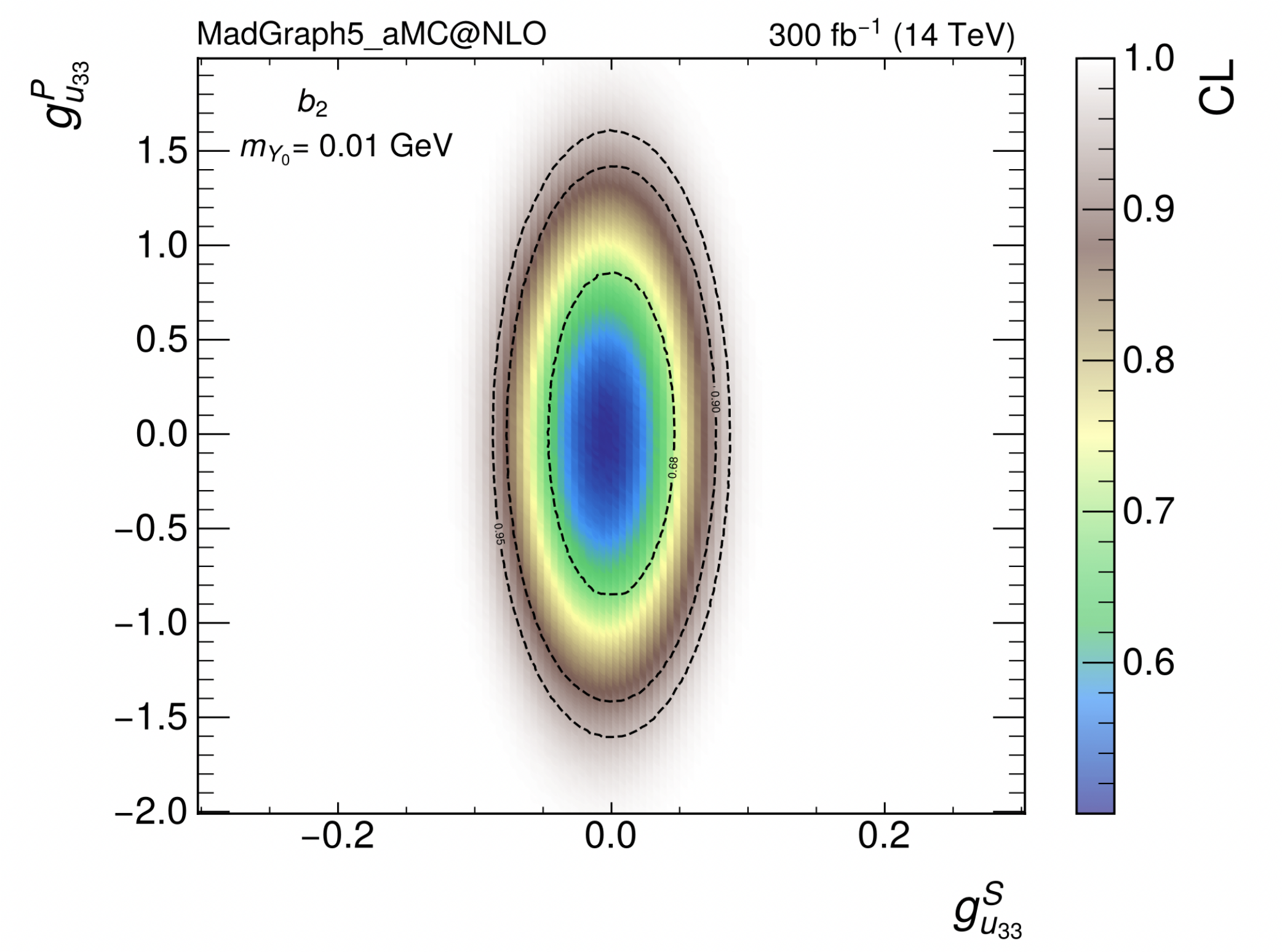} 
\hspace*{-3mm}\includegraphics[height=5cm]{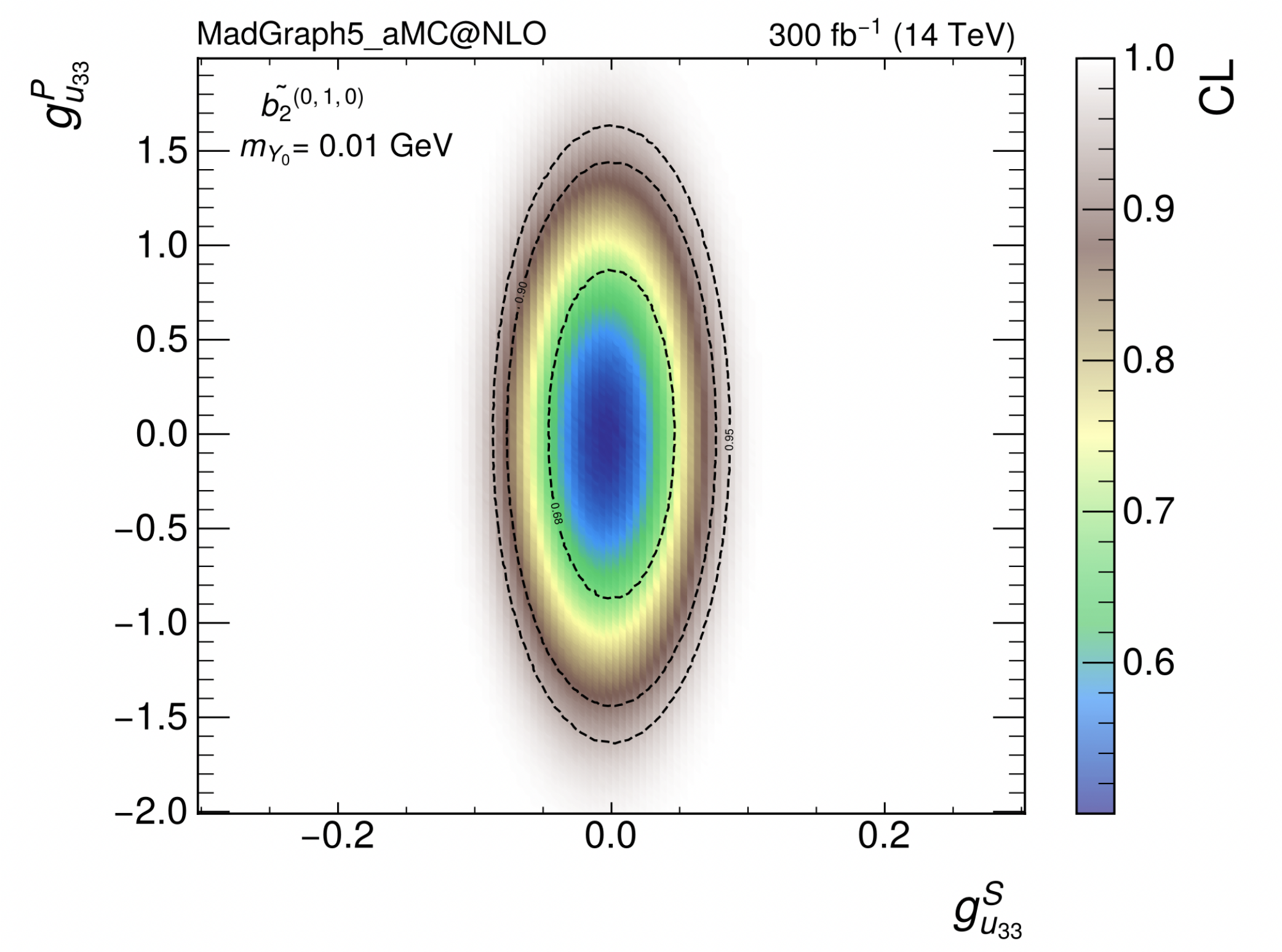} 
\hspace*{-3mm}\includegraphics[height=5cm]{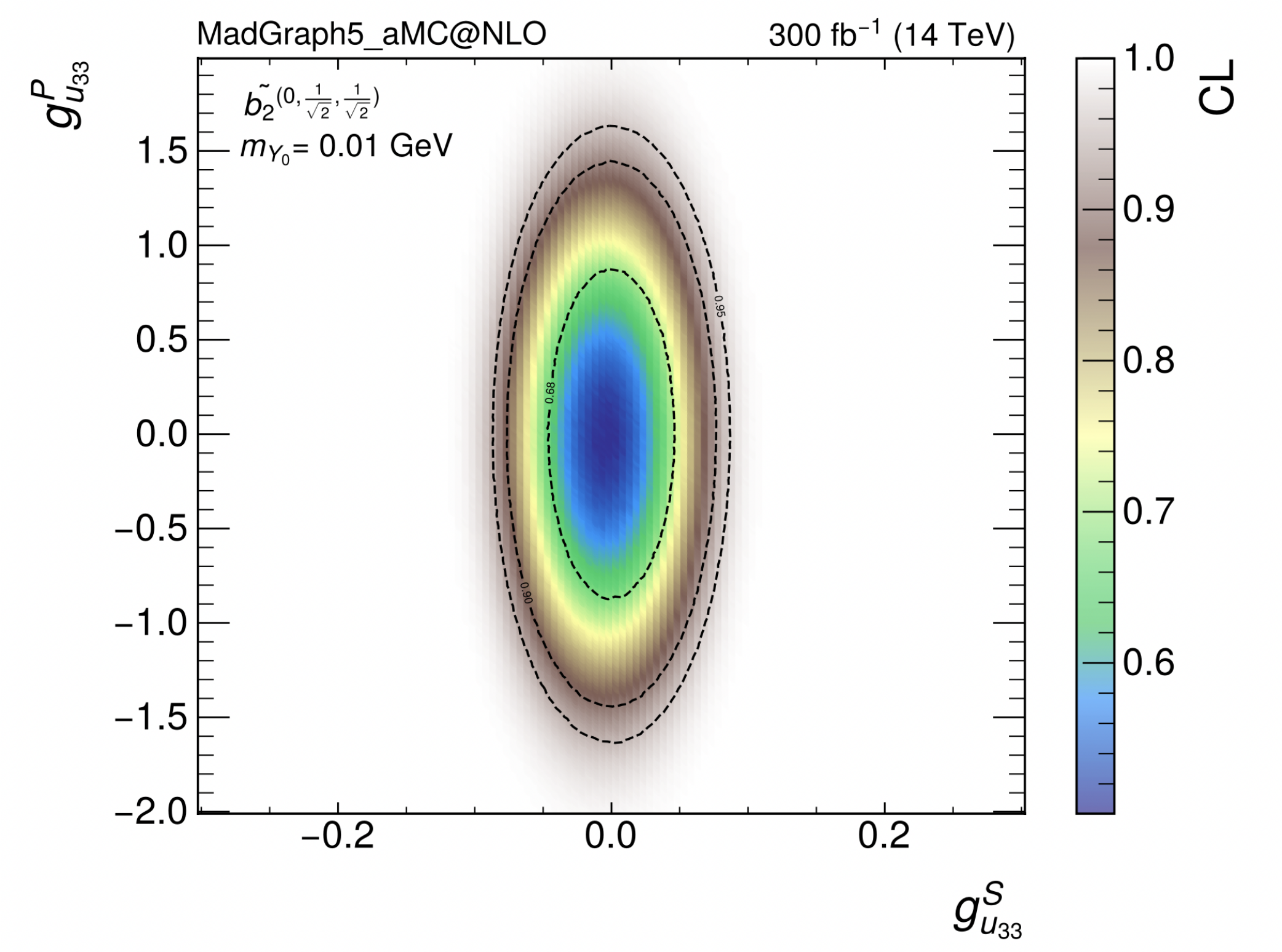}
\end{tabular}
\end{center}
\caption{Exclusion of the SM with an almost massless scalar DM mediator ($Y_0$), including NLO effects with mixed scalar and pseudoscalar Yukawa-like couplings with the top quarks, against the SM as null hypothesis, for the $b_{2}$ observable (left), $\tilde{b}_{2}^{\widehat{y}}$ (center), $\tilde{b}_{2}^{\widehat{d}}$ (right). Limits are shown for a luminosity of $L=300$~fb$^{-1}$.}
\label{fig:comparison_observables_1}
\end{figure}

\begin{figure}[H]
\begin{center}
\begin{tabular}{ccc}
\hspace*{0mm}\includegraphics[height=5.0cm]{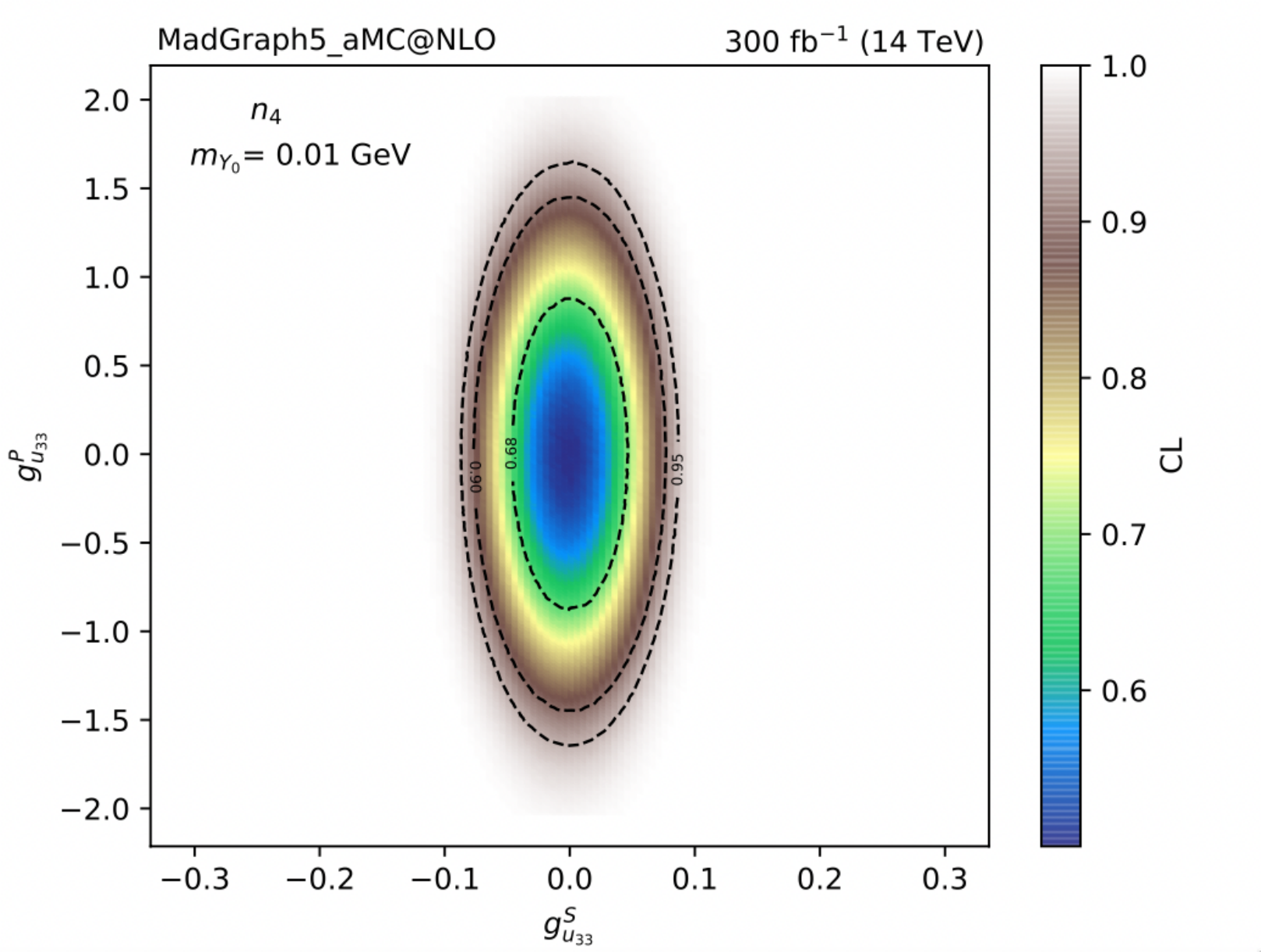} 
\hspace*{0mm}\includegraphics[height=5.22cm]{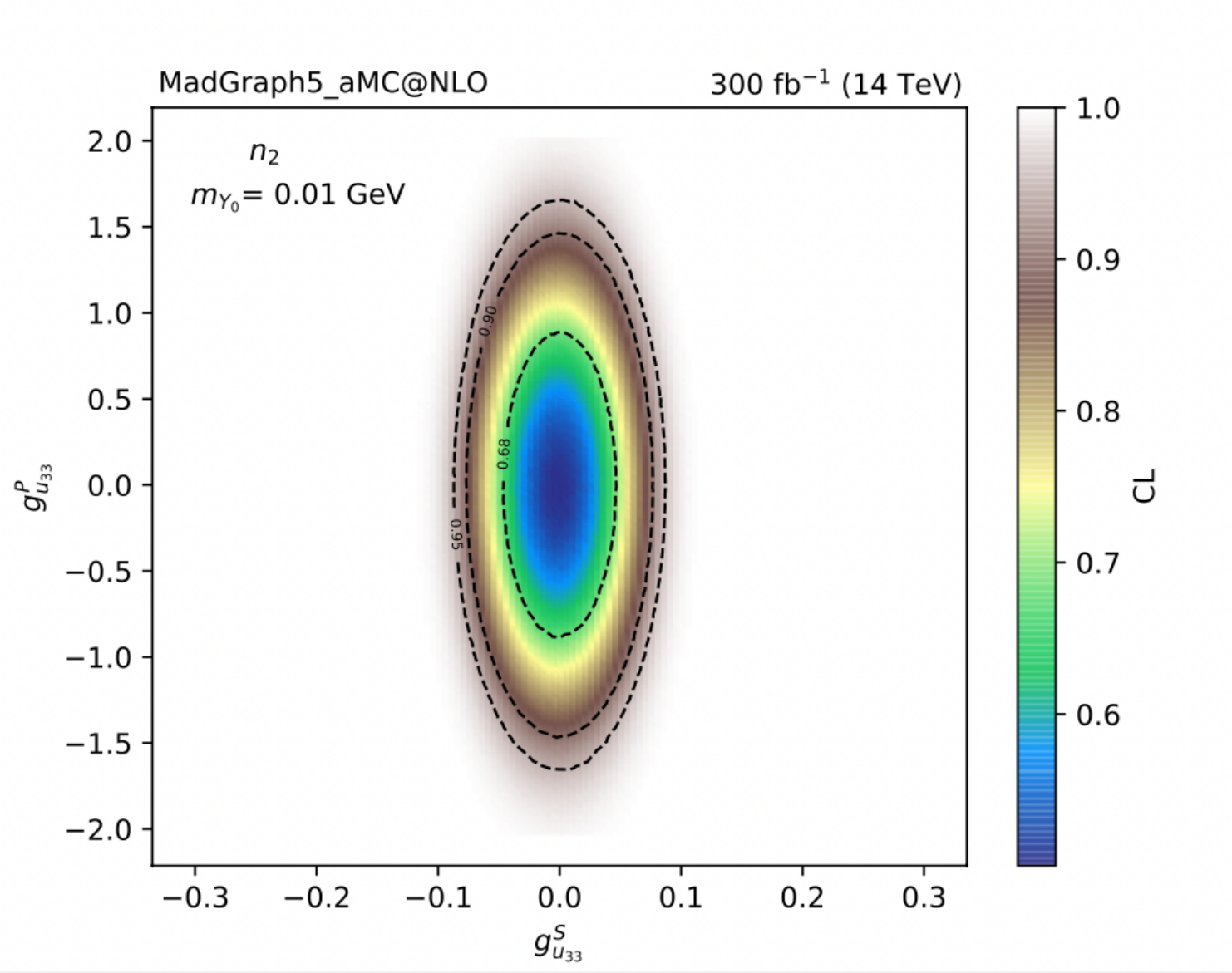}
\end{tabular}
\end{center}
\caption{Exclusion of the SM with an almost massless scalar DM mediator ($Y_0$), including NLO effects with mixed scalar and pseudoscalar Yukawa-like couplings with the top quarks, against the SM as null hypothesis, for both spin observables: $n_{4}$ (left), and $n_{2}$ (right). Limits are shown for a luminosity of $L=300$~fb$^{-1}$.}
\label{fig:comparison_observables_2}
\end{figure}

\cleardoublepage
\section{1D Exclusion Limits}\label{app:B}

In tables~\ref{tab:exclusion_limits_b2},~\ref{tab:exclusion_limits_b2Y},~\ref{tab:exclusion_limits_b2KD},~\ref{tab:exclusion_limits_n4}, and~\ref{tab:exclusion_limits_n2}, we provide a more detailed examination of the 1D limits on the $g^{S/P}_{33}$ coupling constant's, covering a broader range of mediator masses and luminosities for the different observables proposed in this paper.

\begin{table}[H]
  \centering
\begin{ruledtabular}
      \begin{tabular}{cccccc}
  \multicolumn{6}{c}{\textbf{Exclusion Limits from $b_2$}} \\
    \multicolumn{2}{c}{} & \multicolumn{2}{c}{300~fb$^{-1}$} & \multicolumn{2}{c}{3000~fb$^{-1}$} \\
    $m_{Y_0}$ & & \texttt{(68\% CL)} & \texttt{(95\% CL)} & \texttt{(68\% CL)} & \texttt{(95\% CL)} \\
    \hline
    \multirow{2}{*}{$10^{-2}$~GeV} & $g_{u_{33}}^S \in$ & $[-0.0475, 0.0425]$ & $[-0.0875, 0.0875]$ & $[-0.0225, 0.0225]$ & $[-0.0475, 0.0475]$ \\
    & $g_{u_{33}}^P \in$ & $[-0.87, 0.87]$ & $[-1.63, 1.63]$ & $[-0.485, 0.475]$ & $[-0.92, 0.91]$ \\

    \multirow{2}{*}{$1$~GeV} & $g_{u_{33}}^S \in$ & $[-0.0725, 0.0725]$ & $[-0.1425, 0.1425]$ & $[-0.0425, 0.0425]$ & $[-0.0775, 0.0775]$ \\
    & $g_{u_{33}}^P \in$ & $[-0.87, 0.87]$ & $[-1.65, 1.63]$ & $[-0.485, 0.49]$ & $[-0.92, 0.925]$ \\

    \multirow{2}{*}{$10$~GeV} & $g_{u_{33}}^S \in$ & $[-0.1375, 0.1375]$ & $[-0.2575, 0.2625]$ & $[-0.0775, 0.0775]$ & $[-0.1425, 0.1475]$\\
    & $g_{u_{33}}^P \in$ & $[-0.87, 0.89]$ & $[-1.65, 1.67]$ & $[-0.50, 0.49]$ & $[-0.935, 0.94]$  \\

    \multirow{2}{*}{$100$~GeV} & $g_{u_{33}}^S \in$ & $[-0.76, 0.74]$ & $[-1.41, 1.415]$ & $[-0.4125, 0.4275]$ & $[-0.7875, 0.7875]$ \\
    & $g_{u_{33}}^P \in$ & $[-1.25, 1.27]$ & $[-2.37, 2.35]$  & $[-0.70, 0.70]$ & $[-1.325, 1.325]$ \\

    \multirow{2}{*}{$125$~GeV} & $g_{u_{33}}^S \in$ & $[-1.01, 1.015]$ & $[-1.91, 1.915]$ & $[-0.5775, 0.5775]$ & $[-1.0725, 1.0725]$ \\
    & $g_{u_{33}}^P \in$ & $[-1.37, 1.35]$  & $[-2.57, 2.59]$  & $[-0.775, 0.775]$ & $[-1.45, 1.45]$ \\
    \end{tabular}
\end{ruledtabular}
    \caption{Exclusion limits for the $t\Bar{t} Y_0$ CP-couplings, considering the CP-even component against irreducible backgrounds, are depicted for various $Y_0$ masses under fixed luminosities of 300~fb$^{-1}$ and 3000~fb$^{-1}$. The limits are presented at 68\% and 95\% confidence levels for the $b_2$ observable.}
    \label{tab:exclusion_limits_b2}
\end{table}

\begin{table}[H]
  \centering
\begin{ruledtabular}
      \begin{tabular}{cccccc}
  \multicolumn{6}{c}{\textbf{Exclusion Limits from $\tilde{b}_{2}^{\widehat{y}}$}} \\
    \multicolumn{2}{c}{} & \multicolumn{2}{c}{300~fb$^{-1}$} & \multicolumn{2}{c}{3000~fb$^{-1}$} \\
    $m_{Y_0}$ & & \texttt{(68\% CL)} & \texttt{(95\% CL)} & \texttt{(68\% CL)} & \texttt{(95\% CL)} \\
    \hline
    \multirow{2}{*}{$10^{-2}$GeV} & $g_{u_{33}}^S \in$ & $[-0.0425, 0.0425]$ & $[-0.0875, 0.0875]$ & $[-0.0225, 0.0225]$ & $[-0.0475, 0.0475]$ \\
    & $g_{u_{33}}^P \in$ & $[-0.87, 0.87]$ & $[-1.65, 1.65]$ & $[-0.485, 0.49]$ & $[-0.935, 0.925]$ \\

    \multirow{2}{*}{$1$~GeV} & $g_{u_{33}}^S \in$ & $[-0.0725, 0.0725]$ & $[-0.1425, 0.1425]$ & $[-0.0425, 0.0425]$ & $[-0.0775, 0.0775]$ \\
    & $g_{u_{33}}^P \in$ & $[-0.89, 0.89]$ & $[-1.67, 1.67]$ & $[-0.50, 0.49]$ & $[-0.935, 0.94]$ \\

    \multirow{2}{*}{$10$~GeV} & $g_{u_{33}}^S \in$ & $[-0.1375, 0.1375]$ & $[-0.2575, 0.2625]$ & $[-0.0775, 0.0775]$ & $[-0.1475, 0.1425]$\\
    & $g_{u_{33}}^P \in$ & $[-0.89, 0.89]$ & $[-1.69, 1.69]$ & $[-0.50, 0.505]$ & $[-0.95, 0.955]$  \\

    \multirow{2}{*}{$100$~GeV} & $g_{u_{33}}^S \in$ & $[-0.76, 0.765]$ & $[-1.435, 1.44]$ & $[-0.4275, 0.4275]$ & $[-0.8025, 0.8025]$ \\
    & $g_{u_{33}}^P \in$ & $[-1.37, 1.39]$ & $[-2.57, 2.59]$ & $[-0.775, 0.775]$ & $[-1.45, 1.45]$ \\

     \multirow{2}{*}{$125$~GeV} & $g_{u_{33}}^S \in$ & $[-1.06, 1.04]$ & $[-1.985, 1.965]$ & $[-0.5925, 0.5925]$ & $[-1.1025, 1.1025]$ \\
    & $g_{u_{33}}^P \in$ & $[-1.57, 1.55]$ & $[-2.93, 2.91]$ & $[-0.875, 0.875]$ & $[-1.625, 1.625]$ \\
    \end{tabular}
\end{ruledtabular}
    \caption{Exclusion limits for the $t\Bar{t} Y_0$ CP-couplings, considering the \textit{Scenario 1}, are depicted for various $Y_0$ masses under fixed luminosities of 300~fb$^{-1}$ and 3000~fb$^{-1}$. The limits are presented at 68\% and 95\% confidence levels for the $\tilde{b}_{2}^{\widehat{y}}$ observable.}
    \label{tab:exclusion_limits_b2Y}
\end{table}

\begin{table}[H]
  \centering
\begin{ruledtabular}
     \begin{tabular}{cccccc}
  \multicolumn{6}{c}{\textbf{Exclusion Limits from $\tilde{b}_{2}^{\widehat{d}}$}} \\
    \multicolumn{2}{c}{} & \multicolumn{2}{c}{300~fb$^{-1}$} & \multicolumn{2}{c}{3000~fb$^{-1}$} \\
    $m_{Y_0}$ & & \texttt{(68\% CL)} & \texttt{(95\% CL)} & \texttt{(68\% CL)} & \texttt{(95\% CL)} \\
    \hline
   \multirow{2}{*}{$10^{-2}$~GeV} & $g_{u_{33}}^S \in$ & $[-0.0475, 0.0425]$ & $[-0.0875, 0.0875]$ & $[-0.0225, 0.0225]$ & $[-0.0475, 0.0475]$ \\
    & $g_{u_{33}}^P \in$ & $[-0.89, 0.87]$ & $[-1.65, 1.65]$ & $[-0.485, 0.49]$ & $[-0.935, 0.925]$ \\

    \multirow{2}{*}{$1$~GeV} & $g_{u_{33}}^S \in$ & $[-0.0725, 0.0725]$ & $[-0.1425, 0.1425]$ & $[-0.0425, 0.0425]$ & $[-0.0775, 0.0775]$ \\
    & $g_{u_{33}}^P \in$ & $[-0.89, 0.91]$ & $[-1.67, 1.67]$ & $[-0.50, 0.49]$ & $[-0.935, 0.94]$ \\

   \multirow{2}{*}{$10$~GeV} & $g_{u_{33}}^S \in$ & $[-0.1375, 0.1375]$ & $[-0.2625, 0.2575]$ & $[-0.0775, 0.0775]$ & $[-0.1425, 0.1475]$\\
    & $g_{u_{33}}^P \in$ & $[-0.89, 0.91]$ & $[-1.69, 1.69]$ & $[-0.50, 0.505]$ & $[-0.95, 0.955]$  \\

    \multirow{2}{*}{$100$~GeV} & $g_{u_{33}}^S \in$ & $[-0.76, 0.765]$ & $[-1.435, 1.44]$ & $[-0.4275, 0.4275]$ & $[-0.8025, 0.8025]$ \\
    & $g_{u_{33}}^P \in$ & $[-1.37, 1.39]$ & $[-2.61, 2.59]$ & $[-0.775, 0.775]$ & $[-1.45, 1.45]$ \\

    \multirow{2}{*}{$125$~GeV} & $g_{u_{33}}^S \in$ & $[-1.035, 1.04]$ & $[-1.96, 1.965]$ & $[-0.5925, 0.5925]$ & $[-1.1025, 1.1025]$ \\
    & $g_{u_{33}}^P \in$ & $[-1.53, 1.55]$ & $[-2.89, 2.91]$ & $[-0.875, 0.875]$ & $[-1.625, 1.625]$ \\
   \end{tabular}
\end{ruledtabular}
    \caption{Exclusion limits for the $t\Bar{t} Y_0$ CP-couplings, considering the \textit{Scenario 1}, are depicted for various $Y_0$ masses under fixed luminosities of 300~fb$^{-1}$ and 3000~fb$^{-1}$. The limits are presented at 68\% and 95\% confidence levels for the $\tilde{b}_{2}^{\widehat{y}}$  observable.}
    \label{tab:exclusion_limits_b2KD}
\end{table}

\begin{table}[H]
  \centering
\begin{ruledtabular}
      \begin{tabular}{cccccc}
  \multicolumn{6}{c}{\textbf{Exclusion Limits from $n_4$}} \\
   
    \multicolumn{2}{c}{} & \multicolumn{2}{c}{300~fb$^{-1}$} & \multicolumn{2}{c}{3000~fb$^{-1}$} \\
    
    $m_{Y_0}$ & & \texttt{(68\% CL)} & \texttt{(95\% CL)} & \texttt{(68\% CL)} & \texttt{(95\% CL)} \\
    \hline
    \multirow{2}{*}{$10^{-2}$~GeV} & $g_{u_{33}}^S \in$ & $[-0.0425, 0.0425]$ & $[-0.0875, 0.0875]$ & $[-0.0225, 0.0225]$ & $[-0.0475, 0.0475]$ \\
    & $g_{u_{33}}^P \in$ & $[-0.8700, 0.8900]$ & $[-1.6700, 1.6700]$ & $[-0.5025, 0.4875]$  & $[-0.9375, 0.9375]$ \\

    \multirow{2}{*}{$1$~GeV} & $g_{u_{33}}^S \in$ & $[-0.0725, 0.0725]$ & $[-0.1425, 0.1425]$ &$[-0.0425, 0.0425]$ & $[-0.0775, 0.0775]$ \\
    & $g_{u_{33}}^P \in$ & $[-0.8700, 0.8900]$ & $[-1.6500, 1.6500]$ & $[-0.4875, 0.4875]$ &  $[-0.9225, 0.9225]$\\

    \multirow{2}{*}{$10$~GeV} & $g_{u_{33}}^S \in$ & $[-0.1375, 0.1375]$ & $[-0.2575, 0.2575]$ & $[-0.0775, 0.0775]$ & $[-0.1475, 0.1425]$ \\
    & $g_{u_{33}}^P \in$ & $[-0.9100, 0.8900]$ & $[-1.7100, 1.6900]$ & $[-0.5025, 0.5025]$ & $[-0.9525, 0.9525]$ \\

    \multirow{2}{*}{$100$~GeV} & $g_{u_{33}}^S \in$ & $[-0.7600, 0.7650]$ & $[-1.4350, 1.4400]$ & $[-0.4275, 0.4275]$&  $[-0.8025, 0.8025]$ \\
    & $g_{u_{33}}^P \in$ & $[-1.3700, 1.3500]$ & $[-2.5700, 2.5900]$ & $[-0.7750, 0.7750]$ &  $[-1.4500, 1.4500]$\\   

    \multirow{2}{*}{$125$~GeV} & $g_{u_{33}}^S \in$ & $[-1.0600, 1.0400]$ & $[-1.9850, 1.965]$ & $[-0.5925, 0.5925]$ & $[-1.1175, 1.1175]$ \\
    & $g_{u_{33}}^P \in$ & $[-1.5300, 1.5500]$ & $[-2.8900, 2.8700]$ & $[-0.8500, 0.8500]$ &  $[-1.6250, 1.6250]$\\   
    \end{tabular}
\end{ruledtabular}
    \caption{Exclusion limits for the $t\Bar{t} Y_0$ CP-couplings, considering the CP-mixed case against irreducible backgrounds, are depicted for various $Y_0$ masses under fixed luminosities of 300~fb$^{-1}$ and 3000~fb$^{-1}$. The limits are presented at 68\% and 95\% confidence levels for the $n_4$ observable.}
    \label{tab:exclusion_limits_n4}
\end{table}

\begin{table}[H]
  \centering
\begin{ruledtabular}
      \begin{tabular}{cccccc}
  \multicolumn{6}{c}{\textbf{Exclusion Limits from $n_2$}} \\
    \multicolumn{2}{c}{} & \multicolumn{2}{c}{300~fb$^{-1}$} & \multicolumn{2}{c}{3000~fb$^{-1}$} \\
    $m_{Y_0}$ & & \texttt{(68\% CL)} & \texttt{(95\% CL)} & \texttt{(68\% CL)} & \texttt{(95\% CL)} \\
    \hline
    \multirow{2}{*}{$10^{-2}$~GeV} & $g_{u_{33}}^S \in$ & $[-0.0425, 0.0475]$ & $[-0.0875, 0.0875]$ & $[-0.0225, 0.0225]$ & $[-0.0475, 0.0475]$ \\
    & $g_{u_{33}}^P \in$ & $[-0.8900, 0.8900]$ & $[-1.6700, 1.6700]$ & $[-0.4875, 0.4875]$ & $[-0.9375, 0.9375]$ \\

    \multirow{2}{*}{$1$~GeV} & $g_{u_{33}}^S \in$ & $[-0.0725, 0.0725]$ & $[-0.1425, 0.1425]$ & $[-0.0425, 0.0425]$ & $[-0.0775, 0.0775]$ \\
    & $g_{u_{33}}^P \in$ & $[-0.8900, 0.8900]$ & $[-1.6700, 1.6700]$ & $[-0.5025, 0.4875]$ &  $[-0.9375, 0.9375]$ \\

    \multirow{2}{*}{$10$~GeV} & $g_{u_{33}}^S \in$ & $[-0.1375, 0.1375]$ & $[-0.2575, 0.2575]$ & $[-0.0775, 0.0775]$ & $[-0.1425, 0.1425]$ \\
    & $g_{u_{33}}^P \in$ & $[-0.9100, 0.9100]$ & $[-1.7100, 1.7100]$ & $[-0.5025, 0.5175]$  & $[-0.9525, 0.9675]$ \\

    \multirow{2}{*}{$100$~GeV} & $g_{u_{33}}^S \in$ & $[-0.7600, 0.7650]$ & $[-1.4350, 1.4400]$ & $[-0.4275, 0.4275]$& $[-0.8025, 0.8025]$ \\
    & $g_{u_{33}}^P \in$ & $[-1.4100, 1.3900]$ & $[-1.4100, 1.3900]$ & $[-0.7750, 0.8000]$ &  $[-1.4750, 1.4750]$\\

    \multirow{2}{*}{$125$~GeV} & $g_{u_{33}}^S \in$ & $[-1.0350, 1.0400]$ & $[-1.9850, 1.990]$ & $[-0.5925, 0.5925]$&  $[-1.1025, 1.1175]$ \\
    & $g_{u_{33}}^P \in$ & $[-1.5700, 1.5900]$ & $[-2.9700, 2.9500]$ & $[-0.8750, 0.8750]$ & $[-1.6750, 1.6750]$ \\
    \end{tabular}
\end{ruledtabular}
    \caption{Exclusion limits for the $t\Bar{t} Y_0$ CP-couplings, considering the CP-mixed case against irreducible backgrounds, are depicted for various $Y_0$ masses under fixed luminosities of 300~fb$^{-1}$ and 3000~fb$^{-1}$. The limits are presented at 68\% and 95\% confidence levels for the $n_2$ observable.}
    \label{tab:exclusion_limits_n2}
\end{table}

\section{Forward-Backward asymmetries}\label{app:C}

In tables~\ref{tab:exclusion_limits_b2FB_appendix},~\ref{tab:exclusion_limits_b2YFB},~\ref{tab:exclusion_limits_b2KDFB},~\ref{tab:exclusion_limits_b2KDFB2_appendix}, and~\ref{tab:exclusion_limits_b2YFB2}, we show the $g^{S/P}_{33}$ coupling constant's exclusion limits computed using the 2-binned distributions used for the calculation of the forward-backward asymmetries. 

\begin{table}[H]
  \centering
\begin{ruledtabular}
      \begin{tabular}{cccccc}
  \multicolumn{6}{c}{\textbf{Exclusion Limits from $b_2$ Asymmetries FB}} \\
    \multicolumn{2}{c}{} & \multicolumn{2}{c}{300~fb$^{-1}$} & \multicolumn{2}{c}{3000~fb$^{-1}$} \\
    $m_{Y_0}$ & & \texttt{(68\% CL)} & \texttt{(95\% CL)} & \texttt{(68\% CL)} & \texttt{(95\% CL)} \\
    \hline
    \multirow{2}{*}{$10^{-2}$~GeV} & $g_{u_{33}}^S \in$ & $[-0.0425, 0.0475]$ & $[-0.0875, 0.0875]$ & $[-0.0225, 0.0225]$ & $[-0.0475, 0.0475]$ \\
    & $g_{u_{33}}^P \in$ & $[-0.83, 0.83]$ & $[-1.57, 1.57]$ & $[-0.4725, 0.4575]$ & $[-0.8775, 0.8925]$ \\

    \multirow{2}{*}{$10$~GeV} & $g_{u_{33}}^S \in$ & $[-0.1375, 0.1375]$ & $[-0.2575, 0.2625]$ & $[-0.0775, 0.0775]$ & $[-0.1425, 0.1475]$\\
    & $g_{u_{33}}^P \in$ & $[-0.85, 0.85]$ & $[-1.61, 1.61]$ & $[-0.4725, 0.4725]$ & $[-0.8925, 0.8925]$  \\

    \multirow{2}{*}{$125$~GeV} & $g_{u_{33}}^S \in$ & $[-1.01, 1.015]$ & $[-1.885, 1.89]$ & $[-0.5625, 0.5625]$ & $[-1.0575, 1.0575]$ \\
    & $g_{u_{33}}^P \in$ & $[-1.29, 1.27]$  & $[-2.41, 2.43]$  & $[-0.725, 0.725]$ & $[-1.35, 1.375]$ \\
    \end{tabular}
\end{ruledtabular}
    \caption{Exclusion limits for the $t\Bar{t} Y_0$ CP-couplings, considering the CP-mixed case against irreducible backgrounds, are depicted for various $Y_0$ masses under fixed luminosities of 300~fb$^{-1}$ and 3000~fb$^{-1}$. The limits are presented at 68\% and 95\% confidence levels.}
    \label{tab:exclusion_limits_b2FB_appendix}
\end{table}

\begin{table}[H]
  \centering
\begin{ruledtabular}
      \begin{tabular}{cccccc}
  \multicolumn{6}{c}{\textbf{Exclusion Limits from $\tilde{b}_{2}^{\widehat{y}}$ Asymmetries FB}} \\
    \multicolumn{2}{c}{} & \multicolumn{2}{c}{300~fb$^{-1}$} & \multicolumn{2}{c}{3000~fb$^{-1}$} \\
    $m_{Y_0}$ & & \texttt{(68\% CL)} & \texttt{(95\% CL)} & \texttt{(68\% CL)} & \texttt{(95\% CL)} \\
    \hline
    \multirow{2}{*}{$10^{-2}$~GeV} & $g_{u_{33}}^S \in$ & $[-0.0425, 0.0425]$ & $[-0.0875, 0.0875]$ & $[-0.0225, 0.0225]$ & $[-0.0475, 0.0475]$ \\
    & $g_{u_{33}}^P \in$ & $[-0.87, 0.87]$ & $[-1.65, 1.67]$ & $[-0.4875, 0.4875]$ & $[-0.9375, 0.9225]$ \\

    \multirow{2}{*}{$10$~GeV} & $g_{u_{33}}^S \in$ & $[-0.1375, 0.1375]$ & $[-0.2575, 0.2625]$ & $[-0.0775, 0.0775]$ & $[-0.1475, 0.1475]$\\
    & $g_{u_{33}}^P \in$ & $[-0.89, 0.91]$ & $[-1.71, 1.69]$ & $[-0.5025, 0.5025]$ & $[-0.9525, 0.9525]$  \\

     \multirow{2}{*}{$125$~GeV} & $g_{u_{33}}^S \in$ & $[-1.06, 1.065]$ & $[-1.985, 1.99]$ & $[-0.5925, 0.5925]$ & $[-1.1175, 1.1175]$ \\
    & $g_{u_{33}}^P \in$ & $[-1.57, 1.55]$  & $[-2.93, 2.91]$  & $[-0.875, 0.875]$ & $[-1.65, 1.65]$ \\
    \end{tabular}
\end{ruledtabular}
    \caption{Exclusion limits for the $t\Bar{t} Y_0$ CP-couplings, considering the CP-mixed case against irreducible backgrounds, are depicted for various $Y_0$ masses under fixed luminosities of 300~fb$^{-1}$ and 3000~fb$^{-1}$. The limits are presented at 68\% and 95\% confidence levels.}
    \label{tab:exclusion_limits_b2YFB}
\end{table}

\begin{table}[H]
  \centering
\begin{ruledtabular}
      \begin{tabular}{cccccc}
  \multicolumn{6}{c}{\textbf{Exclusion Limits from $\tilde{b}_{2}^{\widehat{d}}$ Asymmetries FB}} \\
    \multicolumn{2}{c}{} & \multicolumn{2}{c}{300~fb$^{-1}$} & \multicolumn{2}{c}{3000~fb$^{-1}$} \\
    $m_{Y_0}$ & & \texttt{(68\% CL)} & \texttt{(95\% CL)} & \texttt{(68\% CL)} & \texttt{(95\% CL)} \\
    \hline
    \multirow{2}{*}{$10^{-2}$~GeV} & $g_{u_{33}}^S \in$ & $[-0.0422, 0.0474]$ & $[-0.0876, 0.0876]$ & $[-0.0225, 0.0225]$ & $[-0.0475, 0.0475]$ \\
    & $g_{u_{33}}^P \in$ & $[-0.87, 0.87]$ & $[-1.67, 1.67]$ & $[-0.4875, 0.4875]$ & $[-0.9375, 0.9375]$ \\

    \multirow{2}{*}{$10$~GeV} & $g_{u_{33}}^S \in$ & $[-0.138, 0.138]$ & $[-0.263, 0.263]$ & $[-0.078, 0.078]$ & $[-0.148, 0.148]$\\
    & $g_{u_{33}}^P \in$ & $[-0.92, 0.92]$ & $[-1.71, 1.71]$ & $[-0.5025, 0.5025]$ & $[-0.9525, 0.9525]$  \\

    \multirow{2}{*}{$125$~GeV} & $g_{u_{33}}^S \in$ & $[-1.06, 1.065]$ & $[-1.985, 1.99]$ & $[-0.5925, 0.5925]$ & $[-1.1175, 1.1175]$ \\
    & $g_{u_{33}}^P \in$ & $[-1.57, 1.55]$  & $[-2.93, 2.95]$  & $[-0.875, 0.875]$ & $[-1.65, 1.65]$ \\
     \end{tabular}
\end{ruledtabular}
    \caption{Exclusion limits for the $t\Bar{t} Y_0$ CP-couplings, considering the CP-mixed case against irreducible backgrounds, are depicted for various $Y_0$ masses under fixed luminosities of 300~fb$^{-1}$ and 3000~fb$^{-1}$. The limits are presented at 68\% and 95\% confidence levels.}
    \label{tab:exclusion_limits_b2KDFB}
\end{table}

\begin{table}[H]
  \centering
\begin{ruledtabular}
      \begin{tabular}{cccccc}
  \multicolumn{6}{c}{\textbf{Exclusion Limits from $n_4$ Asymmetries FB}} \\
    \multicolumn{2}{c}{} & \multicolumn{2}{c}{300~fb$^{-1}$} & \multicolumn{2}{c}{3000~fb$^{-1}$} \\
    $m_{Y_0}$ & & \texttt{(68\% CL)} & \texttt{(95\% CL)} & \texttt{(68\% CL)} & \texttt{(95\% CL)} \\
    \hline
    \multirow{2}{*}{$10^{-2}$~GeV} & $g_{u_{33}}^S \in$ & $[-0.0450, 0.0450]$ & $[-0.0875, 0.0875]$ & $[-0.0225, 0.0225]$ & $[-0.0475, 0.0475]$ \\
    & $g_{u_{33}}^P \in$ & $[-0.8900, 0.8900]$ & $[-1.6700, 1.6700]$ & $[-0.5025, 0.5025]$  & $[-0.9375, 0.9375]$ \\
    
    \multirow{2}{*}{$10$~GeV} & $g_{u_{33}}^S \in$ & $[-0.135, 0.135]$ & $[-0.255, 0.255]$ & $[-0.075, 0.075]$ & $[-0.1475, 0.1425]$ \\
    & $g_{u_{33}}^P \in$ & $[-0.9100, 0.8900]$ & $[-1.7100, 1.6900]$ & $[-0.5025, 0.5025]$ & $[-0.9525, 0.9525]$ \\
    
    \multirow{2}{*}{$125$~GeV} & $g_{u_{33}}^S \in$ & $[-1.05, 1.05]$ & $[-1.965, 1.965]$ & $[-0.593, 0.593]$ & $[-1.118, 1.118]$ \\
    & $g_{u_{33}}^P \in$ & $[-1.53, 1.55]$ & $[-2.89, 2.87]$ & $[-0.85, 0.85]$ &  $[-1.63, 1.63]$\\
    \end{tabular}
\end{ruledtabular}
    \caption{Exclusion limits for the $t\Bar{t} Y_0$ CP-couplings, considering the CP-mixed case against irreducible backgrounds, are depicted for various $Y_0$ masses under fixed luminosities of 300~fb$^{-1}$ and 3000~fb$^{-1}$. The limits are presented at 68\% and 95\% confidence levels for the $n_4$ observable.}
    \label{tab:exclusion_limits_b2KDFB2_appendix}
\end{table}

\begin{table}[H]
  \centering
\begin{ruledtabular}
      \begin{tabular}{cccccc}
  \multicolumn{6}{c}{\textbf{Exclusion Limits from ${n}_2$ Asymmetries FB}} \\
    \multicolumn{2}{c}{} & \multicolumn{2}{c}{300~fb$^{-1}$} & \multicolumn{2}{c}{3000~fb$^{-1}$} \\
    $m_{Y_0}$ & & \texttt{(68\% CL)} & \texttt{(95\% CL)} & \texttt{(68\% CL)} & \texttt{(95\% CL)} \\
    \hline
    \multirow{2}{*}{$10^{-2}$~GeV} & $g_{u_{33}}^S \in$ & $[-0.0425, 0.0475]$ & $[-0.0875, 0.0875]$ & $[-0.0225, 0.0225]$ & $[-0.0475, 0.0475]$ \\
    & $g_{u_{33}}^P \in$ & $[-0.8900, 0.8900]$ & $[-1.6700, 1.6700]$ & $[-0.4875, 0.4875]$ & $[-0.9375, 0.9375]$ \\
    \multirow{2}{*}{$10$~GeV} & $g_{u_{33}}^S \in$ & $[-0.1375, 0.1375]$ & $[-0.2575, 0.2575]$ & $[-0.0775, 0.0775]$ & $[-0.1425, 0.1425]$ \\
    & $g_{u_{33}}^P \in$ & $[-0.9100, 0.9100]$ & $[-1.7100, 1.7100]$ & $[-0.5025, 0.5175]$  & $[-0.9525, 0.9675]$ \\
     \multirow{2}{*}{$125$~GeV} & $g_{u_{33}}^S \in$ & $[-1.0350, 1.0400]$ & $[-1.9850, 1.990]$ & $[-0.5925, 0.5925]$&  $[-1.1025, 1.1175]$ \\
    & $g_{u_{33}}^P \in$ & $[-1.5700, 1.5900]$ & $[-2.9700, 2.9500]$ & $[-0.8750, 0.8750]$ & $[-1.6750, 1.6750]$ \\
    \end{tabular}
\end{ruledtabular}
    \caption{Exclusion limits for the $t\Bar{t} Y_0$ CP-couplings, considering the CP-mixed case against irreducible backgrounds, are depicted for various $Y_0$ masses under fixed luminosities of 300~fb$^{-1}$ and 3000~fb$^{-1}$. The limits are presented at 68\% and 95\% confidence levels for the $n_2$ observable.}
    \label{tab:exclusion_limits_b2YFB2}
\end{table}
\end{widetext}

\end{document}